\documentclass[english, 12pt, a4paper]{article}
\usepackage[T1]{fontenc}
\usepackage[utf8]{inputenc}
\usepackage{natbib}
\usepackage[english]{babel}
\usepackage{lmodern}
\usepackage[top=1.25in, bottom=1.25in, left=1.25in, right=1.25in]{geometry}
\usepackage[pdftex]{graphicx}
\usepackage{verbatim}
\usepackage{amsmath,amssymb,amsopn,dsfont,mathrsfs,bm}
\usepackage{float,placeins,flafter,longtable,array,booktabs}
\usepackage{color}
\usepackage{booktabs}
\usepackage{threeparttable}
\usepackage{blindtext}
\usepackage[hidelinks]{hyperref}

\usepackage{xr}
\makeatletter
\newtheorem{thm}{Theorem}

\newtheorem{prop}{Proposition}
\newtheorem{lem}{Lemma}
\newtheorem{hyp}{Assumption}

\newcommand{\ind}[1]{\mathds{1}\left\{#1\right\}}
\renewcommand{\S}{\mathcal{S}}
\newcommand{\B}{\mathcal{B}}

\newcommand{\E}{E}

\newcommand{\R}{\mathbb R}

\newcommand{\eps}{\varepsilon}

\newcommand{\Deriv}[2]{\frac{\partial #1}{\partial #2}}

\newcommand{\sgn}{\text{sgn}}

\newcommand{\Cov}{\text{Cov}}
\newcommand{\Corr}{\text{Corr}}
\newcommand{\Supp}{\text{Supp}}
\newcommand{\indep}{\perp \!\!\! \perp}
\def\CI{\text{CI}_{1-\alpha}}

\newcommand{\convP}{\stackrel{P}{\longrightarrow}}

\newcommand{\convNor}[1]{\stackrel{d}{\longrightarrow} \mathcal{N}\left(0,#1\right)}

\newcommand{\diam}{\mathrm{diam}}
\newcommand{\conv}{\mathrm{conv}}

\allowdisplaybreaks

\date{}

\begin{document}

\title{Linear Regressions with Combined Data\thanks{First Version: December 6, 2024. We thank Pat Bayer, Christian Bontemps, Stephen Hansen, Marc Henry, Toru Kitagawa, Matt Masten, Robert Moffitt, David Pacini, Davide \mbox{Viviano}, Daniel Wilhelm, and participants at the Encounters in Econometric Theory 2024, the Munich Econometrics Workshop 2024, 2024 ESEM, LMU, Penn State University, the Aarhus Workshop in Econometrics V, the 35th (EC$)^2$ conference in Amsterdam, Toulouse School of Economics, University of Geneva (Statistics), University of Glasgow, University of Gothenburg, University of Manchester, Berlin Applied Micro seminar, the Workshop on Optimal Transport in Econometrics (Carlo Alberto, 2025), the 2025 IAAE Annual Conference, the 2025 Parisian Econometrics Workshop and the 2026 Optimization-Conscious Econometrics Conference (Rice University in Paris). We thank Lavinia Kinne and Ludger Woessmann who kindly shared the PISA sample used in our paper, and Yizhi Su and Haonan Ye for capable research assistance.}}

\author{Xavier D'Haultf{\oe}uille\thanks{CREST-ENSAE and Institut Polytechnique de Paris, xavier.dhaultfoeuille@ensae.fr.} \and Christophe Gaillac\thanks{University of Geneva, GSEM-IEE, christophe.gaillac@unige.ch.} \and Arnaud Maurel\thanks{Duke University, NBER, CESifo and IZA, arnaud.maurel@duke.edu.}}

\maketitle
\begin{abstract}
We study linear regressions in a context where the outcome of interest and some of the covariates are observed in two different datasets that cannot be matched. Traditional approaches obtain point identification by relying, often implicitly, on exclusion restrictions. We show that without such restrictions, coefficients of interest can still be partially identified, with the sharp bounds taking a simple form. We obtain tighter bounds in the presence of variables observed in both datasets but not included in the regression of interest, without imposing additional restrictions on these variables. We develop computationally simple and asymptotically normal estimators of the bounds. Finally, we apply our methodology to estimate racial disparities in patent approval rates and to evaluate the effect of patience and risk-taking on educational performance. 

\smallskip


\end{abstract}

\newpage

\section{Introduction}

It is  often impossible to run the ideal regressions one would like to consider. A common reason for this is that the outcome $Y$ and covariates $X$ of interest are not observed in the same dataset. For instance, in many intergenerational studies (e.g., intergenerational income or wealth mobility), one cannot link parents' and children's outcomes. Even if the outcome and covariates of interest do appear in the same dataset, key control variables are routinely missing. For instance, when measuring the wage returns to education, one may wish to control for a measure of cognitive skills, but such a measure may not be available in the main labor market dataset, even though it appears in another data source. In this paper, we develop new econometric tools allowing empirical researchers to learn about linear regression coefficients in such environments. 

\medskip
To better explain our contribution in this context, we first detail our setup. We assume that $X$ includes two sets of covariates: ``outside'' regressors $X_o$, which only appear in a separate dataset from that including the outcome $Y$, and ``common'' regressors $X_c$, which appear in both datasets. We also consider auxiliary variables, $W_a$, which researchers do not seek to include in the regression but that also appear in both datasets. These types of auxiliary variables are often available to empirical researchers. For instance, if a common variable is a proxy for a variable of interest $X_o$, or a so-called ``bad control'', it seems preferable to focus on the regression of $Y$ on $X_o$, without controlling for that variable. We denote by $W$ the set of common variables, included or not in the regression, so that $W=(X_c', W_a')'$. 

\medskip
In this context,  empirical researchers have traditionally relied on imputation methods. The most common, which corresponds to two-sample two-stage least squares (TSTSLS), consists in first predicting $X_o$ by $W_a$ in the ``$X_o$ dataset'', and then using this prediction in the ``$Y$ dataset''. Importantly, even if such methods do not aim at identifying causal effect, but simply regression coefficients, they still implicitly rely on exclusion restrictions. For instance, imputation based on TSTSLS requires the coefficient of $W_a$ in the infeasible regression of $Y$ on $X$ and $W_a$ to be 0. The resulting estimator is therefore potentially sensitive to their violation. 

\medskip

In this paper, we take another route and instead ask what can be learned about the regression coefficients without imposing such exclusion restrictions. We study identification and develop tools for tractable estimation and inference in this setting.

\medskip
We first derive sharp bounds on these coefficients in the absence of common variables. To do so, we combine the Frisch-Waugh-Lovell theorem with the Cambanis-Simons-Stout inequality. Establishing sharpness in this setting is nontrivial when $X_o$ is multivariate, and constitutes a first theoretical contribution of the paper. We then extend this result to account for common variables. Sharp bounds still take a simple form in this case. Moreover, by leveraging the variation in $W$, one may be able to identify the sign of regression coefficients, and even obtain point identification in special cases. 

\medskip
Importantly for practical purposes, we also show that it is possible to construct a test for, and in some cases reject the exclusion restriction underlying the widely used imputation based on TSTSLS, even when the dimension of $W_a$ does not exceed that of $X_o$. This may seem surprising in light of the well-known non-testability of instrument validity in linear regressions with endogeneity. The key point, however, is that the two setups differ not only in the observed data but also in the parameters of interest. In any event, this means that our bounds should also be useful for researchers using TSTSLS, providing a way to test the exclusion restriction underlying this estimation method.

\medskip
We next turn to the estimation of the sharp bounds and inference on the regression coefficients. We propose simple plug-in estimators and establish their asymptotic normality. For reasons detailed below, this is a challenging problem, which we address by leveraging results on $L$-statistics and on the statistical optimal transport literature. Our proof relies in particular on a refinement of the convergence rates in \cite{fournier2015rate}, obtained by extending a result of \cite{boucheron2015}. We also provide simple confidence intervals for the regression coefficients and establish their asymptotic validity. Simulation results indicate that our inference method performs well in finite samples, while being implementable at a modest computational cost. 

\medskip

Finally, we apply our methodology to two distinct contexts of data combination. In our first application, we revisit racial disparities in U.S. patent approval and show that conclusions about such disparities hinge on the validity of the exclusion restrictions underlying the TSTSLS estimation strategy. When relaxing these restrictions, our bounds are generally wide, pointing to the lack of robustness of the conclusions based on the standard TSTSLS approach. In our second application, we evaluate the relationship between students' patience and risk-taking and educational performance across countries. In contrast to our first application, our bounds are informative and, for some of the specifications, exclude the TSTSLS estimates. Two main conclusions emerge from these applications. First, we can reject the exclusion restrictions underlying the TSTSLS estimator in practice, not just in theory. Second, sharp bounds that are not based on such restrictions can remain informative in empirical applications. 

\medskip
\paragraph{Related literature.} Our paper belongs to a very active literature on data combination problems in econometrics and statistics. See, in particular, \cite{RM2007} for a survey of this literature and recent contributions by \citeauthor{FSS16} (\citeyear{fan2014identifying}, \citeyear{FSS16}), \citeauthor{DGM21} (\citeyear{DGM21}, \citeyear{d2024partially}), \cite{BLL16}, \cite{BFM24}, \cite{meango2025combining} and, in the context of experimental data \cite{ACI20}, \cite{ACIK24} (under a surrogacy assumption) and \cite{rambachan2024program}. Several of these papers impose restrictions that entail point identification. Following the seminal contribution of \cite{cross2002regressions} and subsequent article by \cite{MP06}, our aim is to obtain bounds on parameters of interest under weak restrictions. An important distinction between our work and these last two papers is that we consider different parameters: the best linear parameter in our case versus conditional expectation in theirs. Also, they do not consider the use of auxiliary variables ($W_a$). We show that these variables, which very frequently arise in applications, actually play an important role in this context.

\medskip
Particularly relevant for us is previous work by \cite{pacini2019two}, who also studies partial identification in linear regressions with data combination. Our paper differs from  his in three key dimensions. First, we allow some of the common variables to not be common regressors ($W\ne X_c$), a case that we expect to happen in most applications. We also show that this can drastically reduce the identified sets. Second, we show that, in fact, Pacini's bounds are not sharp when $X_o$ is multivariate. Moreover, the difference relative to the sharp bounds can be large. Finally and crucially for the applicability of our proposed method, we develop estimation and inference. \cite{hwang2022bounding} also relates to our work. While she maintains the restriction that $W= X_c$ and does not improve upon Pacini's bound in his setup, she also considers the case where some regressors are only available in the $Y$ dataset, which we do not study here.\footnote{ 
There are additional data combination cases that we do not consider here. \cite{kitagawa2023} consider a setup where one observes $(Y,X_1, X_c)$ in one dataset and $(Y, X_2, X_c)$ in another. Yet another possibility, considered by \cite{moon2024partial} when $X_1, X_2, X_c$ has finite support, is to observe $(Y, X_c)$, $(X_1, X_c)$ and $(X_2, X_c)$ separately.} Finally, a third related study is \cite{FPPS25}, which was developed contemporaneously
with ours and first circulated later. A fundamental difference is that their analysis is limited to identification, whereas estimation, inference and empirical applicability are central to our paper. On the other hand, this paper complements ours by studying identification in a more general setup. In particular, they derive sharper bounds than \cite{hwang2022bounding} in cases where some regressors are observed only in the $Y$ dataset. 

\medskip
Our paper is also related to our previous work \citep{d2024partially}. There are, however, several key differences between the two. First and foremost, we imposed there a partially linear model, namely $E[Y|X]=X_o'\beta_o + f(X_c)$. In contrast, we focus in this paper on the coefficients of the best linear predictor without imposing functional-form restrictions on the conditional expectation. This is an important distinction for applied work, where researchers typically use linear regressions because of their simplicity rather than because they believe that conditional expectations are (partially) linear. Second, in our previous work, we did not consider the identifying power of auxiliary variables $W_a$. Third, from a technical viewpoint, the restriction on the conditional expectation implies that we relied on entirely different optimal transport results, both for identification and inference. 

\medskip
From a technical viewpoint, our analysis builds on optimal transport tools \citep[see][for a recent review of optimal transport methods in econometrics]{GalichonHenry26}. In particular, our first identification result relies on the Cambanis-Simons-Stout inequality, see \cite{cambanis1976inequalities}. Our asymptotic normality result relates to the asymptotic normality of the so-called Wasserstein-2 distance of empirical measures, recently studied in the statistical literature \citep[see, e.g.][]{del2019central, berthet2020}. Notably, up to a mild strengthening of moment conditions (from order 4 to $4+\eps$ for some $\eps>0$), our result implies asymptotic normality of the Wasserstein-2 distance under weaker conditions than those obtained in the literature \citep{berthet2020}. We also extend this literature by dealing with generated, as opposed to directly observed, variables. This raises challenges because measurement errors modify the ordering of a variable in a complex way, and ordering plays a key role in Wasserstein distances.

\medskip
Finally, our paper also speaks to a large and rapidly growing empirical literature that deals with data combination problems similar to the one considered here, and in most cases by using linear regressions as well. One important example is voting: given the anonymity of ballots, researchers typically regress average votes on average voter characteristics (e.g., income, hours watching Fox News per week) at the county level \citep[see, e.g.,][]{martin2017bias}. This approach implicitly relies on a TSTSLS strategy, where counties play the role of $W_a$, thereby imposing an exclusion restriction. Another leading example is intergenerational income mobility, which often faces the unavailability of linked income data across generations and similarly relies on exclusion restrictions \citep[see, e.g.,][for a survey]{SantavirtaStuhler22}. Data combination issues are also pervasive in consumption research, where income and consumption are often measured in separate datasets \citep[see in particular][]{CLP22}. Similar data combination problems frequently arise in various other subfields, including the economics of education and returns to skill estimation \citep{PP16,garcia2016life,hanushek2020culture}, health \citep{Manski18} and labor \citep{ACI20}. 

\medskip
The methods we devise in this paper are broadly applicable in these different contexts, allowing empirical researchers to relax the exclusion restrictions that are typically maintained to achieve point identification. By applying our method to racial disparities in patent approval \citep{Dossi23} and the effect of preferences on skill differences \citep{hanushek2020culture}, our paper also adds to the empirical literature on these questions.

\paragraph{Outline.} Section \ref{sec:setup} introduces the setup and discusses three broad cases for which our analysis is relevant. Section \ref{sec:identification} presents our identification results. Section \ref{sec:inference} develops estimators of the sharp bounds, establishes their asymptotic normality and develops inference on the regression coefficients. Section \ref{sec:simulations} examines the finite sample properties of our estimators and confidence intervals through Monte Carlo simulations. We provide in Section \ref{sec:applications} two applications, to racial disparities in patent approval and the effect of preferences on skill differences. Finally, Section \ref{sec:conclu} concludes. The appendix includes in particular a discussion of the sharpness of the bounds of \cite{pacini2019two} and gathers all the proofs of our identification results; the proofs of our inference results appear in the online appendix. Finally, our method can be implemented using our companion R package, \texttt{RegCombinBLP}.\footnote{
This package is available on GitHub at \href{https://github.com/cgaillac/RegCombinBLP}{https://github.com/cgaillac/RegCombinBLP} with a user-friendly guide on how to use it.}

\section{Set-up and motivation} 
\label{sec:setup}

\subsection{Set-up}

We seek to identify the best linear predictor $EL(Y|X)$ of $Y$ by $X\in\R^p$, with $X=(X'_o,X'_c)'$. To this end, we assume to have access to two separate datasets that cannot be matched. The first one includes $(Y, W')$, whereas the second one includes $(X_o',W')$; here $W=(W_a',X_c')\in \R^q$. We call $X_o$ the ``outside regressors'',  $X_c$ the ``common regressors'', $W$ the ``common variables'' and $W_a$ the ``auxiliary variables''. The latter are variables that the researcher does not include in the regression of interest, but that may still help for identification since they are available in both datasets. Importantly, they should not be seen as instruments, in the sense that we do not impose below any restrictions on them.

\medskip
In order for the best linear prediction to be well-defined, we maintain the following assumption hereafter:

\begin{hyp}
	$E(Y^2+\|X_o\|^2+\|W\|^2)<\infty$ and $E(XX')$ and $E(WW')$ are nonsingular.
	\label{hyp:mom}
\end{hyp}

Let $b^0=(b^{0,1},...,b^{0,p})\in\R^p$ be such that $EL(Y|X)=X'b^0$. Usually, researchers are interested in specific components of $b^0$, rather than in the whole vector $b^0$. Therefore, in the following we seek to (partially) identify and estimate $b_d:=d'b^0$, for some $d\in \R^p$. For instance, if we focus on $b^{0,2}$, the second component of $b^0$, we let $d=(0,1,0,...,0)$. 

\subsection{Motivation}

Our setup includes at least three cases of broad interest.

\paragraph{Proxies for the covariate of interest.} In this case, we are interested in the relationship between a covariate of interest $X_o$ and $Y$. However, we do not observe $X_o$ in the $Y$ dataset, but only proxies $W_a$ of $X_o$. These proxies are also observed in the $X_o$ dataset. This type of situation arises very frequently in empirical microeconomics. A standard strategy in this case is to use two-sample two-stage least squares (TSTSLS). Namely, one first regresses $X_o$ on $W_a$ in the $X_o$ dataset. Then, we regress $Y$ on the predicted $X_o$ in the $Y$ dataset. Importantly though, this strategy implicitly relies on the following exclusion restriction:
\begin{equation}
EL(Y|X_o, W_a)=EL(Y|X_o).
    \label{eq:excl_restr_TSTSLS}
\end{equation}
This assumption is often restrictive. In our first application, for instance, $Y$ corresponds to patent approval, $X_o$ is the vector of race dummies and $W_a$ is the vector of applicants' last names. Given that patent reviewers always observe last names but typically do not observe race directly, \eqref{eq:excl_restr_TSTSLS} seems unlikely to hold. The method we develop in this paper allows us to (partially) identify $EL(Y|X_o)$ without imposing \eqref{eq:excl_restr_TSTSLS}.  

\paragraph{Missing controls.} In this case, we are interested in recovering the effect of $X_c$ on $Y$ using data from a first dataset. However, one or several key control variables ($X_o$) are missing from this dataset. Our setup applies to situations where the control variables are observed in a second dataset, together with $X_c$. In this sense, our framework complements a growing literature that investigates how credible unconfoundedness is, by allowing researchers to rely on unconfoundedness in a broad range of data combination environments \citep[see, e.g.,][]{AET05,oster2019unobservable, DMP22}. 

\medskip
As above, researchers in this context may also have access to auxiliary variables, $W_a$, which are not included as covariates in the main regression, e.g., because they would be ``bad controls''. As shown below, these variables may still carry informational content regarding the regression coefficients of interest.  

\paragraph{Mediation analysis.} In this case, researchers are interested in the effect of $X_o$ on a given outcome $Y$. As above, we do not observe $X_o$ and $Y$ in the same dataset. A possible and frequent reason is that $Y$ is a long-run outcome, which is not observed in the data including $X_o$. On the other hand, both datasets may include other outcomes $W_a$, such as short-run outcomes. 

\medskip
To identify in this environment the causal effect of $X_o$ on $Y$ (which is $b^0$ under suitable randomization conditions on $X_o$), a common strategy is to rely on a surrogacy assumption \citep[see, e.g.,][]{prentice1989surrogate}. In our setup, this corresponds to 
\begin{equation}
EL(Y|X_o, W_a)=EL(Y|W_a).
    \label{eq:surrog}
\end{equation}
In other words, one assumes that the effect of $X_o$ on $Y$ is entirely mediated by $W_a$.\footnote{One may also include additional covariates $X_c$ observed in both datasets, in which case \eqref{eq:surrog} becomes $EL(Y|X, W_a)=EL(Y|X_c, W_a)$.} However, Condition \eqref{eq:surrog} typically is a strong restriction. For instance, it is reasonable to assume that long-run earnings ($Y$) depend on human capital, even conditional on short-run earnings ($W_a$). Then, if job training ($X_o$) affects human capital, \eqref{eq:surrog} will fail to hold in general. Our results below imply that one can still obtain simple, sharp bounds on $b^0$, without relying on  \eqref{eq:surrog}. 


\section{Identification} 
\label{sec:identification}

Before presenting our identification results, we introduce additional notation. We denote by $\B_d$ the identified set of $b_d$ and let $\overline{b}_d$ and $\underline{b}_d$ be its sharp upper and lower bounds, namely
$$\overline{b}_d =\sup\{d'b: \; b\in \B\}, \quad \underline{b}_d =\inf\{d'b: \; b\in \B\},$$
where $\B$ is the identified set of $b^0$. We focus in the following solely on $\overline{b}_d$, which is without loss of generality since $\underline{b}_d=-\overline{b}_{-d}$. 

\medskip
For any random variables $A$ and $B$, we let $F_A$ denote the cumulative distribution function (cdf) of $A$, $f_A$ its density, $\Supp(A)$ its support and $F_{A|B}$ the cdf of $A$ given $B$. We also let $F_A^{-1}(t):=\inf\{x:F_A(x)\ge t\}$ denote the quantile function of $A$; we denote similarly by $F_{A|B}^{-1}$ the quantile function of $A$ given $B$. For any vector $v$, we let $v_k$ denote its $k$-th element and $v_{-k}$ the vector obtained by removing $v_k$ from $v$. We also let $e_{k,r}$ ($e_k$ in the absence of ambiguity) denote the $k$-th canonical vector of $\R^r$.  For any set $S$, we let $|S|$ denote its cardinality. Finally, we denote by $\mathcal{U}[0,1]$ the uniform distribution over $[0,1]$ and by $\mathcal{N}(\mu,\Sigma)$ the multivariate normal distribution with mean $\mu$ and covariance matrix $\Sigma$.

\subsection{No common variables} 
\label{sub:no_sep}

We first consider a case without nontrivial common variable ($W=X_c=1$), so that $X=(X'_o,1)'\in \R^{p}$. Our main result establishes that $\B$ is convex and compact, and characterizes $\overline{b}_d$ for any $d\in\R^p\backslash\{0\}$. Below, we introduce the variable $\eta_d$ as follows. First, let $(\widetilde{d}_2,...,\widetilde{d}_p)$ be $(p-1)$ vectors in $\R^{p}$ such that $(d,\widetilde{d}_2, ...,\widetilde{d}_p)$ forms a basis of $\R^p$. Let $M$ denote the corresponding matrix and let $T=M^{-1}X$. Then, let
$$\eta_d := T_1 - EL[T_1|T_{-1}].$$
In words, $\eta_d$ is the residual of the (population) regression of $T_1$ on $T_{-1}$. Note that $\eta_d$ does not depend on which exact vectors $(d_2,...,d_p)$ are chosen. Also, if $d=e_{k,p}$, $\eta_d$ is simply the residual of the regression of $X_k$ on $X_{-k}$. Finally, if $p=2$ and $d=(d_1,0)'$, $\eta_d=(X_o-E(X_o))/d_1$.

\begin{thm}
	Suppose that Assumption \ref{hyp:mom} holds and $W=X_c=1$. Then $\B$ is convex, compact, and satisfies $\B\subseteq\mathcal{E}$, with
$$\mathcal{E}:= \{b\in \R^p: \; E[Y]=E[X'b],\; V(Y)\geq V(X'b)\}.$$
Also, letting $U\sim\mathcal{U}[0,1]$, we have, for any $d\in\R^p\backslash\{0\}$, $\B_d=[\underline{b}_d,\overline{b}_d]$, with
	\begin{align}
	\overline{b}_d = & E\left[F^{-1}_{d'E(XX')^{-1}X}(U) F_{Y}^{-1}(U)\right] \label{eq:supp_fct_no_common_1} \\[1.5mm]
	= & \frac{E[F_{\eta_d}^{-1}(U) F_{Y}^{-1}(U)]}{E(\eta_d^2)}.
		\label{eq:supp_fct_no_common}
	\end{align}
    Finally, if $d=(d_1,...,d_{p-1},0)$ and $V(Y)>0$, then $\overline{b}_d>0>-\overline{b}_{-d}$.  
	\label{thm:ident_no_common}
\end{thm}

The first part of the theorem states that $\B$ is a convex, compact set included in the ellipsoid $\mathcal{E}$. This part implies that $\overline{b}_d$ is equal to the so-called support function of $\B$. As a result, the knowledge of $\overline{b}_d$ for all $d\in\R^p\backslash\{0\}$ characterizes $\B$. Also, $(0,\ldots,0,E[Y])'\in\B$: in the absence of common variables, we can always rationalize $Y$ and $X$ as independent. The equality $E[X'b^0]=E[Y]$ implies that if $d=\lambda E[X]$ for some $\lambda\in\R$, $d'b^0=\lambda E[Y]$ is point-identified. On the other hand, if $d$ is not proportional to $E[X]$, the random variable $d'E(XX')^{-1}X$ has strictly positive variance, so  $\overline{b}_d>\underline{b}_d$ as long as $V(Y)>0$. If the parameter of interest also excludes the  intercept ($d_p=0$), as is typically the case, then $\overline{b}_d>0>\underline{b}_d$. 

\medskip
In the case of a single regressor (and the intercept) and $d=(1,0)'$, Equation \eqref{eq:supp_fct_no_common} reduces to
\begin{equation}
\overline{b}_d = \frac{E\left[(F_{X_o}^{-1}(U) - E(X_o)) F_{Y}^{-1}(U)\right]}{V(X_o)}.	
	\label{eq:upper_simple}
\end{equation}
On the other hand, the true coefficient satisfies $b_d=b^{0,1}=E[(X_o - E(X_o))Y]/V(X_o)$. Thus, \eqref{eq:upper_simple} indicates that the sharp upper bound on the unknown term $E[X_oY]$ is $E[F_{X_o}^{-1}(U) F_{Y}^{-1}(U)]$. This is well-known, and corresponds to the so-called Cambanis-Simons-Stout inequality \citep[see][]{cambanis1976inequalities}. The logic is that (i) $F_{X_o}^{-1}(U)$ and $F_{Y}^{-1}(U)$ are distributed as $X_o$ and $Y$, since $U$ is uniformly distributed, and (ii) these two variables exhibit maximal positive dependence. The exact meaning of (ii) is that the copula of $F_{X_o}^{-1}(U)$ and $F_{Y}^{-1}(U)$ corresponds to the Fr\'echet-Hoeffding upper bound.

\medskip
With multiple regressors, \eqref{eq:supp_fct_no_common} cannot be directly deduced from the Cambanis-Simons-Stout inequality. To get some intuition on \eqref{eq:supp_fct_no_common}, suppose that $d=e_{1,p}$. Then, $\eta_d$ is the residual of the linear regression of $X_1$ on $X_{-1}$. If we observed $(Y,X)$, the coefficient of $X_{1}$ in the best linear prediction of $Y$ by $X$ would be $E[\eta_d Y]/E(\eta_d^2)$, by the Frisch-Waugh-Lovell theorem. Now, if we only know the marginal distributions of $\eta_d$ and $Y$, the numerator in \eqref{eq:supp_fct_no_common} is simply the upper bound of $E[\eta_d Y]$. That the sharp upper bound $\overline{b}_d$ satisfies \eqref{eq:supp_fct_no_common} is not obvious, however, because we also know the distribution of $X_{-1}$ conditional on $\eta_d$, in addition to the marginal distribution of $\eta_d$. This could, in principle, lead to $\overline{b}_d < E[F_{\eta_d}^{-1}(U) F_Y^{-1}(U)]/E(\eta_d^2)$. Theorem \ref{thm:ident_no_common}  shows that this is not the case: the conditional distribution of $X_{-1}$ does not carry any additional information about $E[\eta_d Y]$. Although this can be deduced from Lemma 3.3 in \cite{delon2023generalized}, we propose an alternative proof, which has the advantage of being constructive. 

\medskip
\cite{pacini2019two} also obtains bounds on $b_d$, see his Theorem 1. However, it turns out that when $X$ is multivariate, his bound is only an outer bound rather than the sharp bound $\overline{b}_d$ on $b_d$. In Appendix \ref{sub:bounds_in_pacini}, we detail why this is the case, and provide an illustration showing that the sharp bounds given by Theorem \ref{thm:ident_no_common} above can in practice be substantially tighter than Pacini's bounds.


\subsection{Common variables} 
\label{sub:common_regressors}

\subsubsection{Main result} 
\label{ssub:main_results}

Let us now turn to the situation where some covariates are observed in both datasets. An application of Theorem \ref{thm:ident_no_common} conditional on $W$ shows that $\overline{b}_d$ satisfies
\begin{equation}
\overline{b}_d = \frac{E[F^{-1}_{\eta_d|W}(U|W)F^{-1}_{Y|W}(U|W)]}{E(\eta^2_d)}.
    \label{eq:sharp_bound_W}
\end{equation}
When $W$ is finitely supported, this expression can be directly turned to the data. Otherwise, however, the expression involves the conditional quantile functions, which can be difficult to estimate. We then consider simpler, outer bounds. Let us define as above $\eta_d$, with the sole difference that now $X=(X'_o,X'_c)'$. Next, let $\delta_d$ and $\nu_d$ be such that $EL(\eta_d|W)=W'\delta_d$ and  $\nu_d:=\eta_d-W'\delta_d$. Define $\delta_Y$ and $\nu_Y$ similarly, with $Y$ in place of $\eta_d$. 

\begin{thm}
Suppose that Assumption \ref{hyp:mom} holds. Then $\B$ is convex, compact, and for any $d\in\R^p\backslash\{0\}   $, $\B_d=[\underline{b}_d, \overline{b}_d]$, where for any function $g$, $\overline{b}_d\le \overline{b}^g_d$, with
\begin{equation}\label{eq:upper:sigma_w1}
    \overline{b}^g_d := \frac{1}{E(\eta_d^2)}\left\{\delta_d' E(WW') \delta_Y + E[F_{\nu_d|g(W)}^{-1}(U|g(W)) F_{\nu_Y|g(W)}^{-1}(U|g(W))]\right\},
\end{equation}
where $U|W\sim\mathcal{U}[0,1]$, and equality holds if $\nu_Y\indep W|g(W)$ and $\nu_d\indep W|g(W)$.
\label{thm:caract_id_set_common}
\end{thm}

When $g(W)$ has a finite support, the outer bound $\overline{b}^g_d$ is elementary to estimate and does not suffer from any curse of dimensionality. Moreover, this bound is actually sharp when $\nu_Y\indep W|g(W)$ and $\nu_d\indep W|g(W)$, as is the case for instance with $g(W)=1$, if $Y$ and $\eta_d$ follow a linear location model in $W$. On the other hand, without $X_c$, there is no guarantee that $\overline{b}^g_d$ is smaller than the simple bound $E[F_{\eta_d}^{-1}(U) F_{Y}^{-1}(U)]/E(\eta_d^2)$ obtained without $W_a$.

\paragraph{How do common regressors affect identification?}

Even without auxiliary variables $W_a$, the identified interval on the coefficients of $X_o$ may exclude 0 in the presence of common regressors, implying that the sign of these coefficients is identified. To see this, suppose that dim$(X_o)=1$, $X_o=f_1(X_c) + \zeta_o$, $Y = g_1(X_c)+\zeta_Y$ and $\zeta_o| X_c\sim\mathcal{N}(0,\sigma^2_o)$, $\zeta_Y| X_c\sim\mathcal{N}(0,\sigma^2_Y)$. Let also $f(X_c):=f_1(X_c) - EL(f_1(X_c)|X_c)$ and $g(X_c):=g_1(X_c) - EL(g_1(X_c)|X_c)$. Using Equation \eqref{eq:upper:sigma_w1}, the fact that by construction $\delta_d=0$, and the normality of $\zeta_o$ and $\zeta_Y$, we obtain that the bounds on $b^{0,1}$ satisfy
\begin{align*}
\overline{b}_{e_1} = & \frac{E\left[f(X_c)g(X_c)\right] + \sigma_o \sigma_Y}{E(\eta_{e_1}^2)}, \\
\underline{b}_{e_1} = & \frac{E\left[f(X_c)g(X_c)\right] - \sigma_o \sigma_Y}{E(\eta_{e_1}^2)}, 
\end{align*}
where $\eta_{e_1}=f(X_c) + \zeta_o$. In particular, if $|E\left[f(X_c)g(X_c)\right]| > \sigma_o \sigma_Y$, 0 is excluded from the identified set of $b^{0,1}$. This occurs when $X_o$ and $Y$ strongly depend on $X_c$ in a nonlinear way, so that $E\left[f(X_c)g(X_c)\right]$ dominates the contribution from independent terms (namely, $\sigma_o \sigma_Y$). In the extreme case where $X_o$ and $Y$ are deterministic functions of $X_c$, so that $\sigma_o =\sigma_Y=0$, we obtain point identification.

\medskip
That said, the identified interval on the coefficients of $X_o$ may widen in some cases when including common covariates. Even if observing $X_c$ in both datasets does increase the information on the joint distribution of $(Y, X_o)$, the parameter we consider also changes. In particular, the denominator $E[\eta_d^2]$ in \eqref{eq:sharp_bound_W} may substantially decrease, if the $R^2$ of the linear regression of $X_o$ on $X_c$ is large. To illustrate this, suppose that $(X_o,X_c')'\sim\mathcal{N}(0,\Sigma_o)$ and $(Y,X_c')\sim\mathcal{N}(0,\Sigma_Y)$, with
\begin{equation}
\Sigma_o=\begin{pmatrix}
	1 & \rho_o \\
	\rho_o & 1
\end{pmatrix}	\quad \text{ and } \Sigma_Y=\begin{pmatrix}
	1 & \rho_Y \\
	\rho_Y & 1
\end{pmatrix}.	
	\label{eq:def_Sigma}
\end{equation}
Then, some algebra shows that without observing $X_c$, $[\underline{b}_{e_1},\overline{b}_{e_1}]=[-1,1]$. With $X_c$, on the other hand,
$$[\underline{b}_{e_1},\overline{b}_{e_1}]=\left[-\sqrt{\frac{1-\rho^2_Y}{1-\rho^2_o}},\; \sqrt{\frac{1-\rho^2_Y}{1-\rho^2_o}}\right].$$ 
Thus, the interval $[\underline{b}_{e_1},\overline{b}_{e_1}]$ shrinks if $|\rho_Y|>|\rho_o|$ but widens otherwise. 

\medskip
Finally, we may also identify the sign of components of $b_c$, the regression coefficient of $X_c$. In fact, $b_c$ may even be point identified: if $X_c$ and $X_o$ are uncorrelated, $b_c$ is simply the coefficient of the regression of $Y$ on $X_c$.

\paragraph{The role of auxiliary variables.}

By observing auxiliary variables $W_a$ that are not in the regression of interest, we increase the available information without modifying the parameter of interest. Then, the interval $[\underline{b}_{e_1},\overline{b}_{e_1}]$ always shrinks (at least weakly so). This may lead to excluding 0 from $\B$ even without common variables $X_c$, a case that occurs whenever 
\begin{equation}
\delta_d' E(WW') \delta_Y+E\left[F_{\nu_d|W}^{-1}(U|W)F_{\nu_Y|W}^{-1}(U|W)\right]<0
	\label{eq:W_a_0_excl}
\end{equation}
for some $d\in\R^p$. Intuitively, \eqref{eq:W_a_0_excl} requires enough dependence between $Y$ and $W_a$ and between $X_o$ and $W_a$. For instance, if $(W,X_o)\sim\mathcal{N}(0,\Sigma_o)$ and $(W,Y)\sim\mathcal{N}(0,\Sigma_Y)$, with $\Sigma_o$ and $\Sigma_Y$ as in \eqref{eq:def_Sigma}, we obtain
$$[\underline{b}_{e_1},\overline{b}_{e_1}]=\left[\rho_o \rho_Y - \sqrt{(1-\rho_o^2)(1-\rho_Y^2)},\; \rho_o \rho_Y +  \sqrt{(1-\rho_o^2)(1-\rho_Y^2)}\right].$$ 
Then, $0\not\in[\underline{b}_{e_1},\overline{b}_{e_1}]$ if and only if $\rho^2_o \rho^2_Y > (1-\rho_o^2)(1-\rho_Y^2)$. This holds when $W_a$ is sufficiently correlated with $X_o$ and $Y$. For instance, when $\rho_o=\rho_Y$, this occurs if and only if $W_a$ explains more than half of the variance of $X_o$ ($\rho_o^2>1/2$).

\medskip
A leading case with auxiliary variables is the case of surrogates. Recall that in this case, $X_o$ corresponds to the treatment variable, $Y$ is a long-run outcome while $W_a$ denotes short-run outcomes (surrogate variable). Then, Theorem \ref{thm:caract_id_set_common} yields two sets of bounds, sharp and outer, on the effect of $X_o$ on $Y$, without imposing a surrogacy assumption.

\paragraph{Link with TSTSLS.}

Recall that the TSTSLS estimand $b_{\text{TSTSLS}}$ identifies $b^0$ if Equation \eqref{eq:excl_restr_TSTSLS} holds, namely if the coefficient of $W_a$ in the ``long'' regression of $Y$ on $(X,W_a)$ is 0. Now, the discussion above (``How do common regressors affect identification?'') shows that if one views $W_a$ as a common regressor, 0 may not belong to the identified set of the regression coefficient of $W_a$. This implies that the exclusion restriction underlying the TSTSLS estimand can actually be rejected by the data, as happens in our second application below. This could appear to contradict the standard IV setup, where the underlying exclusion restriction is fundamentally not testable, but note that Equation~(1) differs from this exclusion restriction. The following proposition helps us to understand why Equation \eqref{eq:excl_restr_TSTSLS} can be rejected in practice. 

\begin{prop}
    Suppose that Assumption \ref{hyp:mom} holds, $X_c=1$, $X_o\in\R$ and $W_a\in \R$. If $|\Corr(W_a,X_o)|<|\Corr(W_a,Y)|$, we reject \eqref{eq:excl_restr_TSTSLS}, the exclusion restriction underlying $b_{\text{TSTSLS}}$.
    \label{prop:TSTSLS}
\end{prop}

The intuition behind this result is that if Equation \eqref{eq:excl_restr_TSTSLS} holds, then $W_a$ predicts $Y$ only indirectly, through its effect on $X_o$. As a result, and keeping in mind that $|\Corr(W_a,X_o)| \leq 1$, we reach a contradiction if i) $W_a$ and $X_o$ are weakly correlated, and ii) $Y$ and $W_a$ are strongly correlated. Note that Proposition \ref{prop:TSTSLS} only gives a simple sufficient condition for the TSTSLS estimand to be invalid: $b_{\text{TSTSLS}}$ may not belong to the identified set of $b^0$ even if $|\Corr(W_a,X_o)|\ge|\Corr(W_a,Y)|$.


\subsubsection{Testing and weakening the common population assumption} 
\label{ssub:testing}

We have maintained thus far that the two samples at hand are drawn from the same population. While this is a standard assumption in the data combination literature, it is important to consider the extent to which this can be relaxed. To this end, let us introduce the binary variable $D$, with $D=1$ (resp. $D=0$) if we consider the $Y$ dataset (resp. the $X_o$ dataset). Then, our setup implies that we only observe the distributions of $(W,Y)|D=1$ and $(W,X_o)|D=0$, assuming that $D\indep (W,X_o,Y)$. With common variables, this condition can be tested, since it implies $F_{W|D=1}=F_{W|D=0}$. If this implication is rejected, we can weaken the independence assumption by assuming instead that
\begin{equation}
(X_o,Y) \indep D|W, \quad p:=P(D=1) \text{ is known.}	
	\label{eq:indep_condit}
\end{equation}
In words, the first condition imposes that conditional on $W$, the two datasets are drawn from the same population, while the two populations corresponding to $D=0$ and $D=1$ may differ in their marginal distributions of $W$. The second condition in \eqref{eq:indep_condit} implies that the joint distribution of $(D,W)$, and thus the ``propensity score'' $p(W):=P(D=1|W)$, can be retrieved from the knowledge of the distributions of $W|D=0$ and $W|D=1$.

\medskip
If \eqref{eq:indep_condit} holds, the sharp upper bound $\overline{b}_d$ can be obtained by reasoning as in Theorem \ref{thm:caract_id_set_common}, using an inverse probability weighting scheme. Specifically, to identify $\mu_d:=E[T_{-1}T_{-1}']^{-1} E[T_{-1}T_1]$, and in turn $\eta_d= T_1 - T_{-1}'\mu_d$ (recall our definition of $T$ at the beginning of Section \ref{sub:no_sep}), we cannot directly regress $T_1$ on $T_{-1}$ conditional on $D=0$. Yet, we can recover $\mu_d$ by considering a weighted regression instead, since
$$\mu_d = E\left[\frac{(1-D)T_{-1}T_{-1}'}{1-p(W)}\right]^{-1}E\left[\frac{(1-D)T_{-1}T_1}{1-p(W)}\right].$$
Then, Equation \eqref{eq:upper:sigma_w1} is replaced by:
$$\overline{b}_d=\frac{\delta_d'E(WW')\delta_Y + E\left[F_{\nu_d|W,D=0}^{-1}(U|W)F_{\nu_Y|W,D=1}^{-1}(U|W)\right]}{E\left[(1-D)\eta_d^2/(1-p(W))\right]}.$$

Another point to note is that if the two populations differ, the parameter of interest may correspond to one of the two populations only. For instance, one may consider, instead of $EL(Y|X)$, $EL(Y|X,D=1)$. In this case, $\delta_Y$ is given by $E[WW'|D=1]^{-1} E[WY|D=1]$ and is thus obtained by an unweighted regression, whereas $\delta_d$ (and then $\nu_d$) is obtained by regressing $\eta_d$ on $W$ with weights $p(W)/(1-p(W))$, and the upper bound $\overline{b}_d$ becomes
\begin{equation}
\overline{b}_d=\frac{\delta_d'E(WW'|D=1)\delta_Y + E\left[F_{\nu_d|W,D=0}^{-1}(U|W)F_{\nu_Y|W,D=1}^{-1}(U|W)|D=1\right]}{E\left[(1-D)\eta_d^2p(W)/(1-p(W))\right]/E[D]}.    
\label{eq:bound_subpop}
\end{equation}

Finally, another practically relevant situation is one in which one sample is drawn from a subpopulation of the population from which the other sample is drawn. Then, we identify instead (for instance) the distribution of $(Y,W)$ given $D=1$ and the distribution of $(X,W)$. In this case and if we focus as above on $EL(Y|X,D=1)$, we obtain a similar upper bound on $\overline{b}_d$ as in \eqref{eq:bound_subpop}, with the only difference that $\mu_d$ (and then $\eta_d$) is obtained by regressing $T_1$ on $T_{-1}$ with weights $p(W)$. 
Note that in this case and the one before, we do not require the joint independence condition in \eqref{eq:indep_condit} but only $X_o\indep D|W$.


\subsubsection{Auxiliary, non-common variables} 
\label{ssub:additional_variables}

In practice, one may have access to auxiliary variables that appear in the dataset of $Y$ only, or in the dataset of $X_o$ only. For instance, suppose we identify the distributions of $(W,Y,Z)$ from one dataset and that of $(W,X_o)$ from the other. The following proposition shows that, for identification purposes, knowing the conditional distribution of $Z|W,Y$ provides no additional information. Hereafter, we let $\B_Z$ denote the identified set of $b^0$ when observing some auxiliary non-common variables $Z$.

\begin{prop}
	Suppose that Assumption \ref{hyp:mom} holds. Then $\B_Z=\B$.
	\label{prop:additional_var}
\end{prop}

A similar result clearly holds if we consider instead a variable that appears only in the dataset of $X_o$. The bottom line is that, among variables not included in the regression, only those that are common across the two datasets are relevant for identification. 


\section{Estimation and inference} 
\label{sec:inference}

\subsection{No common variables}\label{sub:no_common}
\subsubsection{Estimation of $\overline{b}_d$} 
\label{ssub:computation_of_the_estimator}

Consider first the simplest situation where we only observe two independent samples, $\mathcal{S}_1:=(Y_i)_{i=1,...,n}$ and $\mathcal{S}_2:=(X_j)_{j=1,...m}$. Let $\widehat{\eta}_{dj}$ denote $j$'s residual in the sample regression of $T_1$ on $T_{-1}$. To ease notation, we let hereafter $F:=F_Y$ and $G:=F_{\eta_d}$, and let $F_n$ and $\widehat{G}_m$ denote the empirical cdfs of $(Y_i)_{i=1,...,n}$ and $(\widehat{\eta}_{dj})_{j=1,...,m}$, respectively. From Theorem \ref{thm:ident_no_common}, we have $\overline{b}_d = \int_0^1 F^{-1}(t)G^{-1}(t)dt/E(\eta_d^2)$. Then, we consider the plug-in estimator of $\overline{b}_d$:
$$\widehat{\overline{b}}_d = \frac{\int_0^1 F_n^{-1}(t)\widehat{G}^{-1}_m(t)dt}{\widehat{E}(\widehat{\eta}_d^2)},$$
where $\widehat{E}(\widehat{\eta}_d^2)$ denotes the empirical variance of $(\widehat{\eta}_{dj})_{j=1,...,m}$. Remark that when $m=n$, we simply have, denoting by $Y_{(i)}$ the $i$-th order statistic of $(Y_i)_{i=1,...,n}$ (similarly for $\widehat{\eta}_{d(i)})$:
$$\int_0^1 F_n^{-1}(t)\widehat{G}^{-1}_m(t)dt = \frac{1}{n}\sum_{i=1}^n Y_{(i)}\widehat{\eta}_{d(i)}.$$
Otherwise, we can still compute the numerator of $\widehat{\overline{b}}_d$ at low cost. To see this, note that for any real-valued variables $U_1$, $U_2$ with finite second moments and cdfs $F_1, F_2$,
\begin{equation}
\int_0^1 F_1^{-1}(t)F_2^{-1}(t)dt = \frac{1}{2}\left[E[U_1^2] + E[U_2^2] - W_2^2(F_1,F_2)\right],	
	\label{eq:link_W2}
\end{equation}
where $W_2$ is the Wasserstein-2 distance, $W_2(F_1,F_2):=(\int (F_2^{-1}(t)-F_1^{-1}(t))^2dt)^{1/2}$. For variables with support size of $n$ and $m$ respectively, as is the case here, we can then compute $W_2(F_1,F_2)$ with algorithms of complexity $O(m + n)$, see, e.g., \cite{rubner2000earth}.


\subsubsection{Asymptotic properties and inference} 
\label{ssub:inference}

We first establish the asymptotic normality of $\widehat{\overline{b}}_d$, under the following assumptions. 

\begin{hyp}
	We observe $(Y_1,...,Y_n)$ and $(X_{o,1},...,X_{o,m})$, two independent samples of i.i.d. variables with the same distribution as $Y$ and $X_o$, respectively.
	\label{hyp:samples}
\end{hyp}

\begin{hyp}\label{hyp:reg} 
One of the following holds: 
\begin{enumerate}
    \item[(i)] $E[|Y|^{2+\eps}]<\infty$ for some $\eps>0$, $|\Supp(\eta_d)|=|\Supp(X)|<\infty$ and $\forall h\in\Supp(\eta_d)$, $F^{-1}$ is continuous at $G(h)$.
\item[(ii)] $E[\|X\|^4]<\infty$, $|\Supp(Y)|<\infty$, $\Supp(\eta_d)$ is an interval and $G$ is continuous.
\item[(iii)] $E(|Y|^{4+\eps}+\|X\|^{4+\eps})<\infty$ for some $\eps>0$, $F^{-1}$ and $G$ are continuous and for either $Z=Y$ or $Z=\eta_d$, the distribution of $Z$ is continuous with respect to the Lebesgue measure and there exists $C_1,C_2>0$ such that for all $z$ in the interior of $\Supp(Z)$,
		\begin{equation}
		\frac{f_Z(z)}{F_Z(z)(1-F_Z(z))}  \ge \min\left(C_1, \frac{C_2}{|z|\ln(1+|z|)^2}\right).	
			\label{eq:condit_hazard}
		\end{equation}
\end{enumerate}
\end{hyp}

We consider in Assumption \ref{hyp:reg} three possibilities, depending on whether $\eta_d$ and $Y$ are finitely supported or not. The first case corresponds to $\eta_d$ being finitely supported. In such a case, $Y$ can be continuous or discrete, as long as, in the latter case, there is no $(h,y)$ such that $F(y)=G(h) \in (0,1)$. The second case corresponds to $Y$ being finitely supported and $\eta_d$ continuous. The third case corresponds to the two variables being, loosely speaking, continuous (actually, case (iii) is compatible with $Y$ having point masses, if we let $Z=\eta_d$). Then, we impose moment conditions but also \eqref{eq:condit_hazard}. The latter holds on $\Supp(Z)\cap[0,\infty)$ for all distributions that have increasing hazard rates, such as log-concave distributions (as their survival function is then log-concave). It also holds for many distributions with decreasing hazard rates, such as Pareto and Weibull distributions. More generally, we expect Condition \eqref{eq:condit_hazard} to be mild, since for any continuous probability measure with cdf $F$, density $f$ and supremum of support equal to $\overline{x}\le \infty$, we have, for all $A<\overline{x}$ satisfying $F(A)>0$,
$$\int_A^{\overline{x}} \frac{f(x)}{F(x)(1-F(x))}dx \ge \int_A^{\overline{x}} (-\ln[1-F(x)])'dx =\infty.$$
On the other hand, for any $C_1,C_2>0$,
$$\int_A^{\overline{x}} \min\left(C_1,\frac{C_2}{|x|\ln(1+|x|)^2}\right)dx < \infty.$$
Thus, one cannot have $f(x)/[F(x)(1-F(x))]\le \min[C_1, C_2/(|x|\ln(1+|x|)^2)]$ for all $x$ large enough; and similarly one cannot have $f(x)/[F(x)(1-F(x))]\le \min[C_1, C_2/(|x|\ln(1+|x|)^2)]$ for all $x$ small enough.

\medskip
To define the asymptotic distribution, we introduce additional objects. First,  let $h(x):= \int_0^1 F^{-1}[G(x^-)+u(G(x)-G(x^-))]du$ and 
\begin{align*}
\psi_1 & := - \overline{b}_d (\eta_d^2 - E[\eta_d^2]),\\
\psi_2 & := -E[h(\eta_d) T_{-1}']E[T_{-1}T_{-1}']^{-1} T_{-1} \eta_d, \\
\psi_3 & := -\int [\ind{\eta_d \le t} - G(t)]F^{-1}\circ G(t)dt, \\
\psi_4 & := -\int [\ind{Y \le t} - F(t)]G^{-1} \circ F(t)dt.
\end{align*}
These four variables correspond to the influence functions of respectively $\sqrt{m}(\widehat{E}(\widehat{\eta}^2_d)- E(\eta^2_d))$, $\sqrt{m} \int_0^1 F^{-1} (\widehat{G}_m^{-1} - G_m^{-1})dt$, $\sqrt{m}\int_0^1 F^{-1}(G^{-1}_m-G^{-1})dt$, and $\sqrt{n}\int_0^1 G^{-1}(F^{-1}_n-F^{-1})dt$, with $G_m$ the empirical cdf of the $(\eta_{dj})_{j=1,...,m}$ (note that $G_m$ cannot be computed in practice, since the $(\eta_{dj})_{j=1,...,m}$ are unobserved). 

\begin{thm}\label{thm:asym_Xo}
	Suppose that $\min(m,n)\to\infty$,  $n/(m+n)\to\lambda\in[0,1]$ and Assumptions \ref{hyp:mom}-\ref{hyp:reg} hold. Then,
$$\sqrt{\frac{mn}{m+n}}\left(\widehat{\overline{b}}_d-\overline{b}_d\right)\convNor{V_d},$$
where $V_d := \left[ \lambda V\left(\psi_1 +\psi_2 + \psi_3\right) + (1-\lambda) V\left(\psi_4\right)\right]/E(\eta^2_d)^2$.
\end{thm}

\paragraph{Remarks on the result.} First, we comment on the assumptions underlying Theorem \ref{thm:asym_Xo}. We allow not only for $\lambda\in (0,1)$, but also for $\lambda=0$ or $\lambda=1$, which corresponds to cases where one sample is much larger than the other. In these cases, the asymptotic variance $V_d$ simplifies. Also, when $\min(|\Supp(X)|,|\Supp(Y)|)<\infty$, we obtain weak convergence under minimal conditions; note that $E[\|X\|^4]<\infty$ is close to being necessary for the OLS estimator $\widehat{\gamma}$ of the regression of $T_1$ on $T_{-1}$ to be $\sqrt{m}-$consistent.

\medskip
When $\min(|\Supp(X)|,|\Supp(Y)|)=\infty$,  the conditions we impose are probably not minimal, but a moment of order 4 for $Y$ and $\eta_d$ seems necessary in view of \eqref{eq:link_W2} and the discussion of Theorem 1 in \cite{del2019central}. Moreover, closely related results in the literature on the asymptotic normality of $W_2(F_n, G_m)$ impose strong restrictions.\footnote{By the proof of Point 2 of Theorem \ref{thm:asym_Xo}, we obtain, under Assumptions \ref{hyp:mom}-\ref{hyp:reg}, the asymptotic normality of $(nm/(n+m))^{1/2} (W_2(F_n, G_m)-W_2(F,G))$.} In particular, instead of Assumption \ref{hyp:reg}-(iii), Proposition 2.3 in \cite{del2019central} imposes strong and high-level conditions (see (2-7)-(2.9) in their paper), while Theorem 14 in \cite{berthet2020} also imposes strong regularity conditions. In particular, because their Assumption (FG) must hold for both the left and right tails of the distributions, one can show that their subconditions (FG1) and (FG3) already imply (up to letting $\eps=0$) Assumption \ref{hyp:reg}-(iii) for both $Z=Y$ and $Z=\eta_d$.\footnote{On the other hand, both \cite{berthet2020} and \cite{del2019central} also consider more general Wasserstein distances than just $W_2$.}

\medskip
\paragraph{Sketch of the proof.} In a first step, we account for the fact that $\eta_d$ and $E[\eta_d^2]$ are estimated. This requires in particular to show that
$$\sqrt{m} \int_0^1 F_n^{-1}(\widehat{G}^{-1}_m - G^{-1}_m)dt = - E[h(\eta_d)T'_{-1}]\sqrt{m}(\widehat{\gamma}-\gamma_0) + o_P\left(1\right),$$
where $\gamma_0$ is the limit in probability of $\widehat{\gamma}$. This result is not obvious; our proof relies in particular, again, on the Cambanis-Simons-Stout inequality. The second step is to study the asymptotic behavior of $(nm/(n+m))^{1/2} \int_0^1 [F_n^{-1}(t)G_m^{-1}(t)-F^{-1}(t) G^{-1}(t)]dt$. Here, we use the decomposition
\begin{align*}
\int_0^1 [F_n^{-1}(t)G_m^{-1}(t)-F^{-1}(t) G^{-1}(t)]dt  = & \int_0^1 F^{-1}(t)(G_m^{-1}(t)-G^{-1}(t))dt \\
& +\int_0^1 G^{-1}(t)(F_n^{-1}(t)-F^{-1}(t))dt  + r_{n,m},	
\end{align*}
where $r_{n,m} := \int_0^1 (F_n^{-1}(t)-F^{-1}(t))(G_m^{-1}(t)-G^{-1}(t))dt$. We prove that the first two terms $T_{1m}$ and $T_{2n}$ are asymptotically linear by adapting results on L-statistics, see in particular Theorem 1 in Chapter 19 of \cite{SW86}. That the remainder term $r_{n,m}$ is negligible if $Y$ (say) is finitely supported follows from the continuity of $G^{-1}$ at the support points of $Y$. Note that if this continuity condition does not hold, we lose asymptotic normality; see \cite{del2024central} for the exact distribution in such cases. If Assumption \ref{hyp:reg}-(iii) holds, we relate instead the remainder term to bounds on the convergence rate of $W_2(F_n,F)$ and $W_2(G_m,G)$. However, existing results on such rates, and in particular Theorem 1 in \cite{fournier2015rate}, are not sufficient for our purpose. Here, we improve upon their bound, which holds under weak restrictions, by leveraging in particular Condition \eqref{eq:condit_hazard}. We do this by linking $W_2(F_n,F)$ to the variance of order statistics, and relying on a lemma similar to  Corollary 2.12 in \cite{boucheron2015}; see Lemma \ref{lem:borne_var} in Online Appendix \ref{sec:additional_lemmas}.

\paragraph{Confidence intervals.}

We construct such intervals on $b_d$ using the asymptotic normality of $\widehat{\overline{b}}_d$ and a plug-in estimator of $V_d$. Specifically, let $\widehat{h}(x)= \int_0^1 F_n^{-1}[\widehat{G}_m(x^-)+u(\widehat{G}_m(x)-\widehat{G}_m(x^-))]du$ and
\begin{align*}
\widehat{\psi}_{1i} & := - \widehat{\overline{b}}_d \left(\widehat{\eta}_{di}^2 - \frac{1}{m}\sum_{j=1}^m \widehat{\eta}_{dj}^2\right),\\
\widehat{\psi}_{2i} & := -\left(\frac{1}{m}\sum_{j=1}^m \widehat{h}(\widehat{\eta}_{dj}) T_{-1j}'\right) \left(\frac{1}{m}\sum_{j=1}^m T_{-1j}T_{-1j}'\right)^{-1} T_{-1i} \widehat{\eta}_{di}, \\
\widehat{\psi}_{3i} & :=- \int [\ind{\widehat{\eta}_{di} \le t} - \widehat{G}_m(t)]F_n^{-1}\circ \widehat{G}_m(t)dt, \\
\widehat{\psi}_{4i} & := -\int [\ind{Y_i \le t} - F_n(t)]\widehat{G}_m^{-1} \circ F_n(t)dt.
\end{align*}
Then, define
$$\widehat{V}_d := \frac{1}{\left(\frac{1}{m}\sum_{j=1}^m \widehat{\eta}_{dj}^2\right)^2}\times \left[\frac{n}{m(n+m)}\sum_{j=1}^m \left(\widehat{\psi}_{1j} + \widehat{\psi}_{2j} + \widehat{\psi}_{3j}\right)^2 + \frac{m}{n(n+m)}  \sum_{i=1}^n  \widehat{\psi}_{4i}^2 \right].$$
Note that $\widehat{V}_d$ depends on $d$; in particular, $\widehat{V}_{-d}$ is the estimator of the asymptotic variance of $\overline{b}_{-d}=-\underline{b}_d$. We then consider the following confidence intervals on $b_d$ with nominal level $1-\alpha$:
$$\text{CI}_{1-\alpha} := \left[-\widehat{\overline{b}}_{-d} - z_{1-\alpha} \sqrt{\frac{n+m}{nm}\widehat{V}_{-d}}, \; \widehat{\overline{b}}_{d} + z_{1-\alpha} \sqrt{\frac{n+m}{nm} \widehat{V}_d}\,\right],$$
where $z_{1-\alpha}$ is the quantile of order $1-\alpha$ of a standard normal distribution. We can replace the usual quantile $z_{1-\alpha/2}$ by $z_{1-\alpha}$ since by Theorem \ref{thm:ident_no_common}, $\overline{b}_d >0>\underline{b}_{d}$ as long as $d'e_{p}=0$ and $V(Y)>0$. 

\begin{thm}\label{thm:asym_CI}
	Suppose that $\min(m,n)\to\infty$,  $n/(m+n)\to\lambda\in[0,1]$, Assumptions \ref{hyp:mom}-\ref{hyp:reg} hold, $d'e_{p}=0$ and $V(Y)>0$. Then,
$$\inf_{b_d\in [\underline{b}_d,\overline{b}_d]} \underset{n\to\infty}{\lim \sup}  \ P\left(b_d\in \CI\right)=1-\alpha.$$
\end{thm}

Once again, the proof of Theorem \ref{thm:asym_CI} is not straightforward. In particular, two difficulties are (i) to prove convergence of $(1/m)\sum_{j=1}^m \widehat{h}(\widehat{\eta}_{dj}) T_{-1j}'$; (ii) to handle the terms including $\widehat{\psi}_{3i}$ and $\widehat{\psi}_{4i}$. For (ii), we rely in particular on an extension of Lemma A.1 in \cite{del2019central}, see Lemma \ref{lem:A.1.modif} in Online Appendix \ref{sec:additional_lemmas}. 


\subsection{Common variables} 
\label{ssub:inf_common_variable}

\subsubsection{Computation of the estimator} 
\label{ssub:common_variables}

Let us turn to the case where common variables are observed. Then, we observe $\mathcal{S}_1:=\{(Y_1, W^{(1)}_1),....,(Y_n,W^{(1)}_n)\}$ and $\mathcal{S}_2:=\{(X_{o1}, W^{(2)}_1),....,(X_{om},W^{(2)}_m)\}$, where $W^{(1)}_i$ and $W^{(2)}_j$ are both distributed as $W$. We add the exponents $(\ell)$ to indicate that $W^{(\ell)}\in \mathcal{S}_\ell$. 

\medskip
When $W$ has a finite support $\{w_1,...,w_K\}$, we simply consider the plug-in version of \eqref{eq:sharp_bound_W}:
\begin{equation}
\widehat{\overline{b}}_d := \frac{1}{\widehat{E}(\widehat{\eta}_d^2)} \sum_{k=1}^K \widehat{p}_k \int_0^1 F_{n|k}^{-1}(u)\widehat{G}_{m|k}^{-1}(u)du,
    \label{eq:def_est_finite_W}
\end{equation}
where $\widehat{p}_k$ is an estimator of $P(W=w_k)$, $F_{n|k}$ (resp. $\widehat{G}_{m|k}$) is the empirical cdf of $Y$ (resp. $\widehat{\eta}_d$) on the subsample of $\mathcal{S}_1$ satisfying $W^{(1)}_i=w_k$ (resp., the subsample of $\mathcal{S}_2$ satisfying $W^{(2)}_j=w_k$). The estimator $\widehat{p}_k$ can be obtained by combining the two samples:
$$\widehat{p}_k:=\frac{1}{m+n}\left[\sum_{i=1}^n \ind{W^{(1)}_i=w_k} + \sum_{j=1}^m \ind{W^{(2)}_j=w_k}\right].$$

When $W$ does not have a finite support, or there are too few observations for a given value of $W$, the sharp bound involves the nonparametric functions $F^{-1}_{\nu_d|W}$ and $F^{-1}_{\nu_Y|W}$. To avoid the complications associated to their estimation, or very imprecise estimators, we exploit Theorem \ref{thm:caract_id_set_common} and consider the outer bound $\overline{b}^g_d$ for a function $g$ taking  finitely many values $(g_1,...,g_K)$. Nevertheless, we will show in Theorem \ref{thm:cons} that under suitable conditions, our estimator can still consistently estimate the sharp bound. We have 
$$\overline{b}^g_d = \frac{1}{E(\eta_d^2)}\left\{\delta_d' E(WW') \delta_Y + \sum_{k=1}^K p_k \E\left[F_{|k}^{-1}(U) G_{|k}^{-1}(U)\right]\right\},$$
where, with a slight abuse of notation, we let $p_k:=P(g(W)=g_k)$, $F_{|k}:=F_{\nu_Y|g(W)}(.|g_k)$ and $G_{|k}:=F_{\nu_d|g(W)}(.|g_k)$. Again, we consider a plug-in estimator of $\overline{b}^g_d$:
\begin{equation}
\widehat{\overline{b}}{}^g_d = \frac{1}{\widehat{E}(\widehat{\eta}_d^2)}\left\{\widehat{\delta}_d' \widehat{E}(WW') \widehat{\delta}_Y + \sum_{k=1}^K \widehat{p}_k \int_0^1 \widehat{F}_{|k}^{-1}(u)\widehat{G}_{|k}^{-1}(u)du\right\},
    \label{eq:est_with_g}
\end{equation}
where $\widehat{F}_{|k}$ (resp. $\widehat{G}_{|k}$) is the empirical cdf of $\widehat{\nu}_Y$ (resp. $\widehat{\nu}_d$) on the subsample of $\mathcal{S}_1$ satisfying $g(W^{(1)}_i)=g_k$ (resp., the subsample of $\mathcal{S}_2$ satisfying $g(W^{(2)}_j)=g_k$). As above, $\widehat{E}(WW')$ and $\widehat{p}_k$ are computed on the two samples.

\paragraph{Choice of $g(.)$.} Recall from Theorem \ref{thm:caract_id_set_common} that $\overline{b}^g_d$ is sharp if $\nu_Y$ $\indep W|g(W)$ and $\nu_d\indep W|g(W)$. Hence, we can expect tight bounds if $g(W)$ captures most of the dependence between $(\nu_Y,\nu_d)$ and $W$. Since $\nu_Y$ and $\nu_d$ are already residuals, we seek to capture possible heteroskedasticity by regressing $|\nu_Y|$ and $|\nu_d|$ linearly on $W$. This yields two indices, $W'\widehat{\varsigma}_Y$ and $W'\widehat{\varsigma}_d$; we let $\varsigma_Y$ and $\varsigma_d$ denote the corresponding population parameters and $g_0(W)=(W'\varsigma_Y,W'\varsigma_d)$. If $Y$ and $\eta_d$ satisfy a linear location-scale model, namely $Y= W'\delta_Y+(W'\varsigma_Y) \xi_Y$, $\xi_Y\indep W$, $W'\varsigma_Y>0$, where, without loss of generality, we normalize $\E|\xi_Y|=1$, and analogously for $\eta_d$, then $\nu_Y\indep W|g_0(W)$ and $\nu_d\indep W|g_0(W)$. The variable $g_0(W)$ is not finitely supported in general, however. Thus, we perform $K$-means clustering on $(W'\widehat{\varsigma}_Y, W'\widehat{\varsigma}_d)$. This yields a function $\widehat{g}(W)$ taking $K$ values only. The choice of $K$ is discussed in Section \ref{sec:simulations} below. 


\subsubsection{Asymptotic properties and inference}
\label{ssub:as_prop_common}

\paragraph{Finitely supported $W$.}

The estimator $\widehat{\overline{b}}_d$ defined in \eqref{eq:def_est_finite_W}  is asymptotically normal under similar conditions as in the case without common variables. Online Appendix \ref{app:as_var_common} gives the exact expression of the asymptotic variance. When including a finer $W$, the bounds become tighter. To maintain size control, we rely on \cite{stoye2020simple} to construct confidence intervals. This method has the appealing features of not requiring any tuning parameter, being simple to compute, and relying on mild conditions, beyond the joint asymptotic normality of the lower and upper bounds. We implement this inference method  in the applications (Section~\ref{sec:applications}). 

\paragraph{Other cases.}

The estimator $\widehat{\overline{b}}{}^g_d$ defined in \eqref{eq:est_with_g} is more complicated to analyze, in particular when $g(.)$ is estimated, as suggested above. Nevertheless, we show that under some conditions, it converges to $\overline{b}^{g_0}_d \ge \overline{b}_d$, where we recall that  $g_0(W) = (W'\varsigma_Y, W'\varsigma_d)$. In other words, under the conditions below, $\widehat{\overline{b}}{}^{\,\widehat{g}}_d$ always converges to an outer bound on $b_d$. Moreover, under location-scale models on $Y$ and $\eta_d$, since $\nu_Y \indep W \mid g_0(W)$ and $\nu_d \indep W \mid g_0(W)$, we have, in view of Theorem \ref{thm:caract_id_set_common}, $\overline{b}^{\,g_0}_d = \overline{b}_d$. Therefore, in such cases, $\widehat{\overline{b}}{}^{\,\widehat{g}}_d$ converges to the sharp bound $\overline{b}_d$. 

\medskip
Let $K_{n,m}$ denote the number of cluster centers used in the K-means algorithm, with the dependence on $n$ and $m$ made explicit here. We assume the following:

\begin{hyp}~
    \begin{enumerate}
        \item\label{kmeans-cond1} $W$ has compact support. There exist finite constants $L_Y,L_d$ such that, for every $v,v'\in\Supp(V)$,
        \begin{align}
            W_2(F_{\nu_Y|g_0(W)=v},F_{\nu_Y|g_0(W)=v'}) &\le L_Y\|v-v'\|,  \label{eq:w2-lipschitz-g0_0} \\
            W_2(F_{\nu_d|g_0(W)=v},F_{\nu_d|g_0(W)=v'}) & \le L_d\|v-v'\|.
            \label{eq:w2-lipschitz-g0_1}
        \end{align}
        \item\label{kmeans-cond2} \emph{(a)} There exist a neighborhood $\mathcal N$ of $\varsigma:=(\varsigma_Y,\varsigma_d)$ and constant $0<\overline f<\infty$ such that, for every $\widetilde\varsigma\in\mathcal N$, the random vector
        $Z(\widetilde\varsigma):=W'\widetilde\varsigma$ admits a density $f_{Z(\widetilde\varsigma)}$ on a compact support $\mathcal X_{\widetilde\varsigma}\subset\R^{2}$ satisfying
        \begin{equation}
            \sup_{\widetilde\varsigma\in\mathcal N}\sup_{z\in\R^{2}} f_{Z(\widetilde\varsigma)}(z)\le \overline f.
            \label{eq:uniform-projected-density}
        \end{equation}

        \medskip
        \emph{(b)} There exists $\eps>0$ such that
        \begin{equation}
            \sup_{w\in\Supp(W)}\E\!\left[|\nu_Y|^{2+\eps}+|\nu_d|^{2+\eps}\mid W=w\right]<\infty.
            \label{eq:moment_consistency}
        \end{equation}
        \item\label{kmeans-cond3}  We have $\min(m,n)\to\infty$, $n/(m+n)\to\lambda\in(0,1)$, $K_{n,m}\to\infty$, and
        \begin{equation}
            \frac{K_{n,m}^{5}\log K_{n,m}}{\min(n,m)}\to0.
            \label{eq:K-rates-consistency}
        \end{equation}
    \end{enumerate}
    \label{hyp:consistency_ghat}
\end{hyp}

Assumption \ref{hyp:consistency_ghat}.\ref{kmeans-cond1} is a regularity condition satisfied for instance in the location-scale model. To see this, note that then $\nu_k=(W'\varsigma_k)\xi_k$ (with $k=Y$ or $k=d$), and by considering the true copula of $(v_{r(k)}\xi_k,v_{r(k)}'\xi_k)$ with $r(Y) =1$, $r(d)=2$, we obtain, for any $v=(v_1,v_2)$ and $v'=(v_1',v_2')$,
\[
 W_2(F_{\nu_k|g_0(W)=v},F_{\nu_k|g_0(W)=v'})\le \E[\xi_k^2]^{1/2}\,|v_{r(k)}-v_{r(k)}'|\le \E[\xi_k^2]^{1/2}\,\|v-v'\|.
\]

Assumption \ref{hyp:consistency_ghat}.\ref{kmeans-cond2}  has two parts. Part (a) holds for instance if (i) given the other components of $W$, two components of $W$ admit a bounded density in $\R^2$ and (ii) their vector of coefficients in $g_0(W)$, $(\varsigma_{Y,c_1},\varsigma_{Y,c_2})'$ and $(\varsigma_{d,c_1},\varsigma_{d,c_2})'$, say, are not collinear in $\R^2$. Condition (b) holds if $\nu_Y$ and $\nu_d$ have $2+\eps$ moments and $w \mapsto \E[|\nu_Y|^{2+\eps}+|\nu_d|^{2+\eps}\mid W=w]$ is continuous (since $W$ has compact support). This latter condition holds in particular in the location-scale model.

\medskip
Finally, Assumption \ref{hyp:consistency_ghat}.\ref{kmeans-cond3} imposes that $K_{n,m}$ tends to infinity, ensuring that we can approximate $g_0(W)$ arbitrarily well, but also that it does not grow too fast, ensuring that the number of units in each cell tends to infinity and thus $\int_0^1 \widehat{F}_{|k}^{-1}(u)\widehat{G}_{|k}^{-1}(u)du$ converges for each $k$.

\begin{thm}[Consistency]\label{thm:cons}
	Suppose Assumptions \ref{hyp:mom}-\ref{hyp:samples} and Assumption \ref{hyp:consistency_ghat} hold. Then,  
	$$\widehat{\overline b}{}_d^{\,\widehat g_{K_{n,m}}} \;\convP \overline b_d^{\,g_0}.$$
	Moreover, if, for $U\in\{Y,\eta_d\}$, $U= W'\delta_U+(W'\varsigma_U) \xi_U$ with $\xi_U\indep W$, then $\overline b_d^{\,g_0}=\overline b_d$.
\end{thm}

The proof is long, but uses three key ingredients: (i) a bound on the complexity of Voronoi cells in $\R^q$; (ii) convergence of the empirical $K$-means solution, which is not straightforward here because $K_{n,m}$ tends to infinity and (iii) Zador's theorem \citep[Theorem~6.2]{GrafLuschgy2000}, which implies that the error due to replacing \(g_0(W)\) by a finite number of K-means cells converges to zero. A more detailed sketch is provided at the beginning of the proof.

\medskip
Finally, we leave formal results on the asymptotic normality of the estimated bound for future research. When $g$ is not estimated, this should follow from a similar reasoning as in Theorem \ref{thm:asym_Xo}. Note however that the asymptotic variance of $\widehat{\overline{b}}_d$ now includes additional terms due to the estimation of $\nu_Y$, $\nu_d$ and $\delta_d'E[WW']\delta_Y$. Its exact expression, which includes eleven terms instead of four as above, is given in Online Appendix \ref{app:as_var_common}. We also suggest using Stoye's confidence interval in this case. We conjecture that, under suitable regularity conditions, estimating $g$ does not affect the asymptotic distribution of the bound. We investigate this conjecture in the simulations below.



\section{Simulations}\label{sec:simulations}

We now study the finite sample performances of our estimators and inference method. We consider a single DGP encompassing three cases of available data: one in which only $Y$ and $X_o$ are available, one in which $X_c$ is also observed jointly and enters the main regression and one in which $W_a$, in addition to $X_c$, is observed. In the latter case, the parameters remain the same as in the second case. The DGP is as follows. We let $W_a \sim \mathcal{U}[0,1]$, $X_c \sim \mathcal{N}(0, 1)$ and
    \begin{align*}
     X_o = & X_c a_1 + W_a a_2 + (1+ W_a  d_1) \eta, \; \eta | W_a, X_c\sim \mathcal{N}(0, \sigma^2_\eta),\\ 
    Y = & X_o b_1 + X_c b_2 + W_a d_2 + \varepsilon,\; \varepsilon | X, W_a, \eta \sim \mathcal{N}(0, \sigma^2_\varepsilon).   
    \end{align*}
We fix $a_1=1$, $a_2=10$, $d_1=1$, $\sigma_\eta = 1$, $b_1=1$, $b_2=1$, $d_2=0.25$ and $\sigma_\varepsilon = 4$. The true bounds in the first two cases are obtained by simulations, whereas there is a closed-form expression in the last case. We fix $n=m$ and vary it from 400 to 4,800. With $X_c$ or $(X_c, W_a)$, we construct $g(W)$ as described in Subsection \ref{ssub:common_variables}, with $K=\max(2,\lfloor N^{0.15}\rfloor)$ satisfying the theoretically-permissible upper bound \eqref{eq:K-rates-consistency}, where $\lfloor x\rfloor$ denotes the integer part of $x$; we discuss alternative choices of $K$ below.  The results are displayed in Table \ref{table:MC:simus}. We report the average of the estimated bounds (``Bounds'') and the average of the estimated 95\% confidence intervals $\text{CI}_{1-\alpha}$  (``95\% CI'') for $b^{0,1}$. We also report the mean difference between the length of the confidence sets and that of the identified set, see column ``Ex. length'' in the table. Finally, the column ``Covg'' corresponds to the minimum, over $b_1$ in the identified set of $b^{0,1}$, of the estimated probability that $b_1$ belongs to the confidence interval.

\medskip
A couple of remarks are in order. First, as expected, the 95\% confidence intervals shrink with the sample sizes $n$,  approximately at the $n^{-1/2}$ rate in the three cases we consider. This is reflected in the evolution of the excess length across sample sizes. Second, the confidence intervals exhibit satisfactory coverage. In particular, coverages for all panels are generally close to the nominal 95\% level, even for small sample sizes. Coverage rates are generally conservative, but still very close to the nominal level for the specification reported in Panel 3. This is remarkable: one would in principle need to use the continuous variable $g(W)=1+W_ad_1$ to obtain the sharp bounds, by Theorem \ref{thm:caract_id_set_common}, whereas we instead rely on a finitely supported variable $g(W)$ with few points of support (from 2 to 3 when $n$ varies from $400$ to $4,800$). Third, and importantly, the identified set is much tighter in Panel 3 than in Panels 1 and 2. This illustrates the substantial identifying power of the auxiliary variable $W_a$. For this particular DGP, the identifying power - measured by the reduction in the length of the identified set - of $W_a$ is in fact larger than that of the common regressor $X_c$.

\begin{table}[H]
\begin{center}
\begin{tabular}{rccccc}
  \toprule
 & & Bounds & 95\% CI & Ex. length & Covg. \\
\midrule
\textbf{Panel 1:}   & &  \multicolumn{4}{l}{Without $(X_c,W_a)$}  \\
Identified     &  & [-1.625,1.627] & &  &  \\
400 & & [-1.620,1.622] & [-1.743,1.745] & 0.235 & 0.929 \\
800 & & [-1.624,1.626] & [-1.711,1.713] & 0.173 & 0.944 \\
1,200 & & [-1.623,1.625] & [-1.694,1.696] & 0.138 & 0.938 \\
2,400 & & [-1.625,1.627] & [-1.675,1.677] & 0.100 & 0.943 \\
4,800 & & [-1.625,1.627] & [-1.660,1.662] & 0.071 & 0.942 \\
\midrule
\textbf{Panel 2:}   & &  \multicolumn{4}{l}{With $X_c$}  \\
Identified     &  & [-1.583,1.585] & &  &  \\
400 & & [-1.572,1.575] & [-1.704,1.706] & 0.242 & 0.940 \\
800 & & [-1.578,1.581] & [-1.673,1.676] & 0.181 & 0.956 \\
1,200 & & [-1.579,1.581] & [-1.657,1.659] & 0.148 & 0.958 \\
2,400 & & [-1.579,1.582] & [-1.633,1.636] & 0.102 & 0.960 \\
4,800 & & [-1.580,1.583] & [-1.619,1.622] & 0.074 & 0.959 \\
\midrule
\textbf{Panel 3:}   & &  \multicolumn{4}{l}{With $(X_c,W_a)$}  \\
Identified     &  & [0.196,1.405] & &  &  \\
400 & & [0.204,1.405] & [0.056,1.567] & 0.301 & 0.931 \\
800 & & [0.202,1.412] & [0.098,1.526] & 0.219 & 0.938 \\
1,200 & & [0.196,1.404] & [0.111,1.497] & 0.176 & 0.940 \\
2,400 & & [0.198,1.406] & [0.138,1.472] & 0.124 & 0.937 \\
4,800 & & [0.196,1.405] & [0.154,1.451] & 0.089 & 0.948 \\
\bottomrule
\end{tabular}

\end{center}
\vspace{0.1cm}
\footnotesize{Notes: results obtained with 2,000 simulations for each row.  400, 800 etc. correspond to the sizes of the two samples ($n=m$). Column ``Bounds'' reports either the identified set or the average of the estimated bounds over simulations. Column ``95\% CI'' reports the average of the 95\% confidence intervals over simulations. ``Ex. Length'' is the excess length, i.e. the average length of the confidence region minus the length of the identified set.  Column ``Covg.'' displays the minimum, over $b_1\in \B_{e_1}$, of the estimated probability that $b_1\in\CI(b^{0,1})$. In Panel 1,  the true coefficients of $(X_o, 1)$ are  $(1.103,-0.393)$, while in Panels 2-3, the true coefficients of $(X_o, X_c, 1)$ are $( 1.019,0.981,0.026)$.} 
\caption{Monte Carlo simulation results on the confidence intervals for $b^{0,1}$}
\label{table:MC:simus}
\end{table}

\medskip
Table \ref{table:MC_time} reports the computational time needed to compute the estimated bounds and associated confidence intervals. When $W_a$ is observed, this time also includes the $K$-means clustering we perform to compute $g(W)$. The main takeaway is that our procedure is very fast: it takes less than 2 seconds when observing $(X_c,W_a)$ with $n=m=12,000$, and about 15 seconds even with $n=m$ as large as 120,000.

\begin{table}[H]
\begin{center}
\begin{tabular}{rccccc}
\toprule
$n(=m)$ &  Without $(X_c,W_a)$& With $X_c$ & With $(X_c,W_a)$ \\
\midrule
1,200 & 0.003 & 0.087 & 0.106 \\
12,000 & 0.033 & 0.920 & 1.174 \\
120,000 & 0.21 & 10.46 & 14.93 \\
\bottomrule
\end{tabular}
\end{center}
\vspace{0.1cm}
\footnotesize{Notes: these times are obtained on the same DGP as above, taking the average over 100 replications and using our companion R package \texttt{RegCombinBLP}. We  parallelize the computation over 20 CPUs on an Intel Xeon Gold 6130 CPU 2.10GHz with 382Gb of RAM.} 
\caption{Time (in s.) for computing the point estimates and confidence intervals.}
\label{table:MC_time}
\end{table}

Finally, we explore the effect of the tuning parameter $K$ on coverage; see Table \ref{table:choice_K} below, where we consider two sample sizes ($n=1,200$ and $n=6,000$). As expected, increasing $K$ decreases the length of the CIs, but also reduces coverage. This probably reflects the fact that the estimated bounds become biased for larger $K$. On the other hand, coverage remains around 95\% for $K=\max(2,\lfloor \min(n,m)^c\rfloor)$ with $c<1/3$, suggesting that our baseline choice $c=0.15$, motivated by the theoretical bound \eqref{eq:K-rates-consistency}, performs well in practice, while larger exponents also appear to yield good results.

\begin{table}[H]
\begin{centering}
\begin{tabular}{rccccc}
\toprule
Number of points $K$ &  & Bounds & 95\% CI & Ex. length & Covg. \\
\midrule
{\bf Panel 1:} & & $n=1,200$ & & &  \\
Identified  &  & [0.196,1.405]  & & &   \\
2 & & [0.198,1.409] & [0.113,1.502] & 0.180 & 0.941 \\
3 & & [0.201,1.407] & [0.117,1.500] & 0.174 & 0.936 \\
4 & & [0.201,1.399] & [0.117,1.492] & 0.166 & 0.928 \\
5 & & [0.204,1.401] & [0.119,1.494] & 0.165 & 0.927 \\
10 & & [0.210,1.392] & [0.126,1.485] & 0.149 & 0.915 \\
20 & & [0.223,1.378] & [0.138,1.471] & 0.123 & 0.860 \\
30 & & [0.236,1.370] & [0.150,1.463] & 0.104 & 0.823 \\
\midrule
{\bf Panel 2:} & & $n=6,000$ & & &  \\
Identified  &  & [0.196,1.405]  & & &   \\
2 & & [0.195,1.408] & [0.157,1.449] & 0.083 & 0.947 \\
3 & & [0.196,1.406] & [0.158,1.448] & 0.081 & 0.947 \\
4 & & [0.198,1.405] & [0.160,1.447] & 0.078 & 0.948 \\
5 & & [0.197,1.404] & [0.159,1.446] & 0.078 & 0.945 \\
10 & & [0.198,1.401] & [0.160,1.443] & 0.074 & 0.938 \\
20 & & [0.201,1.400] & [0.164,1.441] & 0.068 & 0.919 \\
30 & & [0.204,1.397] & [0.167,1.439] & 0.063 & 0.887 \\
\bottomrule
\end{tabular}
\end{centering}
\vspace{0.3cm}

\footnotesize{Notes: same DGP as above, observing $(X_c,W_a)$. 3, 5, 10 etc. are the number of points $K$ taken by $g(\cdot)$. Column ``Bounds'' reports either the identified set or the average of the estimated bounds over simulations. Column ``95\% CI'' reports the average of the 95\% confidence intervals over simulations. ``Ex. Length'' is the excess length, i.e. the average length of the confidence region minus the length of the identified set.  Column ``Covg.'' displays the minimum, over $b_1\in \B_{e_1}$, of the estimated probability that $b_1\in\CI(b^{0,1})$. The results are obtained with 1,000 simulations for each sample size.} 
\caption{Simulation results when varying the number of points $K$ taken by $g$.}

\label{table:choice_K}
\end{table}

\section{Applications}\label{sec:applications}

We now illustrate our approach with two applications. We first study the influence of race on the probability of patent approval in the United States, revisiting recent work on this question \citep{Dossi23}. We then investigate the relationship between students' risk and time preferences and educational achievement across countries \citep{hanushek2020culture}.

\subsection{Race and patent approval}\label{ssec:appli1}

In our first application, we investigate the existence and magnitude of racial and ethnic gaps in science and innovation. This question has attracted much interest in the recent empirical literature \citep[see, e.g.][]{Kerr08,ADQW24,Dossi23}. A key challenge is that datasets typically do not measure race and ethnicity together with the outcome of interest. Using our notation, race/ethnicity is an outside regressor ($X_o$), with successful patent application being the outcome of interest ($Y$). Instead of $X_o$, researchers may observe other characteristics, such as the applicant's name. Then, in other datasets, we may observe these characteristics together with race and ethnicity. A commonly used strategy in this context is to impute race and ethnicity using applicant characteristics observed in both datasets. We take a different route and derive bounds that use both datasets without relying on the exclusion restriction implicit in the imputation approach.

\medskip
Following \cite{Dossi23}, we rely on two datasets. The first is the publicly available dataset released by the United States Patent and Trademark Office (USPTO) covering the universe of patent applications submitted in the United States. We use the Patent Examination research dataset (PatEx), which contains detailed information on all patent applications, including the full names of the applicants \citep{graham2015uspto}. We restrict the sample to applications filed between January 2001 and December 2018 and focus on utility patents.\footnote{Utility patents, also referred to as ``patents for invention'',  constitute 90\% of the patent documents issued by the USPTO in recent years.} We further restrict the sample to applicants based in the United States, and as in \cite{Dossi23}, consider only the first inventor listed on the application.

\medskip
We combine the PatEx dataset with data from the US Census. Namely, we use the information on the aggregate frequency of last names by race and ethnicity from the 2010 Decennial Census Surname Table \citep[][]{comenetz2016frequently}. 6.3 million different last names were recorded for 295 million people.\footnote{In the following, we neglect the statistical uncertainty related to this sample.} Among them, we use the publicly released frequency by race and ethnicity of the 162,254 last names that occur more than 100 times, representing 90.1\% of the overall population. We consider in our analysis five different categories of race and ethnicity, namely: (i) Black or African American (11.99\%), (ii) Asian and native Hawaiian and other Pacific Islander (4.86\%), (iii) Hispanic or Latino (16.29\%), (iv)  American Indian or Alaska Native (1.76\%), and (v) others, which includes White (64.40\%) and those declaring to belong to two or more races (0.69\%), and is used as our reference category.\footnote{Estimation results are robust to splitting the “two or more races” category evenly across the other racial categories.}

\medskip
Estimation results are reported in Table \ref{table:appli1:main}, where the first column presents the TSTSLS point estimates, the second the point estimates of our bounds, and the last column reports the 95\% confidence intervals computed from our asymptotic normality results. We use applicant's last name as an auxiliary variable ($W_a$). Since inventors are a subset of the whole population, the bounds are based on Equation \eqref{eq:bound_subpop} above. We implement the estimators $\widehat{\overline b}{}_d^{\,\widehat g_{K}}$ with $K=\max(2,\lfloor \min(n,m)^{0.15}\rfloor)$, as recommended in the previous section. 

\medskip
Importantly, in contrast to our bounds, the TSTSLS estimates rely on an exclusion restriction. Namely, the applicant's last name is assumed not to be predictive of patent approval conditional on the applicant's race and ethnicity. While this type of name-based exclusion restriction has frequently been used in applied work, its validity is far from obvious in this particular context. Patent examiners always observe applicants' last names but may not observe race, as they only infrequently interview them in person (see \citeauthor{CKS02}, 2002, and \citeauthor{Avivi24}, 2024 for discussions of the USPTO selection process). Thus, one may argue that any racial discrimination in the patent approval process could in fact operate largely through the applicant's last name. This would violate the TSTSLS exclusion restriction.\footnote{This line of reasoning may suggest that the TSTSLS would identify instead the effect of, e.g., having a Black- or Asian-sounding name. It is unclear whether this interpretation is warranted either. First, names could also predict other relevant characteristics. Second, this interpretation would require another exclusion restriction, namely that the coefficients of race in the ``long'' regression are zero, which is arguably strong as well.}

\medskip
Turning to the results, a key takeaway is that the TSTSLS results that are obtained using last names as an exclusion restriction are fragile. Notably, while the TSTSLS estimates point to Black inventors being significantly less likely to be granted a patent, the bounds obtained with our method for this coefficient are wide, with a lower bound as large as $-0.817$ and an upper bound that is positive and large as well ($0.519$). While a similar conclusion holds for Hispanics and American natives, our bounds are somewhat more informative for the coefficient on Asians, with a lower bound of $-0.192$ and an upper bound of $0.259$. At any rate, these results indicate that the conclusions one would reach from the TSTSLS estimates of significant racial differences in the probability of being granted a patent crucially hinge on the underlying exclusion restriction. 

\begin{table}[H]
\begin{center}
		\scalebox{1}{
			\begin{tabular}{rcccc}
\toprule
Coefficient & TSTSLS & Eq.~(11) bounds & 95\% CI \\
\midrule
Black & -0.038 & [-0.817, 0.519] & [-0.828, 0.525] \\
 & (0.006) & & \\
Hispanic & -0.032 & [-0.631, 0.448] & [-0.655, 0.463] \\
 & (0.004) & & \\
Asian & 0.041 & [-0.192, 0.259] & [-0.207, 0.268] \\
 & (0.002) & & \\
American native & -0.047 & [-1.048, 1.032] & [-1.055, 1.053] \\
 & (0.042) & & \\
\bottomrule
\end{tabular}
}
            \end{center}
\vspace{0.1cm}
\footnotesize{Notes:  $W_a = \text{last names}$ and no $X_c$, 2,146,799  patent applications and 91,055 applicants last names. K-means uses $K=8$, $c=0.150$. The CIs and standard errors are obtained clustering at the last name level, following \cite{Dossi23}.}
            \caption{Estimation results for racial inequalities in patent approval}
            \label{table:appli1:main}
\end{table}

Another notable point is that, although the bounds reported in Table~\ref{table:appli1:main} tend to be wide, using last names as $W_a$ does yield substantial improvements over the simple bounds based solely on $Y$ and $X_o$. In particular, without $W_a$, the sharp lower bound for each of the four coefficients equals -1 and is therefore not informative.\footnote{This occurs here because (i) $P(Y=0)$ is larger than $P($race$)$ for all other races than White and multiracial applicants, and ii) $P(Y=1)$ is larger than $P($White$)$.} Hence, even without exclusion restrictions, observing last names in both datasets delivers meaningful informational gains, especially so for the Asian coefficient. We show in Appendix \ref{app:details_1st_app} that bounds based on Equation \eqref{eq:sharp_bound_W} rather than Equation \eqref{eq:upper:sigma_w1} yield further improvements, although the bounds remain wide.

\begin{figure}[H]
{\centering
    \begin{tabular}{cc}
 \includegraphics[width=1\linewidth, height=0.3\textheight]{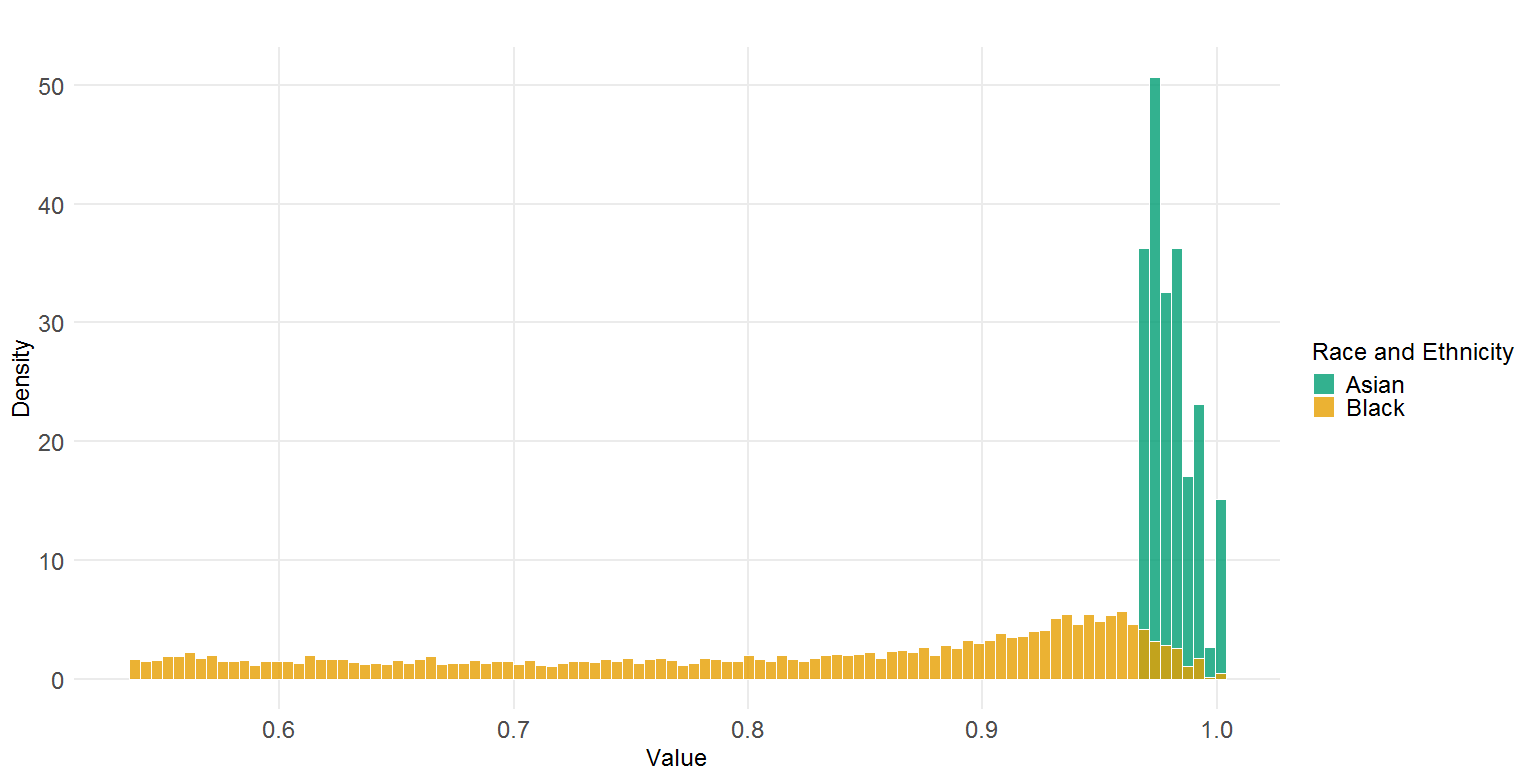}\\
   \end{tabular}}
   \vspace{0.1cm}

\footnotesize{Notes:  10\% of the associated population after sorting by predictability represents 6,669 names for Blacks (dark orange), and  637 for Asians (dark green). We use 100 bins.}
     \caption{Distribution of racial frequencies for a given name for the most predictive names, which cumulatively account for 10\% of the associated population}
   \label{fig:distribution}
\end{figure}

\medskip
Next, we investigate why our bounds are more informative for some races/ethnicities than others. To do so, we report in Figure~\ref{fig:distribution} the distribution of the racial frequencies conditional on last name, focusing on the names that are the most predictive of race and that together account for 10\% of each subpopulation (here, Blacks and Asians - groups for which the bounds are relatively wide and more informative, respectively). This figure shows that last names are highly predictive of being Asian, much more so than for Blacks. This illustrates the connection between the informativeness of our bounds and the extent to which $W_a$ (inventor's last name) is predictive of $X_o$ (race/ethnicity).

\subsection{Preferences and educational achievement}\label{ssec:appli2}

Preference parameters, especially patience and risk taking, play an important role in human capital investment decisions. However, to our knowledge, no single data set jointly measures these preferences and test scores across countries. In the following, we build on the cross-country analysis of \cite{hanushek2020culture} and combine data from the OECD's Programme for International Student Assessment (PISA) with the Global Preference Survey (GPS) to examine how students' time and risk preferences are associated with educational achievement. 

\medskip
PISA assesses achievement in mathematics, science and reading for random samples of 15-year-old students on a three-year cycle, providing repeated cross-sectional data representative of each country-by-wave cell. In the following, we consider as our main ``Y dataset'' the standardized math test scores over the seven waves of PISA testing, covering the period 2000-2018. Over this period, a total of 86 countries participated at least once. We combine these test scores with data from the Global Preference Survey \citep[][]{falk2018global}. The GPS provides scientifically validated data on several preference parameters from representative samples, of around 1,000 respondents in each country surveyed in 2012, measuring patience, risk taking, positive and negative reciprocity, altruism, and trust $(X_{o})$, for 49 different countries. The GPS also records gender for each respondent, which we use as a common regressor $X_c$, together with the country. Restricting the analysis to this subset of 49 countries yields test score data for a total of 1,992,276 students.\footnote{See Appendix \ref{app:details_2nd_app} for more details on the GPS dataset, especially on the measurements of preference parameters. As the PISA and GPS datasets are representative of the same common population after reweighting the observations by the corresponding survey weights, we use the survey weights in our analysis.} 

\medskip
In Table \ref{tab:appli2main}, we compare our bounds on the coefficients of patience and risk taking with the TSTSLS estimates considered by \cite{hanushek2020culture}, where both variables are imputed using country dummies. We consider alternative specifications depending on whether only patience, only risk taking or both are included in the regression. In Panel D, we also include other preference variables (positive and negative reciprocity, altruism and trust) as controls in the regression. Hence, in this last specification, $X_o$ is of dimension 6. In all cases, we compute the bounds $\widehat{\overline{b}}{}^{\,\widehat{g}}_d$, with $\widehat{g}$ obtained by K-means and $K$ determined as in the simulations. 

\medskip
A couple of comments are in order. A first takeaway is that, in contrast to the previous application and despite the absence of any $W_a$ here, our bounds tend to be informative about magnitudes. This holds for both the coefficients of patience and risk-taking, and across all four specifications reported in the table. That the bounds remain informative is particularly noteworthy in Panel D, where we control for four additional preference parameters all included in $X_o$. One might indeed have expected that increasing the dimension of $X_o$ would cause the bounds to widen substantially, yet this is not the case here.

\medskip
Related to this, focusing on Panels C and D, the TSTSLS estimates of the coefficients associated with patience and risk-taking both lie outside of the estimated bounds and the corresponding 95\% confidence intervals. One-sided tests comparing the TSTSLS estimates with the relevant bounds reject equality, for both panels, for the coefficient associated with patience at the 5\% level and for risk-taking at the 10\% level, consistent with a violation of the underlying exclusion restrictions.  Recall that the TSTSLS estimator relies on the arguably strong assumption that countries do not affect test scores beyond their effects through risk aversion and patience. 

\begin{table}[H]
    \begin{center}
        \scalebox{0.76}{
           \begin{tabular}{lcccccccc}
\hline
 & \multicolumn{3}{c}{\textbf{Panel A}} & \multicolumn{3}{c}{\textbf{Panel B}} \\
\cline{2-4} \cline{5-7}
 & \text{TSTSLS} & \text{Eq.(7) bnd.} & \text{CI on } $b^{0,1}$ & \text{TSTSLS} & \text{Eq.(7) bnd.} & \text{CI on } $b^{0,1}$ \\
\text{Patience} & 0.898 & [-0.893,0.896] & [-0.981,0.982] & - & - & - \\
 & (0.125) &  &  & - & - & - \\
\text{Risk-taking} & - & - & - & -0.592 & [-0.945,0.945] & [-1.004,1.004] \\
 & - & - & - & (0.284) &  &  \\
\hline
 & \multicolumn{3}{c}{\textbf{Panel C}} & \multicolumn{3}{c}{\textbf{Panel D}} \\
\cline{2-4} \cline{5-7}
 & \text{TSTSLS} & \text{Eq.(7) bnd.} & \text{CI on } $b^{0,1}$ & \text{TSTSLS} & \text{Eq.(7) bnd.} & \text{CI on } $b^{0,1}$ \\
\text{Patience} & 1.170 & [-0.914,0.916] & [-1.010,1.012] & 1.119 & [-0.915,0.917] & [-1.011,1.012] \\
 & (0.124) &  &  & (0.119) &  &  \\
\text{Risk-taking} & -1.308 & [-0.972,0.972] & [-1.027,1.027] & -1.342 & [-1.007,1.007] & [-1.056,1.056] \\
 & (0.209) &  &  & (0.256) &  &  \\
\text{Additional controls} &  &  &  & X & X & X \\
\cline{2-4} \cline{5-7}
\text{\textbf{Tests equality}} & Stat. & p.value &  & Stat. & p.value &  \\
\cmidrule{2-3} \cmidrule{5-6}
\text{U. bnd. Patience} & 2.158 & 0.015 &  & 1.816 & 0.035 &  \\
\text{L. bnd. Risk-taking} & 1.583 & 0.057 &  & 1.298 & 0.097 &  \\
\hline
\end{tabular}
        }
    \end{center}
    \vspace{0.1cm}
    \footnotesize{Notes: $X_c$ includes countries and gender dummies, no $W_a$ here. Dependent variable: PISA math test score in all PISA waves 2000–2018.  Respectively 1,992,276  and  49,689 observations for the PISA and GPS datasets. Least squares regression weighted by students’ sampling probability. Additional preference controls are positive and negative reciprocity, altruism and trust. The standard errors of the TSTSLS estimators, our confidence intervals and the tests of equality between our lower or upper bounds  and the TSTSLS estimators take into account the clustering at the country level.}
        \caption{Preferences and Student Math Achievement across Countries}
        \label{tab:appli2main}
\end{table}

In terms of magnitudes, focusing on Panel D where we include additional preference controls, our bounds indicate that a 1 standard deviation (SD) increase in patience is at most associated with 0.917 SD increase in math test scores, against 1.119 SD using TSTSLS. Similarly, it follows from our bounds that a 1 SD increase in risk-taking is, at most, associated with a decline of 1.007 SD in student achievement, against a larger decline of 1.342 using TSTSLS.

\medskip
Finally and importantly for practice, our inference method remains computationally tractable even with large datasets and several outside regressors. Even though the datasets used in this analysis contain a very large number of observations (1,992,276 and 49,689), it takes 5 minutes only to reproduce the results of Panels A and B, 9 minutes for Panel C, and 21.5 minutes for Panel D (where $X_o$ is of dimension 6), using our R package \texttt{RegCombinBLP}.\footnote{We parallelize the computation over 15 CPUs on an Intel Xeon Gold 6130 CPU 2.10GHz with 382Gb of RAM.} 

\section{Conclusion} 
\label{sec:conclu}

We study regression coefficients in a context where the outcome of interest and some of the covariates are observed in two different datasets that cannot be matched. This type of data combination environment arises very frequently in empirical work. The usual approach, which consists in imputing the outcome $Y$ or the outside regressors $X_o$ using auxiliary variables $W_a$, hinges on exclusion restrictions that may not hold in practice. We take a different route and derive sharp bounds on the regression coefficients using only the observed distributions. As they take a simple form, these bounds can be estimated at a low computational cost. We establish the asymptotic normality of the resulting estimators and derive asymptotically valid confidence intervals that are easy to compute. Our framework provides a practical way to evaluate the extent to which conclusions based on commonly used imputation methods are sensitive to the underlying exclusion restrictions. 

\medskip
We illustrate our method with two applications. The first studies racial disparities in patent approval, the second the effects of patience and risk-taking on test scores. The first application highlights that in some cases, results based on an imputation approach crucially rely on the underlying exclusion restriction; without it, uncertainty on the true coefficients of interest remains large. The second application shows that our bounds can be informative on the magnitude of the effects, and can also lead to reject the exclusion restrictions underlying the imputation-based approach. 


\newpage
\bibliography{references}

\appendix

\section{Comparison with \cite{pacini2019two}} 
\label{sub:bounds_in_pacini}

\subsection{Sharpness} 
\label{sub:non_sharpness_of_pacini_s_bounds}

\cite{pacini2019two} gives the expression of $\overline{b}_d$, also allowing for common regressors (denoted by $z$ in his paper) but not for additional variables $W_a$. His bounds coincide with ours when $X_o$ is univariate, but not otherwise. In the multidimensional case, his expression of $\overline{b}_d$ is an upper bound of the true bound. This is so because the equality in Lemma 5 of \cite{pacini2019two} should be replaced by an inequality.

\medskip
To see this, first remark that $\mathcal{F}$ there is the set of cdfs $(F_{1y},...,F_{d_x y})$ that are compatible with the distributions of $(x, z)$ and $(y, z)$, with  $F_{ky}$ denoting the joint cdf of $(x_k,y)$. Hence, in the third equality ``$F_{ky} \in \mathcal{F}$'' is not well-defined. A natural fix is then to replace it by ``$F_{ky} \in \mathcal{F}_k$'', where $\mathcal{F}_k$ denotes the set of cdfs $F_{ky}$ compatible with the laws of $(x, z)$ and $(y, z)$. But then, the third equality in the proof of Lemma 5 does not hold, because  $\mathcal{F}$ is not a cartesian product of $\mathcal{F}_k$ in general: it is instead a (strict in general) subset of the cartesian product.


\subsection{Numerical comparison} 
\label{sub:comparison_between_sharp_and_pacini_s_bounds}

\begin{figure}[H]
	\begin{center}
\includegraphics[width=0.75\linewidth, height=0.3\textheight]{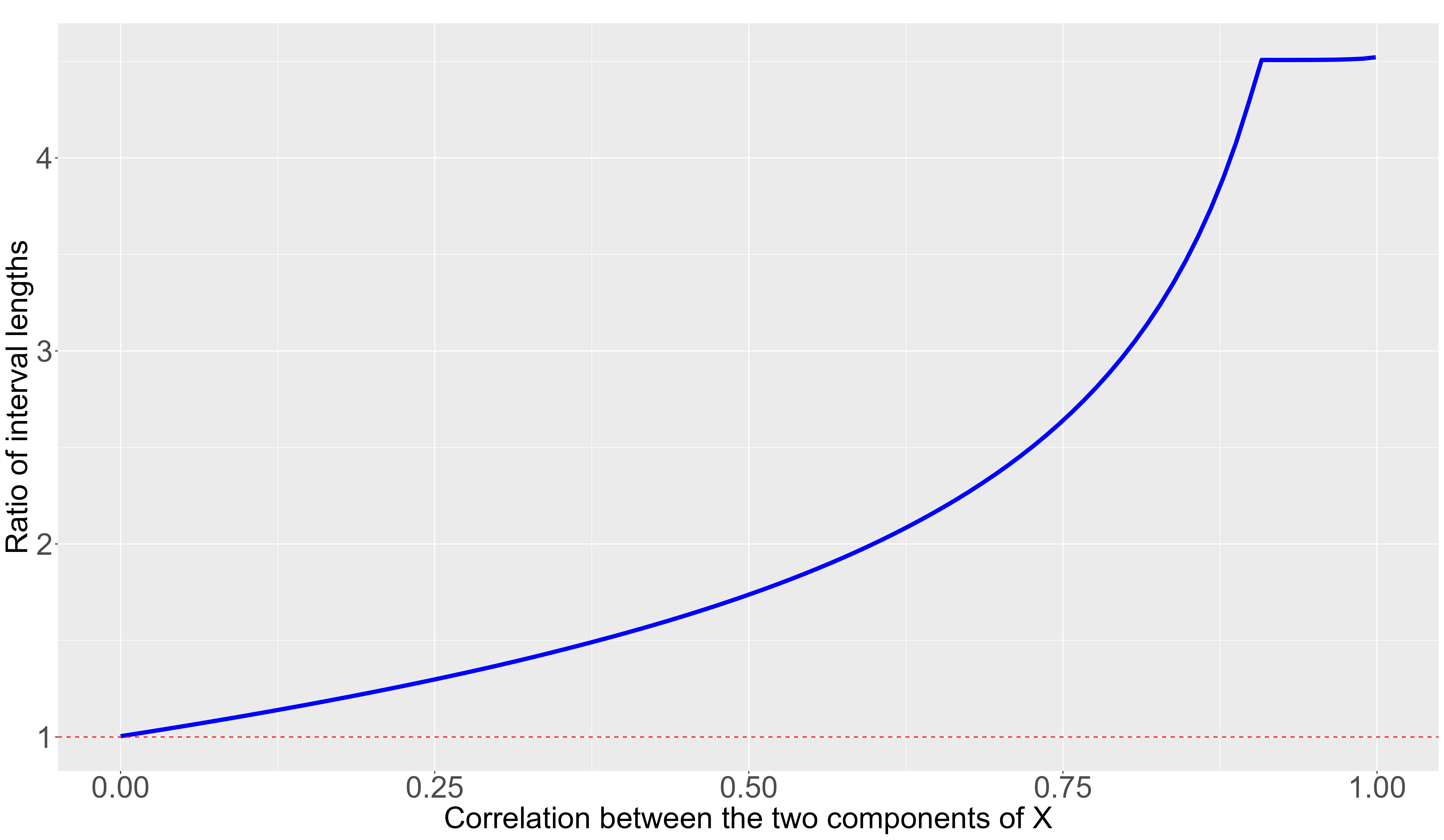}
		\label{fig:norm1}
\end{center}

\footnotesize{Notes: results obtained by approximating the true bounds using a sample of size $10^5$. The ratio of interval lengths is the ratio of the intervals obtained using \cite{pacini2019two} bounds and the sharp bounds.}
	\caption{Comparison between \cite{pacini2019two} bounds and the sharp bounds}
	\label{fig:ratio}
\end{figure}

We illustrate in the following that the bounds provided in \cite{pacini2019two} can in practice be substantially larger than the sharp bounds. To this end, we consider the following class of DGPs, indexed by $\rho$: $\log(Y)\sim \mathcal{N}(0,2)$ and $X=(X_1,X_2) \sim \mathcal{N}(0,\Sigma)$ with $\Sigma$ defined as in \eqref{eq:def_Sigma} (note that $\Sigma$ depends on $\rho$).  To compare the two types of bounds, we consider the following ratio
$$R:= \frac{\overline{b}^p_d - \underline{b}^p_d}{\overline{b}_d - \underline{b}_d},$$
where $d=(1,0)'$ and $(\underline{b}^p_d,\overline{b}^p_d)$ denote Pacini's bounds. Figure \ref{fig:ratio} reports $R$ as a function of $\rho$. When $X_1$ and $X_2$ are independent, the two intervals coincide, but the sharp bounds become tighter as the correlation between $X_1$ and $X_2$ increases. With $\rho\ge 0.88$, the sharp identification interval is more than four times shorter than the one obtained with Pacini's bounds.

\section{Asymptotic variance with common variables}
\label{app:as_var_common}

With common variables, the asymptotic variance takes the form
\begin{align*}
    V^g_d := &  \frac{\lambda}{E[\eta^2_d]^2} V(\psi^g_1+ (1-\lambda)\psi^g_{2,2}+\psi^g_4+\psi^g_5 +(1-\lambda) \psi^g_8 +   \psi^g_{10} ) \\
    & +  \frac{(1-\lambda)}{E[\eta^2_d]^2} V(\lambda\psi^g_{2,1}+\psi^g_3+\psi^g_6 + \psi^g_7 + \lambda\psi^g_9),
\end{align*}	
where $\psi^g_1 := \delta_Y'\left[W^{(2)}\nu_d - E[W^{(2)}T_{-1}']E[T_{-1}T_{-1}']^{-1}T_{-1}\eta_d\right]$, $\psi^g_{2,1} := \delta_d'\left(W^{(1)}W^{(1)}{}' \right.$ $\left.- E[WW']\right)\delta_Y$, $	\psi^g_{2,2} := \delta_d'\left(W^{(2)}W^{(2)}{}'- E[WW']\right)\delta_Y$, $	\psi^g_3 :=  \delta_d'W^{(1)} \nu_Y$, $    \psi^g_{10} := - \overline{b}_d (\eta_d^2 - E[\eta_d^2])$ and
\begin{align*}
	\psi^g_{4} & := - \bigg(
	\sum_{k=1}^K  p_k 
	E\!\left[
	F^{-1}_{|k}\!\circ G_{|k}(\nu_d)
	\,W^{(2)\prime}
	\,\middle|\,
	g(W^{(2)})=g_k
	\right]
	\bigg)E[W^{(2)}W^{(2)}{}']^{-1} \\
    & \qquad \times \left[W^{(2)}\nu_{d} - E[W^{(2)}T_{-1}'] E[T_{-1}T_{-1}']^{-1}T_{-1}\eta_{d}\right],\\
	\psi^g_{5} & := - \sum_{k=1}^K \ind{g(W^{(2)})=g_k} \int [\ind{\nu_{d} \le t} - G_{|k}(t)]F_{|k}^{-1} \circ G_{|k}(t)dt,\\
	\psi^g_{6} & := -\bigg(
	\sum_{k=1}^K  p_k
		E\!\left[
	G^{-1}_{|k}\!\circ F_{|k}(\nu_Y)
	\,W^{(1)\prime}
	\,\middle|\,
	g(W^{(1)})=g_k
	\right]
	\bigg)E[W^{(1)}W^{(1)}{}']^{-1} W^{(1)} \nu_{Y},\\
	\psi^g_{7} & := - \sum_{k=1}^K \ind{g(W^{(1)})=g_k} \int [\ind{\nu_{Y} \le t} - F_{|k}(t)]G_{|k}^{-1} \circ F_{|k}(t)dt,\\
    \psi^g_{8}  &:= \sum_{k=1}^K (\ind{g(W^{(2)})=g_k} - p_k) \int_0^1 F^{-1}_{|k}(u)G^{-1}_{|k}(u)du,\\
    \psi^g_{9}  &:= \sum_{k=1}^K (\ind{g(W^{(1)})=g_k} - p_k) \int_0^1 F^{-1}_{|k}(u)G^{-1}_{|k}(u)du.
\end{align*}

When $W$ is finitely supported on $\{w_1, \ldots, w_K\}$ and estimation is based directly on \eqref{eq:sharp_bound_W}, the OLS decomposition $\eta_d = W'\delta_d + \nu_d$, $Y = W'\delta_Y + \nu_Y$ used to construct the outer bound is bypassed. Accordingly, all influence-function terms associated with the estimation of $\delta_d$, $\delta_Y$, and $\E[WW']$ disappear, hence five of the eleven $\psi$-terms drop out ($\psi^g_{2,1}, \psi^g_{2,2}, \psi^g_3, \psi^g_4, \psi^g_6$). The remaining terms arise are the cell-conditional $L$-statistic pieces ($\psi_5, \psi_7$), multinomial cell-probability terms ($\psi_8, \psi_9$), denominator normalization ($\psi_{10}$), and, as in the no-covariates case, first-stage OLS adjustment $\psi_2$ from the estimation of $\gamma$ in $\eta_d = T_1 - T'_{-1}\gamma$. 
For \(k=1,\ldots,K\), let  $F_{\mid k}:=F_{Y\mid W}(\cdot\mid w_k)$, $G_{\mid k}:=F_{\eta_d\mid W}(\cdot\mid w_k)$, and 
\[ h_k(x) = \int_0^1 F_{\mid k}^{-1} \!\left[ G_{\mid k}(x^-) +u(G_{\mid k}(x)-G_{\mid k} (x^-)) \right]du. \]
The asymptotic variance takes the form
\begin{align*}
	V_d = & \frac{\lambda}{E[\eta_d^2]^2}\,V\left(\psi_2+\psi_5+(1-\lambda)\psi_8+\psi_{10}\right)  + \frac{1-\lambda}{E[\eta_d^2]^2}\,V\left(\psi_7+\lambda\psi_9\right),
\end{align*}
where
\begin{align*}
	\psi_{2}
	& := -\sum_{k=1}^K p_k\E\left[h_k(\eta_d)\,T'_{-1}|W^{(2)}=w_k\right]
	     E[T_{-1}T'_{-1}]^{-1}\,T_{-1}\eta_d, \\
	\psi_{5}
	& := -\sum_{k=1}^K \ind{W^{(2)}=w_k}
	        \int\!\bigl[\ind{\eta_d\le t}- G_{|k}(t)\bigr]\,
	        F^{-1}_{|k}\!\circ  G_{|k}(t)\,dt, \\
	\psi_{7}
	& := -\sum_{k=1}^K \ind{W^{(1)}=w_k}
	        \int\!\bigl[\ind{Y\le t}-F_{|k}(t)\bigr]\,
	         G^{-1}_{|k}\!\circ F_{|k}(t)\,dt, \\
	\psi_{8}
	& := \sum_{k=1}^K \left(\ind{W^{(2)}=w_k}-p_k\right)\int_0^1 F^{-1}_{|k}(u)G^{-1}_{|k}(u)du, \\
	\psi_{9}
	& := \sum_{k=1}^K \left(\ind{W^{(1)}=w_k}-p_k\right)\int_0^1 F^{-1}_{|k}(u)G^{-1}_{|k}(u)du, \\
	\psi_{10}
	& := -\overline{b}_d\left(\eta_d^2-E[\eta_d^2]\right).
\end{align*}

\FloatBarrier

\section{Additional elements on the applications}
\label{app:details_app}

\subsection{Additional results on the first application}
\label{app:details_1st_app}

As an alternative to our main estimator, we consider the estimator based on Equation \eqref{eq:sharp_bound_W}, conditioning on a finite-support transformation $g(W)$ of the common variables. The corresponding upper bound is
\begin{equation}
    \overline{b}^g_d =
    \frac{E\left[
    F^{-1}_{\eta_d|g(W)}(U|g(W))
    F^{-1}_{Y|g(W)}(U|g(W))
    \right]}{E(\eta_d^2)}.
    \label{eq:sharp_bound_gW}
\end{equation}
The lower bound follows by $\underline{b}^g_d=-\overline{b}^g_{-d}$. The resulting bounds are always contained in those obtained without exploiting $W$, since conditioning on $g(W)$ imposes additional restrictions on the admissible dependence between $\eta_d$ and $Y$. Estimation proceeds similarly to our main approach, using empirical conditional quantiles within the groups defined by $g(W)$. Moreover, the influence function formulas derived in Appendix B for Equation \eqref{eq:sharp_bound_W} with finite-support $W$ extend directly to this setting by replacing the conditioning variable $W$ with $g(W)$, while keeping the original Frisch--Waugh--Lovell residual $\eta_d$ unchanged. 

\medskip

This allows us to construct confidence intervals using the same inference approach as for our main estimator. We construct these groups using K-means, with the default choice $K=\lfloor n^c\rfloor$ and $c=0.19$, the largest exponent satisfying our theoretical restrictions. We also evaluate this estimator in DGP3 of our simulations: in this case, the bounds are wider than our preferred ones. However, this may not always be the case: in our first application, the bounds are slightly better than our baseline bounds, as shown in Table \ref{table:appli1:main:Kmeans:eq6} below.  
This alternative estimator is implemented as an option in our companion R package. 

\begin{table}[H]
\begin{center}
\scalebox{1}{
\begin{tabular}{rccc}
\toprule
Coefficient & TSTSLS & Eq.~(6) based bounds & 95\% CI \\
\midrule
Black & -0.038 & [-0.766, 0.418] & [-0.779, 0.428] \\
 & (0.006) & & \\
Hispanic & -0.032 & [-0.553, 0.280] & [-0.573, 0.290] \\
 & (0.004) & & \\
Asian & 0.041 & [-0.131, 0.189] & [-0.143, 0.198] \\
 & (0.002) & & \\
American native & -0.047 & [-0.894, 0.421] & [-0.902, 0.431] \\
 & (0.042) & & \\
\bottomrule
\end{tabular}
}
\end{center}
\vspace{0.1cm}
\footnotesize{Notes: 2,146,799 matched patent applications and 91,055 matched surnames. K-means uses $K=15$, $c=0.190$. Census conditional race shares are treated as fixed.}
\caption{Patent approval: surname-index K-means bounds based on  \eqref{eq:sharp_bound_W}}
\label{table:appli1:main:Kmeans:eq6}
\end{table}

\subsection{Additional details on the second application}
\label{app:details_2nd_app}

The GPS survey \citep[see][]{falk2018global} covers the following countries: Argentina, Australia, Austria, Bosnia and Herzegovina, Brazil, Canada, Switzerland, Chile, Colombia, Costa Rica, Czech Republic, Germany, Algeria, Spain, Estonia, Finland, France, United Kingdom, Georgia, Greece, Croatia, Hungary, Indonesia, Israel, Italy, Jordan, Japan, Kazakhstan, South Korea, Lithuania, Morocco, Moldova, Mexico, Netherlands, Peru, Philippines, Poland, Portugal, Romania, Russia, Saudi Arabia, Serbia, Sweden, Thailand, Turkey, Ukraine, United States, Vietnam, United Arab Emirates. The preference measures therein are based on 12 survey items, which are summarized in Table I in \cite{falk2018global}. We gather them below with their respective weights for completeness: 
\begin{enumerate}
    \item Patience: Intertemporal choice sequence using staircase method (0.712); Self-assessment: willingness to wait (0.288).\\[-1cm]
    \item Risk taking: Lottery choice sequence using staircase method (0.473); Self-assessment: willingness to take risks in general (0.527).\\[-1cm]
    \item Positive reciprocity: Gift in exchange for help (0.515); Self-assessment: willingness to return a favor (0.485).\\[-1cm]
    \item Negative reciprocity: Self-assessment: willingness to take revenge (0.374); Self-assessment: willingness to punish unfair behavior toward self (0.313); Self-assessment: willingness to punish unfair behavior toward others (0.313).\\[-1cm]
    \item Altruism: Donation decision (0.635);
Self-assessment: willingness to give to good causes (0.365).\\[-1cm]
    \item Trust: Self-assessment: people have only the best intentions (1).
\end{enumerate}
The weights endogenously emerged from a preliminary experimental validation procedure \citep[see][]{falk2023preference}.  Each preference measure is standardized at the individual level.

\section{Proofs of the identification results} 
\label{sec:proofs}

\subsection{Theorem \ref{thm:ident_no_common}} 
\label{sub:proof_thm_ident_no_c}

First, if $b\in \B$, then there exists r.v. $(\widetilde{Y},\widetilde{X})$ with $F_{\widetilde{X}}=F_X$, $F_{\widetilde{Y}}=F_Y$ and such that $EL(\widetilde{Y}|\widetilde{X})=\widetilde{X}'b$. Thus, $\widetilde{\eps}:=\widetilde{Y} - \widetilde{X}'b$ satisfies $E(\widetilde{\eps})=0$ and $\Cov(\widetilde{X},\widetilde{\eps})=0$. Hence, $E[\widetilde{Y}]=E[\widetilde{X}'b]$ and
$$V(\widetilde{Y})=V(\widetilde{X}'b)+V(\widetilde{\eps})\ge V(\widetilde{X}'b).$$
As a result, $E(Y)= E(X'b)$, $V(Y)\ge V(X'b)$ and $\B \subseteq \mathcal{E}$. This also implies that $\B$ is bounded.

\medskip
Now, let us prove that $\B$ is closed. This, in turn, will imply that $\B$ is compact. Let $b_n\in\B$ for all $n\ge 1$ with $b_n\to b$ and let us prove that $b\in\B$. Let $(\widetilde{X}_n, \widetilde{Y}_n)$ such that $F_{\widetilde{X}_n}=F_X$, $F_{\widetilde{Y}_n}=F_Y$ and $b_n=E(\widetilde{X}_n\widetilde{X}_n')^{-1}E(\widetilde{X}_n\widetilde{Y}_n)$. Since $E(\widetilde{X}_n\widetilde{X}_n')=E(XX')$, it suffices to prove that there exists $(\widetilde{X},\widetilde{Y})$, with $F_{\widetilde{X}}=F_X$, $F_{\widetilde{Y}}=F_Y$, such that $E[\widetilde{X}\widetilde{Y}]=c:=E(XX')b$. First, note that for all $M$,
    \begin{align*}
    	P\left(\|(\widetilde{X}_n, \widetilde{Y}_n)\|\ge M\right)& \le \frac{E\left[\|(\widetilde{X}_n, \widetilde{Y}_n)\|^2\right]}{M^2} \\
    	& \le \frac{E[\|X\|^2]+E[Y^2]}{M^2}.
    \end{align*}
Hence, $(\widetilde{X}_n, \widetilde{Y}_n)$ is uniformly tight. Then, by Prokhorov's theorem, there exists a subsequence
$(\widetilde{X}_{n_j}, \widetilde{Y}_{n_j})$ that converges in distribution, to $(\widetilde{X}, \widetilde{Y})$ say. Moreover, $F_{\widetilde{X}}=F_X$ and $F_{\widetilde{Y}}=F_Y$. Now, remark that for all $(x,y)\in\R^{+2}$ and all $M>0$, we have
$$xy\ind{xy>M} \le x^2 \ind{x>M^{1/2}}+ y^2 \ind{y>M^{1/2}}.$$
As a result, for all $n\ge 1$ and all $M>0$,
\begin{align*}
& E\left[\|\widetilde{X}_{n_j} \widetilde{Y}_{n_j}\|\ind{\|\widetilde{X}_{n_j} \widetilde{Y}_{n_j}\|>M}\right]
\\ \le & E\left[\|\widetilde{X}_{n_j}\|^2\ind{\|\widetilde{X}_{n_j}\|>M^{1/2}}\right] + E\left[\widetilde{Y}_{n_j}^2\ind{|\widetilde{Y}_{n_j}|>M^{1/2}}\right] \\	
 = & E\left[\|X\|^2\ind{\|X\|>M^{1/2}}\right] + E\left[Y^2\ind{|Y|>M^{1/2}}\right].
\end{align*}
As a result, by the dominated convergence theorem, $\widetilde{X}_{n_j} \widetilde{Y}_{n_j}$ is asymptotically uniformly integrable. This implies \citep[see, e.g.][Theorem 2.20]{van2000asymptotic} that
$$E\left[\widetilde{X}_{n_j} \widetilde{Y}_{n_j}\right] \to E[\widetilde{X} \widetilde{Y}].$$
Because we also have $E\left[\widetilde{X}_{n_j} \widetilde{Y}_{n_j}\right]\to c$, we finally obtain $E[\widetilde{X} \widetilde{Y}]=c$. This proves that $\B$ is closed.

\medskip
Next, we prove that $\B$ is convex. Let $(b_1,b_2)\in \B^2$ and fix $p\in[0,1]$. Then, there exists $(\widetilde{X}_1,\widetilde{Y}_1)$ and $(\widetilde{X}_2,\widetilde{Y}_2)$ rationalizing respectively $b_1$ and $b_2$. Let $D$ following a Bernoulli distribution with probability $p$, $D\sim$Be$(p)$, independent of these random variables and let $(\widetilde{Y},\widetilde{X})=(\widetilde{Y}_1,\widetilde{X}_1)$ if $D=1$, $(\widetilde{Y},\widetilde{X})=(\widetilde{Y}_2,\widetilde{X}_2)$ otherwise. Then, $F_{\widetilde{X}}=F_X$, $F_{\widetilde{Y}}=F_Y$ and
\begin{align*}
E\left[\widetilde{X}\widetilde{Y}\right] = & p E\left[\widetilde{X}_1\widetilde{Y}_1\right] + (1-p) E\left[\widetilde{X}_2\widetilde{Y}_2\right] \\
=&  E(XX') ( p b_1 + (1-p) b_2).
\end{align*}
Hence, $EL(\widetilde{Y}|\widetilde{X})=\widetilde{X}'(pb_1+(1-p)b_2)$, which implies that $\B$ is convex.

\medskip
Now, we prove $\overline{b}_d = E[F_{d'E[XX']^{-1}X}^{-1}(U) F_Y^{-1}(U)]$. We have
\begin{equation}
\overline{b}_d = \max_{\Pi \in \mathcal{M}(F_X, F_Y)} \int \left[d' E[XX']^{-1} x\right]\, y \, d\Pi(x,y),
	\label{eq:OT_prgm}
\end{equation}
where $\mathcal{M}(F,G)$ denotes the set of probability measures with marginal cdfs equal to $F$ and $G$.
Remark that for any $c=(c_1,...,c_p)$ and any $(\widetilde{X},\widetilde{Y})\sim \Pi \in \mathcal{M}(F_X, F_Y)$,
$$(c\widetilde{X},\widetilde{Y})\sim \Pi \in \mathcal{M}(F_{cX}, F_Y).$$
Therefore, letting $X_d:=d'E[XX']^{-1}X$, we obtain
$$\overline{b}_d \leq \max_{\Pi \in \mathcal{M}(F_{X_d}, F_Y)} \int u y d\Pi(u,y).$$
Moreover, by the Cambanis-Simons-Stout inequality \citep[see][]{cambanis1976inequalities},
\begin{equation}\label{eq:CSS}
	\max_{\Pi \in \mathcal{M}(F_{X_d}, F_Y)} \int u y d\Pi(u,y) = E[F_{X_d}^{-1}(U) F_Y^{-1}(U)].
\end{equation}
Hence, $\overline{b}_d \le E[F_{X_d}^{-1}(U) F_Y^{-1}(U)]$.

\medskip
Now, for any $U\sim\mathcal{U}([0,1])$, let $\widetilde{Y}=F_Y^{-1}(U)$. Let also $C$ denote a copula of $M'E[XX']^{-1}X$ (recall the construction of $M$ at the beginning of Section \ref{sub:no_sep}) and let $(U_2,...,U_p)$ be uniform random variables such that $(U,U_2,...,U_p)$ has cdf equal to $C$. Let us define
$$S_d = (F_{X_d}^{-1}(U),F_{\widetilde{d}'_2E[XX']^{-1}X}^{-1}(U_2),...,F_{\widetilde{d}'_pE[XX']^{-1}X}^{-1}(U_p))'.$$
By construction, $S_d\sim M' E[XX']^{-1}X$. Then, let $\widetilde{X}=(M' E[XX']^{-1})^{-1}S_d$, so that $\widetilde{X}\sim X$. Let $\Pi^*$ denote the distribution of $(\widetilde{X}, \widetilde{Y})$. We have
$\Pi^* \in \mathcal{M}(F_X, F_Y)$. Moreover,
$$d'E[XX']^{-1}\widetilde{X}=d'M'{}^{-1}S_d = F_{X_d}^{-1}(U),$$
where the last equality follows since $e_{1,p}' \times M' = d'$. Thus, by definition of $\overline{b}_d$, $\overline{b}_d \ge  E[F_{X_d}^{-1}(U) F_Y^{-1}(U)]$. Equation \eqref{eq:supp_fct_no_common_1} follows.

\medskip
Next, we prove \eqref{eq:supp_fct_no_common}. It suffices to show that $X_d = \eta_d/E(\eta_d^2)$. Remark that
$$d'E(XX')^{-1} X = e_{1,p}' M' E(XX')^{-1} M (M^{-1} X) = e_{1,p}' E(TT')^{-1} T.$$
Moreover, $\eta_d = \gamma' T$, with $\gamma:= [1, \, -E(T_1T_{-1})' E(T_{-1}T_{-1}')^{-1}]'$. Thus,
$$E(\eta_d^2)=\gamma' E(TT')\gamma = E(T_1^2) - E(T_1T_{-1})' E(T_{-1}T_{-1}')^{-1} E(T_1T_{-1}).$$
As a result, $E(TT') \times \gamma/E(\eta_d^2)=e_{1,p}$. The result follows since then,
$$X_d= e_{1,p}' E(TT')^{-1} T= \gamma'T/E(\eta_d^2)=\eta_d/E(\eta_d^2).$$

\medskip
Finally, if $V(Y)>0$, $F_{Y}^{-1}(U)$ is not constant. By Assumption \ref{hyp:mom}, we also have $V(\eta_d)>0$ and thus $F_{\eta_d}^{-1}(U)\sim \eta_d$  is not constant either. Moreover, remark that $E[XX']^{-1}E[X]= E[XX']^{-1}E[XX']e_{p}=e_{p}$. Thus, $d_p=0$ implies $E[\eta_d]=d'e_{p}=0$. Then, by Theorem 1.1 and 1.2 of \cite{jakubowski2021complement}, we obtain
$$E[F_{\eta_d}^{-1}(U)F_{Y}^{-1}(U)]>E[F_{\eta_d}^{-1}(U)]E[F_{Y}^{-1}(U)] = 0.$$
The last point of the theorem follows.


\subsection{Theorem \ref{thm:caract_id_set_common}} 
\label{sub:thm_prop_simple_outer}

Since $b^0 = E(XX')^{-1}E(XY) $, the exact same reasoning as in the proof of Theorem \ref{thm:ident_no_common} shows that the identified set $\B(w)$ of $E(XX')^{-1}E(XY | W=w)$ is convex. By integrating over $w$, $\B$ is thus convex. It is also bounded as a subset of $\mathcal{E}$.

\medskip
Let us now prove that $\B_d=[\underline{b}_d,\overline{b}_d]$, with $\overline{b}_d = \overline{b}^g_d$ satisfying \eqref{eq:upper:sigma_w1} with $g(W) = W$; this will also imply that $\B$ is compact. By the same reasoning as in the proof of Theorem \ref{thm:ident_no_common}, we also have that the identified set $\B_d(w)$ of $b'd$ for $b\in \B(w)$ is the identified set of $E[\eta_dY|W=w]/E(\eta_d^2)$ and that  $\B_d(w) = [\underline{b}_d(w),\overline{b}_d(w)]$, with $\overline{b}_d(w):=\sup\{b'd: b\in \B(w)\}$.  Let $U$ be such that $U|W$ is uniformly distributed on $[0,1]$. Then, $\overline{b}_d(w)$ satisfies
\begin{align*}
\overline{b}_d(w)= & \frac{1}{E(\eta_d^2)}  E\left[F_{W'\delta_d + \nu_d|W}^{-1}(U|W) F_{W'\delta_Y +\nu_Y|W}^{-1}(U|W)|W=w\right]\\
= & \frac{1}{E(\eta_d^2)}  E\left[\left(W'\delta_d + F_{\nu_d|W}^{-1}(U|W)\right) \left(W'\delta_Y + F_{\nu_Y|W}^{-1}(U|W)\right)|W=w\right].
\end{align*}
Next, we have $b_d = E[\eta_dY]/E(\eta_d^2)$  by construction and $ E[\eta_dY]=E[E[\eta_dY|W]]\le E[\overline{b}_d(W)] E(\eta_d^2)$. Moreover, the bound is reached by considering $$(\eta_d,Y)=(F_{W'\delta_d + \nu_d|W}^{-1}(U|W), F_{W'\delta_Y +\nu_Y|W}^{-1}(U|W)).$$ 
To see that this coupling is feasible for the original data-combination problem, conditionally on the constructed $(\eta_d,W)$, draw $X_o$ from its observed conditional distribution given $(\eta_d,W)$, and let $X=(X_o',X_c')'$, where $X_c$ is the corresponding component of $W$. By construction, $(X_o,W)$ and $(Y,W)$ retain their observed distributions, while the induced $\eta_d$ coincides almost surely with the constructed one. 
Thus, $\overline{b}_d=E[\overline{b}_d(W)]$. Since $(W, F_{\nu_\ell|W}^{-1}(U|W))$ (with $\ell\in\{d,Y\}$) has the same distribution as $(W,\nu_\ell)$, we obtain
\begin{align*}
\overline{b}_d = & \frac{1}{E(\eta_d^2)}\left\{\delta_d' E\left[WW'\right]\delta_Y + E\left[F_{\nu_d|W}^{-1}(U|W) W'\delta_Y\right]  + E\left[F_{\nu_Y|W}^{-1}(U|W) W'\delta_d\right]\right\} \\
& + \frac{1}{E(\eta_d^2)}E\left[F^{-1}_{\nu_d|W}(U|W) F^{-1}_{\nu_Y|W}(U|W)\right] \\
= & \frac{1}{E(\eta_d^2)} \left\{ \delta_d' E\left[WW'\right]\delta_Y + E\left[F^{-1}_{\nu_d|W}(U|W) F^{-1}_{\nu_Y|W}(U|W)\right]\right\}.
\end{align*}

To prove that $\overline{b}_d\le \overline{b}^g_d$, remark that
\begin{align*}
E[\eta_dY]=& E\left[(W'\delta_d + \nu_d)(W'\delta_Y + \nu_Y)\right] \\
= & \delta_d' E\left[WW'\right]\delta_Y + E[\nu_d \nu_Y] \\
= & \delta_d'E\left[WW'\right]\delta_Y + E\left[E[\nu_d \nu_Y|g(W)]\right] \\
\le & \delta_d' E\left[WW'\right]\delta_Y + E\left[F^{-1}_{\nu_d|g(W)}(U|g(W)) F^{-1}_{\nu_Y|g(W)}(U|g(W))\right],
\end{align*}
where the last inequality follows by the Cambanis-Simons-Stout inequality. If $\nu_d\indep W|g(W)$ and $\nu_Y\indep W|g(W)$, the last expression is equal to $\overline{b}_d$. The third point of the proposition follows.


\subsection{Proposition \ref{prop:TSTSLS}} 

Let us assume that \eqref{eq:excl_restr_TSTSLS} holds and let us show that Corr$(W_a,Y)\le$Corr$(W_a,X_o)$. First, suppose that $\Cov(W_a,X_o)\ne 0$. Then,
$$\frac{\Cov(X_o,Y)}{V(X_o)} = b_{\text{TSTSLS}},$$
with $b_{\text{TSTSLS}} = \Cov(W_a,Y)/\Cov(W_a,X_o)$. As a result, letting $\rho_{AB}:=\Corr(A,B)$ for any random variables $A$ and $B$, we obtain
\begin{align*}
  \rho_{X_oY}= & \frac{\Cov(W_a,Y)V(X_o)}{\Cov(W_a,X_o)}\frac{1}{V(X_o)^{1/2}V(Y)^{1/2}} = \frac{\rho_{W_aY}}{\rho_{W_aX_o}}.
\end{align*}
Let $R$ denote the correlation matrix of $(W_a,X_o,Y)$. By what precedes,  
$$R = \begin{pmatrix}
    1 & \rho_{W_aX_o} & \rho_{W_aY} \\
    \rho_{W_aX_o} & 1 & \rho_{W_aY}/\rho_{W_aX_o} \\
    \rho_{W_aY} & \rho_{W_aY}/\rho_{W_aX_o}& 1
\end{pmatrix}.$$
Some algebra shows that its non-negative determinant is equal to $1 +  \rho_{W_aY}^2 - \rho_{W_aX_o}^2 - (\rho_{W_aY}/\rho_{W_aX_o})^2$. Then, 
$$(1-\rho_{W_aX_o}^2)\left[1-\left(\frac{\rho_{W_aY}}{\rho_{W_aX_o}}\right)^2\right]\ge 0,$$
Thus, either $\rho_{W_aX_o}^2=1$, in which case $|\rho_{W_aY}| \le |\rho_{W_aX_o}|$, or $|\rho_{W_aY}/\rho_{W_aX_o}|\le 1$, in which case also $|\rho_{W_aY}| \le |\rho_{W_aX_o}|$. So the result holds if $\Cov(X_o,W_a)\ne 0$. 

\medskip
Now, suppose that $\Cov(W_a,X_o)= 0$. The restriction on best linear projections  \eqref{eq:excl_restr_TSTSLS} implies that $\Cov(W_a,Y-X' b^0)=0$. But then, $\Cov(W_a,Y)= 0$. Hence, in this case as well, $|\rho_{W_aY}| \le |\rho_{W_aX_o}|$. The result follows.


\subsection{Proposition \ref{prop:additional_var}} 
\label{sub:proposition_ref_prop_additional_var}

Let us denote by $\B_Z(w)$ the identified set of $E[XY|W=w]$ when observing $Z$, whereas $\B(w)$ still denotes the identified set of $E[XY|W=w]$ without the knowledge of $Z$. Again, the same reasoning as in the proof of Theorem \ref{thm:ident_no_common} shows that the identified set $\B_Z(w)$ of $E[XY|W=w]$ is non-empty, closed and convex. Thus, it is characterized by its so-called support function $\sigma_{\B_Z(w)}(d):=\sup\{b'd:b\in \B_Z(w)\}$. As in \eqref{eq:OT_prgm}, we have
$$\sigma_{\B_Z(w)}(d) = \max_{\Pi_w \in \mathcal{M}(F_{X_o|W=w}, F_{Y,Z|W=w})} \int \left[d' E[XX']^{-1} (x'_o, x_c')' \right]\, y \, d\Pi_w(x_o,y,z),$$
where $w=(x_c', w_a')'$. By Lemma 3.3 of \cite{delon2023generalized},
\begin{align*}
\sigma_{\B_Z(w)}(d) = & \max_{\Pi_w \in \mathcal{M}(F_{X_o|W=w}, F_{Y}|W=w)} \int \left[d' E[XX']^{-1} (x'_o, x_c')' \right]\, y \, d\Pi_w(x_o,y) \\
= & \sigma_{\B(w)}(d),    
\end{align*}
the support function of $\B(w)$ evaluated at $d$.
Hence, by integrating over $w$, we obtain $\sigma_{\B_Z}=\sigma_{\B}$. The result follows since these functions characterize $\B_Z$ and $\B$.


\newpage
\setcounter{page}{1}
\begin{center}
    {\huge Online Appendix}
\end{center}

\section{Proofs of the statistical results}

\subsection{Theorem \ref{thm:asym_Xo}} 
\label{sub:theorem_ref_thm_asym_Xo}

We prove the results in three main steps. The first step obtains linear approximations of terms related to the estimation of $\eta_d$ and $E(\eta_d^2)$. The second step obtains a linear approximation of two other terms. The third step shows that a remainder term is negligible and concludes.

\subsubsection*{1. Linear approximation of the first terms} 
\label{ssub:linear_approximation_of_the_first_terms}

We first show that
\begin{align}
\sqrt{\frac{nm}{n+m}} \left(\widehat{\overline{b}}_d - \overline{b}_d\right) = & \frac{1}{E(\eta_d^2)} \left[\sqrt{\frac{nm}{n+m}}\int_0^1 (F_n^{-1}G_m^{-1} - F^{-1}G^{-1})dt  + \frac{\sqrt{\lambda}}{m^{1/2}}\sum_{i=1}^m (\psi_{1i}+\psi_{2i})\right] \notag \\
&  + o_P(1).
	\label{eq:first_part_asym}
\end{align}
First, remark that
\begin{equation}
\widehat{\overline{b}}_d - \overline{b}_d = \frac{1}{\widehat{E}(\widehat{\eta}_d^2)}\left[\int_0^1 (F_n^{-1}\widehat{G}_m^{-1} - F^{-1}G^{-1})dt  - \overline{b}_d(\widehat{E}(\widehat{\eta}_d^2) - E(\eta_d^2))\right].	
	\label{eq:decomp_bsup_hat}
\end{equation}
Moreover, since $\widehat{\eta}_{di} - \eta_{di} = -T_{-1i}'(\widehat{\gamma}-\gamma_0)$,
\begin{align*}
	\widehat{E}(\widehat{\eta}_d^2) - E(\eta_d^2) & = \frac{1}{m}\sum_{i=1}^m \eta_{di}^2 - E[\eta^2_d] - \frac{2}{m}\sum_{i=1}^m \eta_{di} T_{-1i}'(\widehat{\gamma}-\gamma_0) \\
	& + (\widehat{\gamma}-\gamma_0)'\left(\frac{1}{m}\sum_{i=1}^m T_{-1i}T_{-1i}'\right)(\widehat{\gamma}-\gamma_0) \\
	& =  \frac{1}{m}\sum_{i=1}^m \eta_{di}^2 - E[\eta^2_d] + o_P(m^{-1/2}).
\end{align*}
The last equality follows since $E[\|X\|^4]<\infty$ implies both $\widehat{\gamma}-\gamma_0= O_P(m^{-1/2})$ and $(1/m)\sum_{i=1}^m \eta_{di} T_{-1i} \convP 0$. Combined with \eqref{eq:decomp_bsup_hat}, $n/(n+m)\to \lambda$ and the definition of $\psi_1$, this yields
\begin{align}
\sqrt{\frac{nm}{n+m}} \left(\widehat{\overline{b}}_d - \overline{b}_d\right) = & \frac{1}{E(\eta_d^2)} \left[\sqrt{\frac{nm}{n+m}}\int_0^1 (F_n^{-1}\widehat{G}_m^{-1} - F^{-1}G^{-1})dt  + \frac{\sqrt{\lambda}}{m^{1/2}}\sum_{i=1}^m \psi_{1i}\right] \notag \\
& + o_P(1).
	\label{eq:for_first_part_asym}
\end{align}

\medskip
Let us now prove that
\begin{equation}
\sqrt{m} \int_0^1 F_n^{-1}(\widehat{G}^{-1}_m - G^{-1}_m)dt = - E[h(\eta_d)T'_{-1}]\sqrt{m}(\widehat{\gamma}-\gamma_0) + o_P\left(1\right).	
	\label{eq:remainder}
\end{equation}
When combined with \eqref{eq:for_first_part_asym}, the standard result that
$$\sqrt{m}(\widehat{\gamma}-\gamma_0) = E[T_{-1}T'_{-1}]^{-1} \frac{1}{m^{1/2}} \sum_{i=1}^m T_{-1i}\eta_{di} + o_P(1),$$
and the definition of $\psi_2$, this will entail \eqref{eq:first_part_asym}.

\medskip
Remark that if $\eta_d \sim G$ and $U|X \sim\mathcal{U}[0,1]$, then
$$G(\eta_d^-) + U (G(\eta_d) - G(\eta_d^-)) \sim \mathcal{U}[0,1].$$
As a result, if we let $(U_1,...,U_m)$ be i.i.d., uniform variables, the $\tilde{Y}_i := F^{-1}[G(\eta_{di}^-) + U_i (G(\eta_{di}) - G(\eta_{di}^-))]$ are i.i.d. with cdf $F$. Let $\sigma_1$ denote a permutation on $\{1,...,m\}$ such that $\eta_{\sigma_1(1)}\le ...\le \eta_{\sigma_1(m)}$ and $\tilde{Y}_{\sigma_1(1)} \le ... \le \tilde{Y}_{\sigma_1(m)}$. Let also $\sigma_2$ denote a permutation  on $\{1,...,m\}$ such that $\widehat{\eta}_{\sigma_2(1)} \le ... \le \widehat{\eta}_{\sigma_2(m)}$; if $\sigma_1$ satisfies these inequalities, let $\sigma_2:=\sigma_1$. Finally, let $\lceil \cdot\rceil$ denote the ceiling function. Then, define $Q_m(t):=\eta_{d\sigma_2(\lceil mt\rceil)}$ and $\widehat{Q}_m(t):=\widehat{\eta}_{d\sigma_1(\lceil mt\rceil)}$. By the Cambanis-Simons-Stout inequality,
$$\int_0^1 F_n^{-1}(\widehat{Q}_m - G^{-1}_m)dt \le \int_0^1 F_n^{-1}(\widehat{G}^{-1}_m - G^{-1}_m)dt \le \int_0^1 F_n^{-1}(\widehat{G}^{-1}_m - Q_m)dt.$$
Next, remark that
\begin{align*}
\widehat{Q}_m(t) - G^{-1}_m(t)& =-T_{-1\sigma_1(\lceil mt\rceil)}'(\widehat{\gamma}-\gamma_0),\\	
\widehat{G}^{-1}_m(t) - Q_m(t) & = - T_{-1\sigma_2(\lceil mt\rceil)}'(\widehat{\gamma}-\gamma_0).
\end{align*}
Then, letting $Q_{1m}(t):=T_{-1\sigma_1(\lceil mt\rceil)}$ and $Q_{2m}(t):=T_{-1\sigma_2(\lceil mt\rceil)}$, we obtain
\begin{align}
- \left[\int_0^1 F_n^{-1}Q_{1m}'dt\right](\widehat{\gamma}-\gamma_0) \le & \int_0^1 F_n^{-1}(\widehat{G}^{-1}_m - G^{-1}_m)dt \notag \\
 \le & - \left[\int_0^1 F_n^{-1}Q_{2m}'dt\right](\widehat{\gamma}-\gamma_0).
\label{eq:gendarmes}	
\end{align}

Let $\tilde{F}_m$ denote the empirical cdf of $(\tilde{Y}_i)_{i=1,...,m}$. Then,
\begin{align*}
	\left(\int_0^1 \left[F_n^{-1}-\tilde{F}_m^{-1}\right]^2 dt\right)^{1/2} & = W_2(F_n,\tilde{F}_m) \\
	& \le W_2(F_n,F)+W_2(\tilde{F}_m,F) \\
	& \convP 0.
\end{align*}
The inequality holds since $W_2$ is a distance. The convergence to 0 follows since convergence of the Wasserstein-2 distance is equivalent to weak convergence and convergence of the second moment \citep[see, e.g., Theorem 6.9 in][]{villani2009optimal}, and both $(Y_i)_{i=1,...,n}$ and $(\tilde{Y}_i)_{i=1,...,m}$ are i.i.d. with cdf $F$. Hence, we have, for $k\in\{1,2\}$
\begin{align}
	\left\|\int_0^1 \left(F_n^{-1}-\tilde{F}_m^{-1}\right) Q_{km}'dt\right\| & \le \left(\int_0^1 \left[F_n^{-1}-\tilde{F}_m^{-1}\right]^2 dt\right)^{1/2} \left(\int_0^1 \left\|Q_{km}\right\|^2 dt\right)^{1/2} \notag \\
	& = o_P(1). 	\label{eq:approx_Ytilde}
\end{align}
Next, remark that
	\begin{align*}
\int_0^1 \tilde{F}_m^{-1} Q_{1m}dt	& = \frac{1}{m}\sum_{i=1}^m \tilde{Y}_{\sigma_1(i)}T_{-1\sigma_1(i)} \\
& = \frac{1}{m}\sum_{i=1}^m \tilde{Y}_i T_{-1i} \\
& \convP E[\tilde{Y}_1 T_{-11}].
	\end{align*}
Moreover, by definition of $h$ and $\tilde{Y}$,
$$E[\tilde{Y} T_{-1}]=E[E[\tilde{Y}|\eta_d,T_{-1}] T_{-1}] = E[h(\eta_d)T_{-1}].$$
Together with \eqref{eq:approx_Ytilde}, this proves that
$$\int_0^1 F_n^{-1} Q_{1m}dt \convP E[h(\eta_{d})T_{-1}].$$
Using \eqref{eq:approx_Ytilde} again but with $k=2$ and \eqref{eq:gendarmes}, \eqref{eq:remainder} follows provided that
\begin{equation}
\int_0^1 \tilde{F}_m^{-1} Q_{2m}dt \convP E[h(\eta_{d})T_{-1}].	
	\label{eq:verif_fin}
\end{equation}
Let us now show that \eqref{eq:verif_fin} hold under Assumption \ref{hyp:reg}. First, consider the case where Assumption \ref{hyp:reg}-(i) holds. Because $|\Supp(\eta_d)|=|\Supp(X)|$, we have, by \eqref{eq:conv_unif_eta} in Lemma \ref{lem:eta} and with probability approaching one (wpao),  $\widehat{\eta}_{di}>\widehat{\eta}_{dj}$ that implies $\eta_{di}>\eta_{dj}$ for all $i\ne j$. Thus, because $\eta_{d\sigma_1(i)}\le \eta_{d\sigma_1(i+1)}$, we have $\widehat{\eta}_{d\sigma_1(i)}\le \widehat{\eta}_{d\sigma_1(i+1)}$ for all $i=1,...,m-1$. By construction of $\sigma_2$, this implies that $\sigma_2 =\sigma_1$, wpao. Since $\int_0^1 \tilde{F}_m^{-1} Q_{2m}dt=\int_0^1 \tilde{F}_m^{-1} Q_{1m}dt $ in this case, we obtain \eqref{eq:verif_fin}.

\medskip
Next, assume that Assumption \ref{hyp:reg}-(ii) holds. Let $\Supp(Y)=\{y_1,...,y_K\}$ with $-\infty=:y_0< y_1 < ... < y_K$. Then $\tilde{Y}_i = F^{-1}(G(\eta_{di}))$. Moreover, $F^{-1}$ is constant on $I_k:=(F(y_{k-1}), F(y_k)]$ for all $k\in\{1,...,K\}$. By \eqref{eq:conv_unif_eta} and continuity of $G$, we have
$$P\left(\exists (i,k,\ell) \in\{1,...,m\} \times \{1,...,K\}^2: \, k\ne \ell,\, G(\eta_{di}) \in I_k, G(\widehat{\eta}_{di}) \in I_\ell\right) \to 0.$$
As a result, $\tilde{Y}_i = F^{-1}(G(\widehat{\eta}_{di}))$ for all $i$ wpao. Under this event, $\sigma_2=\sigma_1$ and as above, we obtain \eqref{eq:verif_fin}.

\medskip
Finally, suppose that Assumption \ref{hyp:reg}-(iii) holds.
Then,  $h(x)=F^{-1}[G(x)]$ is continuous and $\tilde{Y}_i = h(\eta_{di})$ for all $i$. Let $\widehat{\tilde{Y}}_i:=h(\widehat{\eta}_{di})$. For any $M>0$, let $f_M(x)=\min(\max(x,-M),M)$. Fix $\delta>0$. By the dominated convergence theorem $$\lim_{K\to\infty}E[h(\eta_{d})^2\ind{|\eta_d|>K}]\to 0.$$ Then, let $K>0$ be such that $$E[h(\eta_{d})^2\ind{|\eta_d|>K}]<\frac{\delta}{6E[\|T_{-1}\|^2]}.$$
Let also $M=\max(h(K), -h(-K))$. Then,
\begin{align*}
\int_0^1 \tilde{F}_m^{-1} Q_{2m}dt & = \frac{1}{m}\sum_{i=1}^m \tilde{Y}_{\sigma_1(i)}T_{-1\sigma_2(i)} \\
& = \frac{1}{m}\sum_{i=1}^m \left[\tilde{Y}_{\sigma_1(i)} - f_M(\tilde{Y}_{\sigma_1(i)})\right]T_{-1\sigma_2(i)} + \frac{1}{m}\sum_{i=1}^m f_M(\widehat{\tilde{Y}}_{\sigma_2(i)})T_{-1\sigma_2(i)} \\
&  + \frac{1}{m}\sum_{i=1}^m [f_M(\tilde{Y}_{\sigma_1(i)})-f_M(\widehat{\tilde{Y}}_{\sigma_2(i)})]T_{-1\sigma_2(i)} \\
& =: T_0 + T_1 + T_2.
\end{align*}
Consider $T_{0}$. By Cauchy-Schwarz inequality,
\begin{equation}
\|T_{0}\|\le \left(\frac{1}{m}\sum_{i=1}^m \left[\tilde{Y}_{i} - f_M(\tilde{Y}_{i})\right]^2\right)^{1/2}\left(\frac{1}{m}\sum_{i=1}^m \|T_{-1i}\|^2 \right)^{1/2}.	
	\label{eq:CS_T0}
\end{equation}
Since $h$ is increasing, $|\eta_{di}|\le K$ implies $|\tilde{Y}_i| \le M$. Then, $\tilde{Y}_{i} - f_M(\tilde{Y}_{i})\ne 0$ implies $|\eta_{di}|> K$. Remark also that $|x - f_M(x)|\le |x|$. Then,
$$|\tilde{Y}_{i} - f_M(\tilde{Y}_{i})|\le |h(\eta_{di})|\ind{|\eta_{di}|>K}.$$
As a result,
$$\frac{1}{m}\sum_{i=1}^m \left[\tilde{Y}_{i} - f_M(\tilde{Y}_{i})\right]^2 \le \frac{1}{m}\sum_{i=1}^m h(\eta_{di})^2\ind{|\eta_{di}|>K}.$$
By the law of large numbers (LLN) and definition of $K$, we obtain, with probability approaching one (wpao),
\begin{equation}
\|T_{0}\|\le \frac{\delta}{5}.	
	\label{eq:T0}
\end{equation}

Next, consider $T_{1}$. We have
\begin{align*}
T_{1} & = \frac{1}{m}\sum_{i=1}^m f_M(\widehat{\tilde{Y}}_{i})T_{-1i} \\
& = \frac{1}{m}\sum_{i=1}^m \left[f_M(\widehat{\tilde{Y}}_{i})-f_M(\tilde{Y}_{i})\right]T_{-1i} + \frac{1}{m}\sum_{i=1}^m \tilde{Y}_{i}T_{-1i} + \frac{1}{m}\sum_{i=1}^m [f_M(\tilde{Y}_{i})-\tilde{Y}_{i}]T_{-1i} \\
& =: T_{11} + T_{12} + T_{13}.
\end{align*}
By the LLN, wpao,
\begin{equation}
\left\|T_{12}-E[h(\eta_{d1})T_{-11}]\right\|\le \frac{\delta}{5}.
	\label{eq:T12}
\end{equation}
By Cauchy-Schwarz inequality, we obtain for $T_{13}$ the same inequality as \eqref{eq:CS_T0}. Thus, wpao,
\begin{equation}
\|T_{13}\|\le \frac{\delta}{5}.	
	\label{eq:T13}
\end{equation}
Turning to $T_{11}$, we have, by Cauchy-Schwarz inequality,
\begin{equation}
\|T_{11}\|\le \left\{\frac{1}{m}\sum_{i=1}^m \left[f_M(\widehat{\tilde{Y}}_{i}) - f_M(\tilde{Y}_{i})\right]^2\right\}^{1/2}\left\{\frac{1}{m}\sum_{i=1}^m \|T_{-1i}\|^2\right\}^{1/2}.	
	\label{eq:T11}
\end{equation}
Remark that $\min(|\widehat{\eta}_{di}|,|\eta_{di}|)\ge K$ implies $\min(|\widehat{\tilde{Y}}_{i}|,|\tilde{Y}_{i}|)\ge M$, and then $f_M(\widehat{\tilde{Y}}_{i}) = f_M(\tilde{Y}_{i})$. Then,
\begin{align*}
	\left|f_M(\widehat{\tilde{Y}}_{i}) - f_M(\tilde{Y}_{i})\right| \le & \left|\widehat{\tilde{Y}}_{i} - \tilde{Y}_{i}\right| \ind{\min(|\widehat{\eta}_{di}|,|\eta_{di}|)< K} \\
		= & \left|h(\widehat{\eta}_{di})-h(\eta_{di})\right| \ind{\min(|\widehat{\eta}_{di}|,|\eta_{di}|)< K}.
\end{align*}
Because $I:=[-K,K]$ is compact, there exists $\nu>0$ such that for all $(x,y)\in I^2$, $|x-y|<\nu$ implies $|h(x)-h(y)|<\delta/\{6E[\|T_{-1}\|^2]\}$. Given \eqref{eq:conv_unif_eta} in Lemma \ref{lem:eta}, \eqref{eq:T11} and the LLN, we have, wpao
\begin{equation}
\|T_{11}\|\le \frac{\delta}{5}.	
	\label{eq:T11_fin}
\end{equation}

Finally, consider $T_2$. First, by Cauchy-Schwarz inequality,
\begin{equation}
\|T_2\|\le \left\{\frac{1}{m}\sum_{i=1}^m \left[f_M(\tilde{Y}_{\sigma_1(i)})- f_M(\widehat{\tilde{Y}}_{\sigma_2(i)}) \right]^2\right\}^{1/2} \left\{\frac{1}{m}\sum_{i=1}^m \|T_{-1i}\|^2\right\}^{1/2}.	
	\label{eq:ineq_T2}
\end{equation}
By the rearrangement inequality, because $f_M \circ h$ is increasing,
$$\sum_{i=1}^m f_M(\tilde{Y}_{\sigma_1(i)})f_M(\widehat{\tilde{Y}}_{\sigma_2(i)}) \ge \sum_{i=1}^m f_M(\tilde{Y}_{i})f_M(\widehat{\tilde{Y}}_{i}).$$
Thus, by what precedes, we have, wpao,
\begin{align*}
\frac{1}{m}\sum_{i=1}^m \left[f_M(\tilde{Y}_{\sigma_1(i)})- f_M(\widehat{\tilde{Y}}_{\sigma_2(i)})\right]^2 & \le \frac{1}{m}\sum_{i=1}^m \left[f_M(\tilde{Y}_i)- f_M(\widehat{\tilde{Y}}_i)\right]^2 \\
& \le \frac{\delta^2}{6E[\|T_{-1}\|^2]}.	
\end{align*}
When combined with \eqref{eq:ineq_T2} and the LLN, we have, wpao
\begin{equation}
\|T_2\|\le \frac{\delta}{5}.		
	\label{eq:T2}
\end{equation}
Finally, by combining \eqref{eq:T0}--\eqref{eq:T11} and \eqref{eq:T2}, we obtain that wpao,
$$\left\|\frac{1}{m}\sum_{i=1}^m \tilde{Y}_{\sigma_1(i)}T_{-1\sigma_2(i)} - E[h(\eta_d)T_{-1}]\right\| \le \delta.$$
Because $\delta$ was arbitrary, \eqref{eq:verif_fin}, and in turn \eqref{eq:remainder}, follows.

\subsubsection*{2. Linear approximation of the other terms} 
\label{ssub:part_2}

\medskip

Consider the following decomposition
\begin{align*}
\int_0^1 [F_n^{-1}(t)G_m^{-1}(t)-F^{-1}(t) G^{-1}(t)]dt= & \int_0^1 F^{-1}(G_m^{-1}-G^{-1})dt + \int_0^1 G^{-1}(F_n^{-1}-F^{-1})dt  + r_{n,m},	
\end{align*}
where $r_{n,m} := \int_0^1 (F_n^{-1}-F^{-1})(G_m^{-1}-G^{-1})dt$. We prove that the first two terms $T_{1m}$ and $T_{2n}$ are asymptotically linear. We prove below that the last term is asymptotically negligible.

\medskip
First, consider $T_{2n}=\int_0^1 G^{-1}(F_n^{-1}-F^{-1})dt$. We can always construct i.i.d. uniform random variables $\xi_i$ such that $Y_i = F^{-1}(\xi_i)$, see e.g. Eq. (55) p.57 in \citeauthor{SW86} (1986, SW hereafter). Now, we apply Theorem 19.1 in SW, combined with their Remark 2 p.667. Remark that their $\tilde{T}_n$ defined in their Eq. (56) corresponds to our $\int_0^1 G^{-1}F_n^{-1}dt$, with their $h$ being the identity function so that their $g(\mathbb{G}_n^{-1})$ is our $F_n^{-1}$ and their $J$ is our $G^{-1}$. Given that their (58) is the same as their (11), with just $\Psi_n=\Psi$, we can replace in their Theorem 19.1-(i), provided that their Assumptions 19.1 and 19.2 hold, $T_n-\mu_n$ by their $\tilde{T}_n-\mu$, which is our $T_{2n}$.

\medskip
We first check that SW's Assumption 19.1 holds. First, assume that Assumption \ref{hyp:reg}-(i) holds. Then, by Remark 19.1 in SW, we have $|F^{-1}(t)| \le M_1/[t(1-t)]^{1/(2+\eps)}$ for some $M_1$. Moreover, $|G^{-1}(t)| \le \max(\Supp(\eta_d))$. Thus, (16) and (19) in SW's Assumption 19.1 holds, with their $(b_1,b_2,d_1,d_2)$ satisfying $b_1=b_2=0$ and $d_1=d_2=1/(2+\eps)$ and thus their $a$ satisfying $a<1/2$. If instead Assumption \ref{hyp:reg}-(ii) holds, we reason similarly but instead use $b_1=b_2=1/4$ and $d_1=d_2=0$. Finally, assume that Assumption \ref{hyp:reg}-(iii) holds. Using again Remark 19.1 in SW, (16) and (19) in their Assumption 19.1 holds, with their $(b_1,b_2,d_1,d_2)$ satisfying $b_1=...=d_2=1/(4+\eps)$ and thus their $a$ satisfying again $a<1/2$. 

\medskip
Now, let us check that SW's Assumption 19.2 holds. Since $J_n=J$ in $\tilde{T}_n$, their assumption reduces in our context to the continuity of $G^{-1}$ except on a set of $\mu$-measure 0, where $\mu$ is the measure associated with $F^{-1}$. If Assumption \ref{hyp:reg}-(i) holds, $G^{-1}$ is continuous except at $G(h)$ for $h\in\Supp(\eta_d)$. But since $F^{-1}$ is continuous at $G(h)$, $\mu(\{G(h)\})=0$. If, instead, Assumption \ref{hyp:reg}-(ii) holds, the fact that $\Supp(\eta_d)$ is an interval implies that $G^{-1}$ is continuous. Finally, if Assumption \ref{hyp:reg}-(iii) holds, because $G^{-1}$ is monotone, its set of discontinuities $\mathcal{D}_{G^{-1}}$ is countable. Since $F^{-1}$ is continuous, $\mu(\{x\})=0$ for each $x \in \mathcal{D}_{G^{-1}}$. Hence, in all cases, SW's Assumption 19.2 holds.

\medskip 
Then, by Theorem 19.1 in SW and their equation just above (13),
$$\sqrt{n} T_{2n} =  - \frac{1}{n^{1/2}}\sum_{i=1}^n \int_0^1 [\ind{\xi_i \le t} -t] G^{-1}(t)dF^{-1}(t) + o_P(1).$$
Using $m/(n+m)\to (1-\lambda)$, Lemma \ref{lem:change_var} below and the definition of $\psi_4$, we  obtain
\begin{equation}
\sqrt{\frac{nm}{n+m}} T_{2n} =  (1-\lambda)^{1/2} \frac{1}{n^{1/2}}\sum_{i=1}^n \psi_{4i} + o_P(1). \label{eq:lin_Lstat2}
\end{equation}

We reason similarly for $T_{1m}$. Note that Assumption \ref{hyp:reg} is not symmetric in $F$ and $G$ so care should be taken when checking Assumption 19.2 in SW (their Assumption 19.1 holds by just reverting the choices of $(b_1,b_2)$ and $(d_1,d_2)$ considered above). Now, we must check the continuity of $F^{-1}$ except on a set of $\mu$-measure 0, where $\mu$ is the measure associated with $G^{-1}$. If Assumption \ref{hyp:reg}-(i) holds,  note that the set of discontinuity points $\mathcal{D}_{F^{-1}}$ of $F^{-1}$ is countable. Moreover, by assumption, $ \mathcal{D}_{F^{-1}} \cap G(\Supp(\eta_d))=\emptyset$. Hence, $\mu(\{x\})=0$ for each $x \in \mathcal{D}_{F^{-1}}$, implying that $\mu(\mathcal{D}_{F^{-1}})=0$. If Assumption \ref{hyp:reg}-(ii) holds, the same holds because $G^{-1}$ is continuous.  Finally, if Assumption \ref{hyp:reg}-(iii) holds, $F^{-1}$ is continuous so Assumption 19.2 in SW directly holds. Hence, at the end of the day, we obtain
\begin{equation}
\sqrt{\frac{nm}{n+m}} T_{1m}  =  \lambda^{1/2} \frac{1}{m^{1/2}}\sum_{i=1}^m \psi_{3i} + o_P(1). \label{eq:lin_Lstat1} 	
\end{equation}

\subsubsection*{3. Asymptotically negligible remainder term}

We now show that $R_{n,m}:=\sqrt{nm/(n+m)} r_{n,m} =o_P(1)$. Combined with \eqref{eq:first_part_asym}, \eqref{eq:lin_Lstat2} and \eqref{eq:lin_Lstat1}, this implies
$$\sqrt{\frac{nm}{n+m}} \left(\widehat{\overline{b}}_d - \overline{b}_d\right) = \frac{1}{E(\eta_d^2)} \left[\frac{\sqrt{\lambda}}{m^{1/2}}\sum_{i=1}^m (\psi_{1i}+\psi_{2i}+\psi_{3i}) + \frac{\sqrt{1-\lambda}}{n^{1/2}}\sum_{i=1}^n \psi_{4i} \right] + o_P(1). $$
The result then follows by $E[\psi_j]=0$, $E(\psi_j^2)<\infty$ for all $j=1,...,4$ and the central limit theorem.

\medskip
\paragraph{Case 1: Assumption \ref{hyp:reg}-(i) or (ii) holds.}
Let us assume that Assumption \ref{hyp:reg}-(ii) holds. Then $G^{-1}$ is continuous on $(0,1)$, and in particular on $F(\Supp(Y))$. Let $\Supp(Y)=\{y_1,...,y_K\}$ and $y_0=-\infty$. Remark that for any function $q$,
\begin{align*}
\int_0^1 F^{-1}(t)q(t)dt & = \int_0^1 \sum_{k=1}^K y_k \ind{F(y_{k-1})<t\le F(y_k)}q(t)dt,\\
\int_0^1 \widehat{F}^{-1}(t)q(t)dt & = \int_0^1 \sum_{k=1}^K y_k \ind{\widehat{F}(y_{k-1})<t\le \widehat{F}(y_k)}q(t)dt.
\end{align*}
Let $I_j:= (\widehat{F}(y_j),F(y_j)]$ if $\widehat{F}(y_j)<F(y_j)$, $I_j:= (F(y_j),\widehat{F}(y_j)]$ otherwise. Then,
\begin{align}
& \int_0^1 (\widehat{F}^{-1}(t) - F^{-1}(t)) q(t)dt \notag \\
= & \int_0^1 \sum_{k=1}^K y_k \bigg(\sgn(\widehat{F}(y_k)-F(y_k)) \ind{t\in I_k} - \sgn(\widehat{F}(y_{k-1})-F(y_{k-1})) \notag \\ 
& \hspace{1.7cm} \times \ind{t\in I_{k-1}}\bigg) q(t)dt \notag \\
= & \int_0^1 \sum_{k=1}^{K-1} (y_k-y_{k+1})\sgn(\widehat{F}(y_k)-F(y_k))\ind{t\in I_k} q(t)dt. \label{eq:nouv_expr}
\end{align}
Now, we have, for $k=1,...,K-1$,
\begin{align*}
\widehat{G}^{-1}\left(\min(\widehat{F}(y_k), F(y_k))\right) \le & \frac{1}{|\widehat{F}(y_k)-F(y_k)|}\int_0^1 \ind{t\in I_k} \widehat{G}^{-1}(t)dt \\
\le & \widehat{G}^{-1}\left(\max(\widehat{F}(y_k), F(y_k))\right).    
\end{align*}
By continuity of $G^{-1}$  at $F(y_k)$, $\widehat{G}^{-1}(F(y_k))\convP G^{-1}(F(y_k))$. Thus,
$$\frac{1}{|\widehat{F}(y_k)-F(y_k)|}\int_0^1 \ind{t\in I_k} \widehat{G}^{-1}(t)dt \convP G^{-1}(F(y_k)).$$
The same result holds replacing $\widehat{G}^{-1}$ by $G^{-1}$. Hence,
$$\frac{1}{|\widehat{F}(y_k)-F(y_k)|}\int_0^1 \ind{t\in I_k} (\widehat{G}^{-1}(t)-G^{-1}(t))dt= o_P(1).$$
Then, replacing $q$ by $\widehat{G}^{-1}(t)-G^{-1}(t)$ in \eqref{eq:nouv_expr}, we obtain
\begin{align*}
	& \int_0^1 (\widehat{F}^{-1}(t) - F^{-1}(t))(\widehat{G}^{-1}(t)-G^{-1}(t)) dt \\
	= & \sum_{k=1}^{K-1} (y_k-y_{k+1})\sgn(\widehat{F}(y_k)-F(y_k))\int_0^1 \ind{t\in I_k}(\widehat{G}^{-1}(t)-G^{-1}(t))dt \\
	= & \sum_{k=1}^{K-1} (y_k-y_{k+1})(\widehat{F}(y_k)-F(y_k)) \times o_P(1).
\end{align*}
Then, because $\sqrt{nm/(n+m)}(\widehat{F}(y_k)-F(y_k))=O_P(1)$, we obtain $R_{n,m}=o_P(1)$. 

\medskip
Now, if Assumption \ref{hyp:reg}-(i) holds, the reasoning is the same as above, just reverting the roles of $F$ and $G$, once we note that by assumption, $F^{-1}$ is continuous at  $G(\Supp(\eta_d))$.

\medskip
\paragraph{Case 2: Assumption \ref{hyp:reg}-(iii) holds.}

We have, by Cauchy-Schwarz inequality,
$$R_{n,m}^2 \le \frac{nm}{n+m} W^2_2(F_n,F) W^2_2(G_m,G).$$
Hence,  by independence,
$$E\left[R_{n,m}^2\right] \le \frac{nm}{n+m} E\left[W^2_2(F_n,F)\right]E\left[W^2_2(G_m,G)\right].$$

Now, assume that $Z=\eta_d$ in Assumption \ref{hyp:reg}-(iii); the proof is the same if, instead, $Z=Y$. Theorem 1 in \cite{fournier2015rate} shows that
$$E\left[W^2_2(F_n,F)\right] \lesssim n^{-1/2},$$
where ``$\lesssim$'' means that the inequality holds up to a number independent of $(n,m)$. We now prove that
\begin{equation}
	E\left[W^2_2(G_m,G)\right] = o(m^{-1/2}),
	\label{eq:controle_main_term}
\end{equation}
which implies that $E[R_{n,m}^2]=o(1)$ and concludes the proof by Markov inequality. First, remark that by Theorem 4.3 of \cite{bobkov2019one},
\begin{equation}
E\left[W^2_2(G_m,G)\right] \le \frac{2}{m}\sum_{i=1}^{m} V(\eta_{d(i)}),	
	\label{eq:link_W2_OST}
\end{equation}
where $\eta_{d(1)} < ... < \eta_{d(m)}$ denotes the order statistic of an i.i.d. sample $(\eta_{d1},...,\eta_{dm})$ from $G$. Then, by Condition \eqref{eq:condit_hazard} and Lemma \ref{lem:borne_var}, we have
\begin{align}
	\sum_{i=1}^m V(\eta_{d(i)} ) & \lesssim E\left[\sum_{i=1}^m \frac{1}{i\wedge (m+1-i)}\left(\frac{1}{C^2} \vee \frac{\eta_{d(i)}^2\ln(1+|\eta_{d(i)}|)^4}{K^2} + \eta_{d(i)}^2\right)\right] \notag \\
	& \lesssim \left(E[Z_m^2] + E[Z_m^2\ln(1+Z_m)^4] \right) \sum_{i=1}^{\left[\frac{m+1}{2}\right]} \frac{1}{i} \notag \\
	&  \lesssim E[Z_m^{2+\eps/3}] \left[1+\ln(m)\right], \label{eq:ineg_OST}
\end{align}
where $Z_m=\max_{i=1,...,m}(|\eta_{di}|)$ and $[x]$ denotes the integer part of $x$. Now,
\begin{equation}
	m^{-\frac{2+\eps/3}{4+\eps}} E[Z_m^{2+\eps/3}] \le \left\{m^{-1} E[Z_m^{4+\eps}]\right\}^{\frac{2+\eps/3}{4+\eps}} = o(1),  \label{eq:controle_mom}
\end{equation}
where the inequality is due to Jensen's inequality and the equality holds by, e.g., Exercise 4 in Section 2.3 of \cite{VdV_Wellner} and because $E[|\eta_{d1}|^{4+\eps}]<\infty$. Combining \eqref{eq:link_W2_OST}, \eqref{eq:ineg_OST} and \eqref{eq:controle_mom}, we obtain \eqref{eq:controle_main_term}.



\subsection{Theorem \ref{thm:asym_CI}}
\label{ssub:proof_asym_CI}

Theorem \ref{thm:ident_no_common} ensures that $\overline{b}_d>0>-\overline{b}_{-d}$. Then, by Theorem \ref{thm:asym_Xo}, it suffices to prove the following:
\begin{align}
&\frac{1}{m}\sum_{j=1}^m \widehat{\psi}_{kj}^2 \convP E[\psi_k^2], \ \text{ for } k\in\{1,...,3\}, \ \frac{1}{n}\sum_{i=1}^n \widehat{\psi}_{4i}^2 \convP E[\psi_4^2] \label{eq:var_conv}\\ 
&\frac{1}{m}\sum_{j=1}^m \widehat{\psi}_{kj}\widehat{\psi}_{k'j} \convP E[\psi_k\psi_{k'}]  \ \text{ for } k,k'\in\{1,...,3\}, \ k\neq k'. \label{eq:cov_conv}
\end{align}  
Hereafter, we let, for any $N \subset \R$ and $\eps\ge 0$, $N^\eps:=\{x\in\R:\exists y\in N: |x-y|\le \eps\}$.

\paragraph{Eq. \eqref{eq:var_conv} holds for $k=1$.} We actually prove 
$$\frac{1}{m}\sum_{j=1}^m (\widehat{\psi}_{1j}-\psi_{1j})^2 \convP 0.$$
The result then follows by the triangle inequality and the LLN applied to the $(\psi_{1j}^2)_{j=1,...,m}$. By definition of $\widehat{\psi}_{1j}$ and $\psi_{1j}$ and convexity of $x\mapsto x^2$,
\begin{align*}
(\widehat{\psi}_{1j}-\psi_{1j})^2 \le & 2 \widehat{\overline{b}}_d{}^2 \bigg(\widehat{\eta}_{dj}^2 - \frac{1}{m}\sum_{j=1}^m \widehat{\eta}_{dj}^2 - \eta_{dj}^2 + E[\eta_d^2]\bigg)^2 + 2 (\eta_{dj}^2 - E[\eta_d^2])^2 (\widehat{\overline{b}}_d - \overline{b}_d)^2.
\end{align*}
The sample mean of the second term on the right-hand side converges to 0 in probability by Theorem \ref{thm:asym_Xo} and $E[\eta_d^4]<\infty$. Recall that $\frac{1}{m}\sum_{j=1}^m \widehat{\eta}_{dj}^2 = \frac{1}{m}\sum_{j=1}^m \eta_{dj}^2 + o_P(m^{-1/2})$. Also, $\widehat{\overline{b}}_d=O_P(1)$. Then, using again convexity of $x\mapsto x^2$, it suffices to prove that 
\begin{equation}
\frac{1}{m}\sum_{j=1}^m \left(\widehat{\eta}_{dj}^2 - \eta_{dj}^2\right)^2\convP 0.
	\label{eq:conv_diff_eta2}
\end{equation}
Remark that $(\widehat{\eta}_{dj}^2 - \eta_{dj}^2)^2 = (\widehat{\eta}_{dj}+\eta_{dj})^2(\widehat{\eta}_{dj}-\eta_{dj})^2$. Then, \eqref{eq:conv_diff_eta2} follows from \eqref{eq:conv_unif_eta} in Lemma \ref{lem:eta} and $(1/m)\sum_{j=1}^m (\widehat{\eta}_{dj}+\eta_{dj})^2=O_P(1)$.

\paragraph{Eq. \eqref{eq:var_conv} holds for $k=2$.} Note that $\widehat{\psi}_{2j}=\widehat{\lambda}'\widehat{\delta}_j$, with $\widehat{\delta}_j =T_{-1j}\widehat{\eta}_{dj} $ and $$\widehat{\lambda} =  \left(\frac{1}{m}\sum_{j=1}^m T_{-1j}T_{-1j}'\right)^{-1} \left(\frac{1}{m}\sum_{j=1}^m \widehat{h}(\widehat{\eta}_{dj}) T_{-1j}'\right).$$ 
This implies
$$\frac{1}{m}\sum_{j=1}^m \widehat{\psi}_{2j}^2 = \widehat{\lambda}' \left(\frac{1}{m}\sum_{j=1}^m\widehat{\delta}_j\widehat{\delta}_j'\right)\widehat{\lambda}.$$
It suffices to show convergence of $\widehat{\lambda}$ and the term in parentheses. Regarding the latter, we have
\begin{align*}
\frac{1}{m}\sum_{j=1}^m\widehat{\delta}_j\widehat{\delta}_j' 
 = & \frac{1}{m}\sum_{j=1}^m(T_{-1j}T'_{-1j}) (\widehat{\eta}_{dj} - \eta_{dj})^2 + \frac{2}{m}\sum_{j=1}^m(T_{-1j}T'_{-1j}) \eta_{dj} (\widehat{\eta}_{dj} - \eta_{dj}) \\
 & + \frac{1}{m}\sum_{j=1}^m(T_{-1j}T'_{-1j}) \eta_{dj}^2.
\end{align*}
Moreover,
$$\bigg\|\frac{1}{m}\sum_{j=1}^m(T_{-1j}T'_{-1j}) (\widehat{\eta}_{dj} - \eta_{dj})^2\bigg\|\le \max_j (\widehat{\eta}_{dj}-\eta_{dj})^2 \frac{1}{m}\sum_{j=1}^m \|T_{-1j}T'_{-1j}\|.$$
By Lemma \ref{lem:eta}, the left-hand side is an $o_P(1)$. Similarly, $(1/m)\sum_{j=1}^m(T_{-1j}T'_{-1j}) \eta_{dj} (\widehat{\eta}_{dj} - \eta_{dj})=o_P(1)$. Then, by the LLN,
$$\frac{1}{m}\sum_{j=1}^m\widehat{\delta}_j\widehat{\delta}_j' \convP E[\eta_{dj}^2 T_{-1j}T'_{-1j}].$$ 

Let us turn to $\widehat{\lambda}$. It suffices to prove that 
\begin{equation}
\frac{1}{m}\sum_{j=1}^m \widehat{h}(\widehat{\eta}_{dj})T'_{-1j} \convP E[h(\eta_d)T_{-1}'].	
	\label{eq:conv_gamma_hat}
\end{equation}

Suppose first that $|\Supp(\eta_d)|=|\Supp(X)|<\infty$ and let $(u_1,...,u_K):=\Supp(\eta_d)$. Then, by \eqref{eq:conv_unif_eta}, wpao, $|\{\widehat{\eta}_{d1},...,\widehat{\eta}_{dm}\}|=K$ and there exists a permutation $\sigma$ such that both $\widehat{\eta}_{d\sigma(1)} \le ...\le \widehat{\eta}_{d\sigma(m)}$ and $\eta_{d\sigma(1)} \le ...\le \eta_{d\sigma(m)}$. If so,
$$\frac{1}{m}\sum_{j=1}^m \widehat{h}(\widehat{\eta}_{dj})T'_{-1j} = \sum_{k=1}^K (G_m(u_k)-G_m(u_{k-1})) \overline{Y}_k \overline{T}_{-1k}',$$
with the convention that $u_{0}=-\infty$ and where, letting $m_k:=|\{j:\eta_{dj}=u_k\}|$, $\alpha_{n,m,k} := \lceil n G_m(u_{k-1})\rceil-1$ and $\beta_{n,m,k} := \lceil n G_m(u_{k})\rceil$,
\begin{align*}
\overline{T}_{-1k}:=&\; \frac{1}{m_k}\sum_{j:\eta_{dj}=u_k} T_{-1j}, \\
\overline{Y}_k := & \; \int_0^1 F_n^{-1}\left[G_m(u_{k-1})+u(G_m(u_{k})-G_m(u_{k-1}))\right]du \\
= & \; \frac{1}{\beta_{n,m,k}-\alpha_{n,m,k}}\left[\sum_{i=\alpha_{n,m,k}+1}^{\beta_{n,m,k}} Y_{(i)} - \left(\lambda^1_{n,m,k} Y_{(\alpha_{n,m,k}+1)} + \lambda^2_{n,m,k} Y_{(\beta_{n,m,k})}\right)\right],
\end{align*}
for some $(\lambda^1_{n,m,k},\lambda^2_{n,m,k})\in[0,1]^2$. By the LLN, $\overline{T}_{-1k}\convP E[T_{-1}|\eta_d=u_k]$. Remark that $\alpha_{n,m,k}/n\convP G(u_{k-1})$ and $\beta_{n,m,k}/n\convP G(u_k)$. Then, by e.g. (22) p.681 in \cite{SW86}, we have
$$\frac{1}{\beta_{n,m,k}-\alpha_{n,m,k}}\sum_{i=\alpha_{n,m,k}+1}^{\beta_{n,m,k}} Y_{(i)}\convP E\left[Y|Y\in[F^{-1}\circ G(u_{k-1}), F^{-1}\circ G(u_{k})]\right]= h(u_k)$$
Moreover, $$ \frac{|Y_{(\beta_{n,m,k})}|}{\beta_{n,m,k}-\alpha_{n,m,k}} \le \frac{\max_i |Y_i|/n}{G(u_k) - G(u_{k-1})+o_P(1)}\convP 0. $$ and
$$\frac{|Y_{(\alpha_{n,m,k}+1)}|}{\beta_{n,m,k}-\alpha_{n,m,k}} \le \frac{\max_i |Y_i|/n}{G(u_k) - G(u_{k-1})+o_P(1)}\convP 0.$$
Hence, $\overline{Y}_k\convP h(u_k)$. As a result,
$$\frac{1}{m}\sum_{j=1}^m \widehat{h}(\widehat{\eta}_{dj})T'_{-1j} \convP \sum_{k=1}^K (G(u_k)-G(u_{k-1})) h(u_k) E[T_{-1}'|\eta_d=u_k] = E[h(\eta_d) T_{-1}'].$$

\medskip
Next, let us prove \eqref{eq:conv_gamma_hat} when $G$ is continuous. By the LLN and Cauchy-Schwarz inequality, \eqref{eq:conv_gamma_hat} holds if 
\begin{equation}
\frac{1}{m}\sum_{j=1}^m \left(\widehat{h}(\widehat{\eta}_{dj})-h(\eta_{dj})\right)^2 \convP 0.	
	\label{eq:conv_for_psi2}
\end{equation}
Fix $\delta>0$. By Assumption \ref{hyp:reg} and the dominated convergence theorem, there exists $\overline{\eps}>0$ and a compact set $I\subset (0,1)$ such that (i) $E[\ind{G(\eta_d)\not\in I}]<\delta$; (ii) $I \subset I^{\overline{\eps}}\subset (0,1)$; (iii) $I^{\overline{\eps}}\cap \mathcal{D}_{F^{-1}}=\emptyset$, with $\mathcal{D}_{F^{-1}}$ the set of discontinuity points of $F^{-1}$. Since $F^{-1}$ is continuous on $I^{ \overline{\eps}}$, it is also uniformly continuous on this compact set. Then, there exists $\eps \in (0,\overline{\eps})$ such that for all $(x,y)\in I^{\overline{\eps}\,2}$, $|x-y|\le \eps$ implies $|F^{-1}(x)-F^{-1}(y)|\le \delta^{1/2}$. Moreover, because $G(\eta_{dj})\in I$ implies that $\eta_{dj}$ belongs to a bounded set, by \eqref{eq:conv_unif_Ghat} and its variant in Lemma \ref{lem:eta}, wpao,
$$\max_{j:G(\eta_{dj})\in I}\; \max\left(|\widehat{G}(\widehat{\eta}_{dj})-G(\eta_{dj})|,|\widehat{G}(\widehat{\eta}^-_{dj})-G(\eta_{dj})|\right) \le \eps.$$
Then, under this event,
\begin{align}
&\frac{1}{m}\sum_{j=1}^m \left(\widehat{h}(\widehat{\eta}_{dj})-h(\eta_{dj})\right)^2\ind{G(\eta_{dj})\in I} \\
& \le  \max_{j: G(\eta_{dj})\in I} \int_0^1 |F_n^{-1}\left(\widehat{G}(\widehat{\eta}_{dj}^-) + u (\widehat{G}(\widehat{\eta}_{dj}) - \widehat{G}(\widehat{\eta}_{dj}^-))) \right)-F^{-1}\circ G(\eta_{dj})|^2 du \notag \\
& \le  \max_{j: G(\eta_{dj})\in I} \sup_{u \in [0,1]} |F_n^{-1}\left(\widehat{G}(\widehat{\eta}_{dj}^-) + u (\widehat{G}(\widehat{\eta}_{dj}) - \widehat{G}(\widehat{\eta}_{dj}^-))) \right)-F^{-1}\circ G(\eta_{dj})|^2  \notag \\
& \le 2 \sup_{x\in I^\eps} |F_n^{-1}(x)-F^{-1}(x)|^2  + 2 \sup_{(t,u)\in I_{\eps}^2:|t-u|\le \eps} |F^{-1}(t)-F^{-1}(u)|^2 \notag \\
& \le 2\left[\sup_{x\in I^\eps} |F_n^{-1}(x)-F^{-1}(x)|^2+ \delta\right],
\label{eq:decomp_diff_hhat}
\end{align}
where we have used Jensen's inequality for the first inequality. Since $F^{-1}$ is continuous on $I^\eps\subset (0,1)$, the first term on the right-hand side is smaller than $\delta$ wpao. Then, to prove \eqref{eq:conv_for_psi2}, it suffices to show that wpao,
$$\frac{1}{m}\sum_{j=1}^m \left(\widehat{h}(\widehat{\eta}_{dj})-h(\eta_{dj})\right)^2\ind{G(\eta_{dj})\not\in I} \le q(\delta),$$	
for some $q(\cdot)$ continuous at 0 and such that $q(0)=0$. By the Cauchy-Schwarz inequality, we obtain 
\begin{align*}
&	\left| \frac{1}{m}\sum_{j=1}^m \left(\widehat{h}(\widehat{\eta}_{dj})-h(\eta_{dj})\right)^2\ind{G(\eta_{dj})\not\in I} \right| \\
& \leq 	\left( \frac{1}{m}\sum_{j=1}^m \ind{G(\eta_{dj})\not\in I} \right)^{1/2}	\left( \frac{1}{m}\sum_{j=1}^m \left(\widehat{h}(\widehat{\eta}_{dj})-h(\eta_{dj})\right)^4 \right)^{1/2}
\end{align*}
hence using that $\sum_{j=1}^m \ind{G(\eta_{dj})\not\in I}/m \convP E[\ind{G(\eta_{dj})\not\in I}] < \delta $,  it suffices to prove that
$$\frac{1}{m}\sum_{j=1}^m \widehat{h}(\widehat{\eta}_{dj})^4=O_P(1),\quad \frac{1}{m}\sum_{j=1}^m  h(\eta_{dj})^4 = O_P(1).$$
The second result follows by the LLN, since $h(\eta_{dj})\stackrel{d}{=}Y_j$ and $E[Y^4]<\infty$. For the first, remark that by Jensen's inequality and since $\widehat{G}(\widehat{\eta}_{dj})-\widehat{G}(\widehat{\eta}_{dj}^-)\ge 1/m$,
\begin{align*}
\widehat{h}(\widehat{\eta}_{dj})^4 \le & \; \int_0^1 F_n^{-1}[\widehat{G}(\widehat{\eta}_{dj}^-)+u(\widehat{G}(\widehat{\eta}_{dj}) - \widehat{G}(\widehat{\eta}_{dj}^-))]^4du \\ 
\le & \; m \int_{\widehat{G}(\widehat{\eta}_{dj}^-)}^{\widehat{G}(\widehat{\eta}_{dj})}F_n^{-1}(u)^4du.	
\end{align*}
Hence, by Fubini-Tonelli's theorem,
$$\frac{1}{m}\sum_{j=1}^m \widehat{h}(\widehat{\eta}_{dj})^4\le \int_0^1 F_n^{-1}(u)^4du = \frac{1}{n}\sum_{i=1}^n Y_i^4= O_P(1).$$
The result follows.

\paragraph{Eq. \eqref{eq:var_conv} holds for $k=3$.}
Let $a\in(0,1)$ be a continuity point of $G^{-1}$ and $F^{-1}$. We have  
\begin{align*}
	\psi_3 & = -\int \left[1\{\eta_d \leq u\} - G(u) \right]F^{-1}\circ G(u)du \\
 	& = \int^{G^{-1}(\xi_d)}_{G^{-1}(1)} F^{-1}\circ G(u)du + \int G(u)F^{-1}\circ G(u)du \\
 	& = \int^{G^{-1}(\xi_d)}_{G^{-1}(a)} F^{-1}\circ G(u)du + \int G(u)F^{-1}\circ G(u)du - \int^{G^{-1}(1)}_{G^{-1}(a)} F^{-1}\circ G(u)du,
\end{align*}
for some $\xi_d\sim\mathcal{U}[0,1]$. As a result,
$$E[\psi_3] = \int_0^1 \int_{G^{-1}(a)}^{G^{-1}(t)} F^{-1}(G(s))dsdt + \int G(u)F^{-1}\circ G(u)du - \int^{G^{-1}(1)}_{G^{-1}(a)}F^{-1}\circ G(u)du.$$ 
Since, by Fubini's theorem, we also have $E[\psi_3]=0$, we obtain
\begin{equation}
\psi_3 =  \psi_3 - E[\psi_3] = \overline{c}_2(\xi_d,F,G),	
	\label{eq:psi_3}
\end{equation}
where 
\begin{align}
  \overline{c}_2(t,F,G):= & c_2(t,F,G) - \int_0^1 c_2(u,F,G)du  \label{eq_c2bar}\\
    c_2(t,F,G):= & \int_{G^{-1}(a)}^{G^{-1}(t)} F^{-1}(G(s))ds ,\notag
\end{align}
Moreover, using that $\widehat{G}_m^{-1}(t) = \widehat{\eta}_{d,(i)}$ for $t \in \left((i-1)/m, i/m\right]$ and all $i=1,\dots,m$,
\begin{equation}\label{eq:psi32}
\frac{1}{m}\sum_{j=1}^m\widehat{\psi}_{3j}^2 = \int_0^1 \overline{c}_2^2(t,F_n,\widehat{G}_m)dt.
\end{equation}
By Lemma \ref{lem:A.1.modif} below and the continuous mapping theorem, it suffices to show that $\mathcal{W}_4(\widehat{G}_m,G)\convP 0$ and either $\mathcal{W}_4(F_n,F)\convP 0$ (if Assumption \ref{hyp:reg}-(ii) or (iii) holds), or $\mathcal{W}_2(F_n,F)\convP 0$ (if Assumption \ref{hyp:reg}-(i) holds). The condition on $F_n$ holds by, e.g., Theorem 2.13 in \cite{bobkov2019one}. By the same theorem, $\mathcal{W}_4(G_m,G)$ converge to 0 a.s. Moreover,  
\begin{align*}
\mathcal{W}_4(\widehat{G}_m,G_m) \le & \left[\frac{1}{m}\sum_{i=1}^m (\eta_{di} - \widehat{\eta}_{di})^4\right]^{1/4} \le \max_{i=1,...,m}\left|\eta_{di} - \widehat{\eta}_{di}\right| \convP 0,
\end{align*}
where the first inequality follows by definition of $\mathcal{W}_4$ and the convergence holds by Lemma \ref{lem:eta}. Then, $\mathcal{W}_4(\widehat{G}_m,G)\convP 0$ follows by the triangle inequality.

\paragraph{Eq. \eqref{eq:var_conv} holds for $k=4$.} The reasoning is the same as for $k=3$: we just exchange the roles of $F$ and $G$ and note that Lemma \ref{lem:A.1.modif} still applies then.

\paragraph{Eq. \eqref{eq:cov_conv} holds for $(k,k')=(1,2)$.} We have 
$$\widehat{\psi}_{1j}\widehat{\psi}_{2j}=- \widehat{\overline{b}}_d \left(\widehat{\eta}_{dj}^2 - \frac{1}{m}\sum_{k=1}^m \widehat{\eta}_{dk}^2\right) \widehat{\lambda}'T_{-1j}\widehat{\eta}_{dj}.$$
The result follows by convergences of $\widehat{\lambda}$ and $\widehat{\overline{b}}_d$, and Eq. \eqref{eq:conv_unif_eta} in Lemma \ref{lem:eta}.

\paragraph{Eq. \eqref{eq:cov_conv} holds for $(k,k')=(1,3)$.} It suffices to prove that
$$\frac{1}{m}\sum_{j=1}^m \widehat{\eta}_{dj}^2\widehat{\psi}_{3j} \convP E[\eta_d^2\psi_3].$$
To this end, note that by \eqref{eq:psi_3}
\begin{align*}
	\eta_{d}^2\psi_3 & = (G^{-1}(\xi_d))^2\overline{c}_2(\xi_d,F,G).
\end{align*}
Hence by the same argument as \eqref{eq:psi32},
$$\frac{1}{m}\sum_{j=1}^m\widehat{\eta}_{dj}^2\widehat{\psi}_{3j} = \int_{0}^1 \overline{\widetilde{c}}_2(t,F_n,\widehat{G}_m) dt, $$
where $\overline{\widetilde{c}}_{2}(t,F,G):=   \widetilde{c}_{2}(t,F,G) - \int_0^1 \left(G^{-1}(u)\right)^2du \int_0^1   c_2(u,F,G)du$ and
\begin{align*}
   \widetilde{c}_{k}(t,F,G) &: = \left(G^{-1}(t)\right)^k \int_{G^{-1}(a)}^{G^{-1}(t)} F^{-1}(G(s))ds.
\end{align*}
We obtain the result using Lemma \ref{lem:A.1.modif} below, the continuous mapping theorem, and the fact as shown above,  we have $\mathcal{W}_4(\widehat{G}_m,G)\convP 0$ and either $\mathcal{W}_4(F_n,F)\convP 0$ (if Assumption \ref{hyp:reg}-(ii) or (iii) holds), or $\mathcal{W}_2(F_n,F)\convP 0$ (if Assumption \ref{hyp:reg}-(i) holds). 

\paragraph{Eq. \eqref{eq:cov_conv} holds for $(k,k')=(2,3)$.} 
We have 
\begin{align*}
	 \frac{1}{m}\sum_{j=1}^m \widehat{\psi}_{2j}\widehat{\psi}_{3j} &  = - \frac{1}{m}\sum_{j=1}^m \widehat{\lambda}'T_{-1j}\widehat{\eta}_{dj} \widehat{\psi}_{3j}\\
	  &  = - \frac{1}{m}\sum_{j=1}^m \lambda'T_{-1j}\widehat{\eta}_{dj} \widehat{\psi}_{3j} + o_p(1),
\end{align*} 
using the convergence of $\widehat{\lambda}$. Then, 

\begin{align*}
	 -\frac{1}{m}\sum_{j=1}^m \lambda'T_{-1j}\widehat{\eta}_{dj} \widehat{\psi}_{3j} = & -\frac{1}{m}\sum_{j=1}^m  \lambda'T_{-1j}(\widehat{\eta}_{dj}\widehat{\psi}_{3j} - \eta_{dj} \psi_{3j}) - \frac{1}{m}\sum_{j=1}^m \psi_{2j} \psi_{3j}. 
\end{align*}	
Using the Cauchy-Schwarz inequality, we have 
$$ \left| \frac{1}{m}\sum_{j=1}^m  \lambda'T_{-1j}(\widehat{\eta}_{dj}\widehat{\psi}_{3j} - \eta_{dj} \psi_{3j})\right|\leq \left(\frac{1}{m}\sum_{j=1}^m (\lambda'T_{-1j})^2\right)^{1/2}\left( \frac{1}{m}\sum_{j=1}^m (\widehat{\eta}_{dj}\widehat{\psi}_{3j} - \eta_{dj}\psi_{3j})^2 \right)^{1/2}$$
The first term on the right hand side is bounded in probability by the LLN, since $E[\|X\|^4]<\infty$. By the LLN for the term $\frac{1}{m}\sum_{j=1}^m \psi_{2j} \psi_{3j}$ it suffices to show that   $\frac{1}{m}\sum_{j=1}^m (\widehat{\eta}_{dj}\widehat{\psi}_{3j}  - \eta_{dj}\psi_{3j})^2 \convP 0$. We have, using 
$\eta_{d}\psi_3 =  \widetilde{c}_1(\xi_d,F,G)$ with $\xi_d \sim \mathcal{U}[0,1]$, that 
\begin{align*}
    \frac{1}{m}\sum_{i=1}^m (\widehat{\eta}_{di}\widehat{\psi}_{3i}  - \eta_{di}\psi_{3i})^2 & =\sum_{i=1}^m \int_{(i-1)/m}^{i/m} \left(\widetilde{c}_1(t,F_n,\widehat{G}_m) - \widetilde{c}_1(\xi_{d,(i)},F, G)\right)^2dt  \\ 
   & \leq 2 \sum_{i=1}^m \int_{(i-1)/m}^{i/m} \left(\widetilde{c}_1(t,F_n,\widehat{G}_m) - \widetilde{c}_1(t,F, G)\right)^2 dt \\
   & \quad + 2 \sum_{i=1}^m \int_{(i-1)/m}^{i/m} \left(\widetilde{c}_1(t,F,G) - \widetilde{c}_1(\xi_{d,(i)},F, G)\right)^2 dt \\
    & \leq 2  \int_{0}^{1} \left(\widetilde{c}_1(t,F_n,\widehat{G}_m) - \widetilde{c}_1(t,F, G)\right)^2 dt \\
   & \quad + 2 \sum_{i=1}^m \int_{(i-1)/m}^{i/m} \left(\widetilde{c}_1(t,F,G) - \widetilde{c}_1(\xi_{d,(i)},F, G)\right)^2dt. 
\end{align*}
By Lemma \ref{lem:A.1.modif} and the continuous mapping theorem, $\int_{0}^{1} \left(\widetilde{c}_1(t,F_n,\widehat{G}_m) - \widetilde{c}_1(t,F, G)\right)^2 dt \convP 0$. Next, let us prove that
\begin{equation}
\Delta:=\sum_{i=1}^m \int_{(i-1)/m}^{i/m} \left(\widetilde{c}_1(t,F,G) - \widetilde{c}_1(\xi_{d,(i)},F, G)\right)^2dt\convP 0.
    \label{eq:conv_last_term_var}
\end{equation}
Let $F_{\xi,n}$ denote the empirical cdf of the $(\xi_{d,i})_{i=1,...,m}$. We have \citep[see, e.g.][Eq. (11) p.86]{SW86}
\begin{equation}
\max_{i=1,...,m} \left|\xi_{d,(i)} - \frac{i}{m}\right|\le \sup_{t\in[0,1]}|F^{-1}_{\xi,n}(t)-t| \convP 0.
    \label{eq:conv_unif}
\end{equation}
Fix $\eps>0$ and $M>1$. Because continuous functions are dense in $L^2([0,1])$, there exists a continuous function $\widetilde{c}^c_1$ such that 
\begin{equation}
\Delta_1:=\int_0^1 \left(\widetilde{c}_1(t,F,G) - \widetilde{c}^c_1(t)\right)^2dt<\frac{\eps}{6M}.
    \label{eq:Delta1}
\end{equation}
Moreover, 
\begin{equation}
\Delta \le 3(\Delta_1 +\Delta_2+\Delta_3),
\label{eq:Delta_decomp}
 \end{equation}
 where 
 \begin{align*}
     \Delta_2 := & \sum_{i=1}^m \int_{(i-1)/m}^{i/m} \left(\widetilde{c}^c_1(t) - \widetilde{c}^c_1(\xi_{d,(i)})\right)^2dt, \\
     \Delta_3 := & \frac{1}{m}\sum_{i=1}^m \left(\widetilde{c}^c_1(\xi_{d,i})- \widetilde{c}_1(\xi_{d,i},F,G)\right)^2.
 \end{align*}
 Since $\widetilde{c}^c_1$ is uniformly continuous on $[0,1]$, there exists $\delta>0$ such that for all $(x,y)\in [0,1]^2$, $|x-y|\le \delta$ implies that $|\widetilde{c}^c_1(x)-\widetilde{c}^c_1(y)|\le [\eps/(6M)]^{1/2}$. Combined with \eqref{eq:conv_unif}, this implies that for all $m>2/\delta$,
 \begin{equation}
P\left(\Delta_2 > \frac{\eps}{6M}\right)\le \frac{1}{M}.
     \label{eq:conv_unif_xi}
 \end{equation}
Finally, by Markov's inequality, for all $q>0$,
\begin{equation}
P\left(\Delta_3 > q \eps\right) \le \frac{E[\Delta_3]}{q\eps} = \frac{\Delta_1}{q\eps} < \frac{1}{q\,6M}.
    \label{eq:markov}
\end{equation}
Using \eqref{eq:Delta1}--\eqref{eq:markov}, we finally obtain
\begin{align*}
P(\Delta > \eps)
&\leq P\left(\Delta_2 > \frac{\eps}{6M}\right)
P\left(\Delta_2 \leq \frac{\eps}{6M},
\Delta_3 > \frac{\eps(1-1/(2M))}{3}\right) \\
&\leq \frac{1}{M}
\frac{1}{\left(1-\frac{1}{2M}\right)2M}
= \frac{1}{M}+\frac{1}{2M-1}.
\end{align*}
Eq. \eqref{eq:conv_last_term_var} follows since $\eps>0$ and $M>1$ were arbitrary.


\subsection{Theorem \ref{thm:cons}}

As the proof is long, we first give a sketch of it. Since the empirical partition $\widehat g_{K_{n,m}}$ is the K-means clustering applied to the estimated index $\widehat g_0$ (rather than the population index $g_0$), it depends both on the sampled data (through K-means) and on the first-stage estimator $\widehat g_0$. To make this dependence transparent, we introduce the population $L^2$-optimal $K$-means partition $g^*_{K_{n,m}}$ (built from the population $L^2$-optimal cluster centers $\mathbf v^*$). We can then decompose the estimator's error as
\begin{equation}
	\widehat{\overline b}{}_d^{\,\widehat g_{K_{n,m}}} - \overline b_d^{g_0} \;=\; \underbrace{\big(\widehat{\overline b}{}_d^{\,\widehat g_{K_{n,m}}} - \overline b_d^{\,\widehat g_{K_{n,m}}}\big)}_{=:\,\widetilde S_{n,m}} \;+\; \underbrace{\big(\overline b_d^{\,\widehat g_{K_{n,m}}} - \overline b_d^{\,g^*_{K_{n,m}}}\big)}_{=:\,\widetilde F_{n,m}} \;+\; \underbrace{\big(\overline b_d^{\,g^*_{K_{n,m}}} - \overline b_d^{g_0}\big)}_{=:\,\widetilde B_{n,m}},
	\label{eq:decomp-cons}
\end{equation}
where we still use the same notation $\overline b_d^{\,g}$ when $g$ is random, in which case $\overline b_d^{\,g}$ is random as a function of $g$. We prove in Step 1 below that $\widetilde S_{n,m}$ converges to zero in probability. We prove in Step 2 that \(\widetilde B_{n,m}\) converges (deterministically) to zero. Finally, we prove in Step 3 that \(\widetilde F_{n,m}\) converges to zero in probability.

\medskip
The term \(\widetilde S_{n,m}\) measures the plug-in error at the data-driven partition \(\widehat g_{K_{n,m}}\). The difficulty is that \(\widehat g_{K_{n,m}}\) depends both on the data used to construct the within-cell empirical cdfs and on the first-stage estimator \(\widehat g_0\).
Thus, the within-cell empirical cdfs cannot be analyzed as if the partition
were fixed. To account for both sources of randomness, we consider the class
of all Voronoi cells in \(\mathbb R^q\) that can arise from different
K-means centers and different values of the linear projection defining
\(\widehat g_0\). We prove that its statistical complexity is sufficiently low
(formally, its VC dimension is of order
\(qK_{n,m}\log K_{n,m}\)) to control the empirical cdfs
uniformly over all such cells and show that $\widetilde S_{n,m}\xrightarrow{p}0$.

\medskip
Two sources contribute to the term \(\widetilde F_{n,m}\). The first is that K-means is applied to the estimated values \(\widehat g(W_i)\) rather than to the true
values \(g_0(W_i)\). The second is that the K-means centers are estimated from
the sample rather than chosen from the population distribution. We show that the first error vanishes as \(\widehat g\) converges to \(g_0\), while the second vanishes because the empirical K-means solution approaches the population-optimal one as the sample size increases
\citep[Theorem~4]{Linder2002}. Finally, Condition~1 ensures that these small
changes in the partition translate into small changes in the corresponding
bound, leading to $\widetilde F_{n,m}\xrightarrow{p}0$.

\medskip
The term \(\widetilde B_{n,m}\) is the approximation error from replacing the
continuous index \(g_0(W)\) by a finite number of K-means cells. Condition~1
implies that this error is bounded by a constant times the square root of $E\!\left[
\left\|g_0(W)-v^*_{k(W)}\right\|^2
\right]$, where \(v^*_{k(W)}\) is the population K-means center closest to \(g_0(W)\).
As the number of cells increases, these centers provide an increasingly fine
approximation to \(g_0(W)\). Zador's theorem
\citep[Theorem~6.2]{GrafLuschgy2000} implies that the mean squared distance
above converges to zero, and therefore $\widetilde B_{n,m}\to0$.

\medskip
Hereafter, we let $n_k$ and $m_k$ denote the numbers of observations in the two datasets that fall into cell $k$ of $\widehat g_{K_{n,m}}$. For two distributions $F$ and $G$, we also denote by $L(F, G) := \int_0^1 F^{-1}(u) G^{-1}(u)\,du = \E[Y_F Y_G]$ where $(Y_F, Y_G)$ is the comonotone coupling with margins $F, G$.  Denote by   $N := \min(n,m)$ and  by $\widehat P_{n,m}$ the pooled empirical distribution of the
$W$'s from the two samples.  Denote by $V:=g_0(W)$ and the generic functional $J(v)$ by
\begin{equation}
	J(v):= \int_0^1 F^{-1}_{\nu_Y|g_0(W)}(u|v)\,F^{-1}_{\nu_d|g_0(W)}(u|v)\,du, \qquad v \in \mathcal V.
	\label{eq:J}
\end{equation}
Under Assumption \ref{hyp:consistency_ghat}-(1), $W$ has compact support and we denote by $D_W := \operatorname{diam}(\operatorname{supp}(W)) < \infty$. Combined with Assumption \ref{hyp:consistency_ghat}-(2a), there exist a closed neighborhood $\mathcal N$ of $\varsigma_0$ and a finite constant denoted by $\overline D$ such that on the event $\mathcal E_n := \{\widehat \varsigma \in \mathcal N\}$, $\operatorname{diam}(\mathcal X_{\widehat \varsigma}) \;\le\; \overline D \;<\;\infty$, with $P(\mathcal E_n) \to 1$.

\medskip
\paragraph{Step 1: $\widetilde S_{n,m} \xrightarrow{p} 0$.} We show that the plug-in error at the data-dependent partition vanishes uniformly under the tuning condition $K_{n,m}^{5} \log K_{n,m} = o(N)$. The plug-in estimator is
$$\widehat{\overline b}{}_d^{\,\widehat g_{K_{n,m}}} \,=\, \frac{1}{\widehat \E[\eta_d^2]}\bigg\{\widehat\delta_d'\widehat \E[WW']\widehat\delta_Y \, + \, \sum_{k=1}^{K_{n,m}} \widehat p_k \int_0^1 \widehat F_{n|k}^{-1}(u)\,\widehat G_{m|k}^{-1}(u)\,du\bigg\},$$
where $\widehat F_{n|k}$ is the empirical cdf of $\widehat\nu_Y$ on the subsample $\{i : \widehat g_{K_{n,m}}(W_i^{(1)}) = g_k\}$, $\widehat G_{m|k}$ is defined analogously from $\widehat\nu_d$ on the auxiliary sample, and $\widehat p_k=(n_k+m_k)/(n+m)$. 

The first-stage OLS estimators satisfy $\widehat\delta_d \xrightarrow{p} \delta_d$, $\widehat\delta_Y \xrightarrow{p} \delta_Y$, $\widehat \E[WW'] \xrightarrow{p} \E[WW']$, and $\widehat \E[\eta_d^2] \xrightarrow{p} \E[\eta_d^2]$ by standard OLS consistency. It remains to show
\begin{equation}
    \sum_{k=1}^{K_{n,m}} \widehat p_k L(\widehat F_{n|k},\widehat G_{m|k})
    -\sum_{k=1}^{K_{n,m}} p_k L(F_{|k},G_{|k}) \xrightarrow{p}0,
    \label{eq:cell-plug}
\end{equation}
where $p_k=P(\widehat g_{K_{n,m}}(W)=g_k)$ and
$L(F,G):=\int_0^1F^{-1}(u)G^{-1}(u)\,du$.
By Lemma~\ref{lem:pop-balance}, there exists $c>0$, 
\begin{equation}
\min_k\frac{n_k}{n}\ge cK_{n,m}^{-2},
\qquad
\min_k\frac{m_k}{m}\ge cK_{n,m}^{-2}
\label{eq:emp-balance-used}
\end{equation}
with probability tending to one. Since $\widehat g_{K_{n,m}}$ is data-dependent, cell membership indicators cannot be treated as fixed. We proceed by uniform empirical-process control over the class of all possible $K$-means Voronoi cells.

\medskip

\noindent\emph{Step 1a: VC dimension of the Voronoi cell class in $W$-space.} The empirical partition $\widehat g_{K_{n,m}}$ assigns $w\in\R^q$ to cell $k$ if $\widehat g_0(w) = w'\widehat\varsigma$ falls in the $k$-th Voronoi cell (in $\R^{2}$) induced by the K-means centers $\widehat{\mathbf v} = (\widehat v_1, \ldots, \widehat v_{K_{n,m}})$. We keep the linear-projection parameter $\varsigma \in \R^{q\times 2}$ visible as a parameter. Concretely, let
$$\mathcal C^{2}_{K_{n,m}} \;:=\; \bigl\{\,C_k(\mathbf v) : \mathbf v \in (\R^{2})^{K_{n,m}},\ k \in \{1,\ldots,K_{n,m}\}\,\bigr\}$$
denote the class of $K_{n,m}$-Voronoi cells in $\R^{2}$, where $C_k(\mathbf v) = \bigcap_{j\neq k} H_{kj}(\mathbf v)$ with $H_{kj}(\mathbf v) = \{z\in\R^{2}: \langle z, v_k-v_j\rangle \geq (\|v_k\|^2 - \|v_j\|^2)/2\}$. The relevant class of subsets of $\R^q$ ($W$-space) is
$$\mathcal C_{K_{n,m}} \;:=\; \bigl\{\{w\in\R^q: w'\varsigma \in C\}: \varsigma \in \R^{q\times 2},\ C \in \mathcal C^{2}_{K_{n,m}}\,\bigr\}.$$
Each element of $\mathcal C_{K_{n,m}}$ is an intersection of at most $K_{n,m}-1$ closed half-spaces in $\R^q$: for $C = \bigcap_j H_j$ with $H_j = \{z: z^\top u_j \geq c_j\}$, the preimage is $\{w: w'\varsigma \in \bigcap_j H_j\} = \bigcap_j \{w: w^\top(\varsigma u_j) \geq c_j\}$, and as $(\varsigma, u_j)$ range, $\varsigma u_j$ spans all of $\R^q$. The class $\mathcal H_q$ of closed half-spaces in $\R^q$ has VC dimension $q + 1$ \citep[Corollary~13.1]{DevroyeGyorfiLugosi1996}. The class of intersections of $s$ sets from a class of VC dimension $V$ has VC dimension at most $c_0\,V s \log s$ for a universal constant $c_0$ \citep[Theorem 1.1]{van2009note}. Applied with $V = q + 1$, $s = K_{n,m} - 1$, this yields
\begin{equation}
	V(\mathcal C_{K_{n,m}}) \;\leq\; c_0\,(q+1)\,(K_{n,m} - 1)\,\log(K_{n,m} - 1) \;=\; O\left(q\,K_{n,m} \log K_{n,m}\right).
	\label{eq:VC-Voronoi}
\end{equation}

\medskip

\noindent\emph{Step 1b: uniform empirical-process bound.} With the linear projection $\varsigma$ made explicit, the relevant class of functions on $(w, z) \in \R^q \times \R$, where $z$ denotes the relevant true residual, is
$$\mathcal F_{K_{n,m}} \;:=\; \bigl\{(w, z) \mapsto \ind{w'\varsigma \in C}\,\ind{z \leq t} : \varsigma \in \R^{q\times 2},\ C \in \mathcal C^{2}_{K_{n,m}},\ t \in \R\bigr\}.$$
By \citet[Lemma~2.6.17]{VdV_Wellner}, $V(\mathcal F_{K_{n,m}}) \leq V(\mathcal C_{K_{n,m}}) + 1 = O(q K_{n,m} \log K_{n,m})$, and every function in $\mathcal F_{K_{n,m}}$ is bounded by 1. 

Let $G_n f := \sqrt{n}(\hat P_n f - P f)$ denote the empirical process. By the finite-entropy-integral maximal inequality \citep[Theorem~2.14.1][with $p=1$ and envelope 1]{VdV_Wellner},
\begin{align*}
	\E\!\left[\sup_{f \in \mathcal F_{K_{n,m}}}\,|G_n f|\right] \leq & C\,J\left(1, \mathcal F_{K_{n,m}}\right), \\
	J(1, \mathcal F_{K_{n,m}}, F) := & \sup_Q \int_0^1 \sqrt{1 + \log  N(\epsilon, \mathcal F_{K_{n,m}}, L^2(Q))}\,d\epsilon,
\end{align*}
where $C$ is a universal constant and the supremum is over probability measures $Q$. The VC covering bound \citep[Theorem~2.6.7]{VdV_Wellner} gives $\log N(\epsilon, \mathcal F_{K_{n,m}}, L^2(Q)) \leq c_1V(\mathcal F_{K_{n,m}})\log(1/\epsilon)$ for a constant $c_1$ uniformly in $Q$, so
\begin{align*}
	J(1, \mathcal F_{K_{n,m}}, F) \leq & \int_0^1 \sqrt{1 + c_1\,V(\mathcal F_{K_{n,m}})\log(1/\epsilon)}\,d\epsilon \\
	=  & O\left(\sqrt{V(\mathcal F_{K_{n,m}})}\right) \\
	= & O\left(\sqrt{q K_{n,m} \log K_{n,m}}\right).
\end{align*}
Combining and dividing by $\sqrt n$:
$$
\E\!\left[\sup_{f \in \mathcal F_{K_{n,m}}}\,\bigl|\hat P_n f - P f\bigr|\right] \;=\; O\!\left(\sqrt{\frac{q\,K_{n,m} \log K_{n,m}}{n}}\right),
$$
and by Markov's inequality,
\begin{equation}
	\sup_{f \in \mathcal F_{K_{n,m}}}\,\bigl|\,\hat P_n f - P f\,\bigr| \;=\; O_P\!\left(\sqrt{\frac{q\,K_{n,m} \log K_{n,m}}{n}}\right).
	\label{eq:VC-GC}
\end{equation}
Applied with $f(w, z) = \ind{w'\varsigma \in C}\ind{z \leq t}$, this bounds the joint empirical measure uniformly in $(\varsigma, C, t)$.  Applied with $t=+\infty$, the same argument bounds cell probabilities uniformly for each sample. Since the deviations are $o_P(K_{n,m}^{-2})$ under Assumption~\ref{hyp:consistency_ghat}(3), combining them with \eqref{eq:emp-balance-used} gives, with probability tending to one,
\begin{equation}
\min_k p_k\ge \frac c2K_{n,m}^{-2},
\qquad
\sup_k\left|\frac{n_k/n}{p_k}-1\right|+\sup_k\left|\frac{m_k/m}{p_k}-1\right|=o_P(1).
\label{eq:population-cell-balance}
\end{equation}
In particular, since $\widehat p_k$ is a convex combination of $n_k/n$ and $m_k/m$,
\begin{equation}
\sup_k\frac{\widehat p_k}{n_k/n}+\sup_k\frac{\widehat p_k}{m_k/m}=O_P(1).
\label{eq:weight-ratio-cells}
\end{equation}

\medskip

\noindent\emph{Step 1c: from joint to conditional cdf.}
Let $\widetilde F_{n|C}$ denote the empirical cdf of the true
residual $\nu_Y$ among observations with $W\in C$, and define
$\widetilde G_{m|C}$ analogously for $\nu_d$. For a cell $C$, write
\[
p_C:=P(W\in C),
\qquad
\widehat p_C:=\widehat P_n(W\in C),
\qquad
F_{|C}(t):=P(\nu_Y\le t\mid W\in C).
\]
Then
\begin{align*}
	& \widetilde F_{n|C}(t)-F_{|C}(t) \\
	&=
	\frac{
		\widehat P_n\!\left(
		\ind{W\in C}\ind{\nu_Y\le t}
		\right)
	}{
		\widehat p_C
	}
	-
	\frac{
		P\!\left(
		\ind{W\in C}\ind{\nu_Y\le t}
		\right)
	}{
		p_C
	}\\
	&=
	\frac{
		\left[
		\widehat P_n\!\left(
		\ind{W\in C}\ind{\nu_Y\le t}
		\right)
		-
		P\!\left(
		\ind{W\in C}\ind{\nu_Y\le t}
		\right)
		\right]
		-
		F_{|C}(t)(\widehat p_C-p_C)
	}{
		\widehat p_C
	}.
\end{align*}
Consequently,
\begin{equation}
	\left|
	\widetilde F_{n|C}(t)-F_{|C}(t)
	\right|
	\le
	\frac{
		\left|
		(\widehat P_n-P)
		\left[
		\ind{W\in C}\ind{\nu_Y\le t}
		\right]
		\right|
		+
		|\widehat p_C-p_C|
	}{
		\widehat p_C
	}.
	\label{eq:conditional-cdf-ratio}
\end{equation}

Let $C_1,\ldots,C_{K_{n,m}}$ denote the actual cells of
$\widehat g_{K_{n,m}}$. Since
$\widehat p_{C_k}=n_k/n$ for these cells and that \eqref{eq:population-cell-balance} gives uniformly in $k$,
$n_k/n=p_k\{1+o_P(1)\}$ and
$\min_k p_k\ge (c/2)K_{n,m}^{-2}$ with probability tending to one, we have
\begin{equation}
\max_{1\le k\le K_{n,m}}\frac{1}{\widehat p_{C_k}}
=O_P\!\left(K_{n,m}^{2}\right).
\label{eq:actual-cell-inverse-prob}
\end{equation}
The uniform empirical-process bound \eqref{eq:VC-GC}, which is valid over
the whole enlarged class $\mathcal C_{K_{n,m}}$, implies
\begin{align*}
\sup_{C\in\mathcal C_{K_{n,m}}}\sup_{t\in\R}
\left|
(\widehat P_n-P)
\left[
\ind{W\in C}\ind{\nu_Y\le t}
\right]
\right|
+ &
\sup_{C\in\mathcal C_{K_{n,m}}}
|\widehat p_C-p_C| \\
& =
O_P\!\left(
\sqrt{\frac{qK_{n,m}\log K_{n,m}}{n}}
\right).
\end{align*}
Evaluating these uniform numerator bounds at the random cells
$C_1,\ldots,C_{K_{n,m}}$, and combining
\eqref{eq:conditional-cdf-ratio} with
\eqref{eq:actual-cell-inverse-prob}, therefore gives
\begin{equation}
	\max_{1\le k\le K_{n,m}}\sup_{t\in\R}
	\left|
	\widetilde F_{n|C_k}(t)-F_{|C_k}(t)
	\right|
	=
	O_P\!\left(
	\sqrt{
		a_N
	}
	\right),
	\label{eq:cond-cdf-rate}
\end{equation}
where $N =\min(n,m) $, and $a_N =(
	qK_{n,m}^{5}\log K_{n,m})/n$.
The corresponding bound holds for
$\widetilde G_{m|C_k}$. Both rates are $o_P(1)$ by
Assumption~\ref{hyp:consistency_ghat}(3).

\medskip

\noindent\emph{Step 1d: from cdf convergence to L-statistic convergence.}
We now prove the weighted convergence of the within-cell comonotone functionals. The proof has three substeps. First, we remove the estimation error in the residuals while keeping exactly the same random cells. Second, we truncate the true residuals and compare the truncated empirical and population functionals using the cdf bound from Step~1c. Third, we remove the truncation after averaging over cells. The key point is that only a weighted moment bound is needed; we never need to prove that the empirical $(2+\eps)$-moment is bounded uniformly in every shrinking cell.

For notational convenience, write
\[
\pi_{1k}:=\frac{n_k}{n},
\qquad
\pi_{2k}:=\frac{m_k}{m},
\qquad
\widehat p_k
=
\frac{n}{n+m}\pi_{1k}
+
\frac{m}{n+m}\pi_{2k}.
\]
By \eqref{eq:weight-ratio-cells}, there exists a random sequence $B_N=O_P(1)$ such that, with probability tending to one,
\begin{equation}
	\widehat p_k\le B_N\pi_{1k},
	\qquad
	\widehat p_k\le B_N\pi_{2k},
	\qquad
	k=1,\ldots,K_{n,m}.
	\label{eq:weight-ratio-BN}
\end{equation}

\medskip
\emph{(a) Removing the estimation error in the residuals.}
Fix a realized cell $k$. The empirical measures
$\widehat F_{n|k}$ and $\widetilde F_{n|k}$ assign mass $1/n_k$ respectively to
$\{\widehat\nu_{Yi}:i\in C_k\}$ and $\{\nu_{Yi}:i\in C_k\}$.
Pairing, within the cell, each estimated residual with its own true residual defines a valid coupling. Therefore
\begin{equation}
	W_2^2(\widehat F_{n|k},\widetilde F_{n|k})
	\le
	\frac1{n_k}\sum_{i:W_i^{(1)}\in C_k}
	(\widehat\nu_{Yi}-\nu_{Yi})^2.
	\label{eq:cell-resid-coupling-Y}
\end{equation}
Multiplying by $\widehat p_k$, summing over $k$, and using
\eqref{eq:weight-ratio-BN},
\begin{align}
	\sum_{k=1}^{K_{n,m}}\widehat p_k
	W_2^2(\widehat F_{n|k},\widetilde F_{n|k})
	&\le
	B_N
		\sum_{k=1}^{K_{n,m}}
	\frac{n_k}{n}
	\frac1{n_k}
	\sum_{i:W_i^{(1)}\in C_k}
	(\widehat\nu_{Yi}-\nu_{Yi})^2
	\nonumber\\
	&=
	B_N\frac1n
	\sum_{i=1}^n
	(\widehat\nu_{Yi}-\nu_{Yi})^2
	=o_P(1).
	\label{eq:resid-W2-Y}
\end{align}
The last equality follows from the usual $L^2$ consistency of the first-stage OLS residuals. The identical argument on the auxiliary sample gives
\begin{equation}
	\sum_{k=1}^{K_{n,m}} \widehat p_k
	W_2^2(\widehat G_{m|k},\widetilde G_{m|k})
	=o_P(1).
	\label{eq:resid-W2-d}
\end{equation}

We also need weighted second moments. For example,
\begin{align}
	\sum_{k=1}^{K_{n,m}} \widehat p_k
	\int_{\R} x^2\,d\widehat F_{n|k}(x)
	&\le
	B_N
	\sum_{k=1}^{K_{n,m}}
	\frac{n_k}{n}
	\frac1{n_k}
	\sum_{i:W_i^{(1)}\in C_k}
	\widehat\nu_{Yi}^{\,2}
	\nonumber\\
	&=
	B_N\frac1n\sum_{i=1}^n\widehat\nu_{Yi}^{\,2}
	=
	O_P(1),
	\label{eq:weighted-second-hatY}
\end{align}
and similarly
\begin{equation}
		\sum_{k=1}^{K_{n,m}}\widehat p_k
	\int_{\R} x^2\,d\widetilde F_{n|k}(x)
	+
	\sum_{k=1}^{K_{n,m}}\widehat p_k
	\int_{\R} x^2\,d\widehat G_{m|k}(x)
	+
		\sum_{k=1}^{K_{n,m}}\widehat p_k
	\int_{\R} x^2\,d\widetilde G_{m|k}(x)
	=
	O_P(1).
	\label{eq:weighted-second-all}
\end{equation}

For any four laws $F,F',G,G'$ with finite second moments,
\begin{align}
	|L(F,G)-L(F',G')|
	&\le
	\left|
	\int_0^1
	\left(F^{-1}-F'^{-1}\right)G^{-1}\,du
	\right|
	+
	\left|
	\int_0^1
	F'^{-1}\left(G^{-1}-G'^{-1}\right)\,du
	\right|
	\nonumber\\
	&\le
	W_2(F,F')\,
	\|G^{-1}\|_{L^2(0,1)}
	+
	\|F'^{-1}\|_{L^2(0,1)}\,
	W_2(G,G'),
	\label{eq:L-W2-basic}
\end{align}
where the last inequality is Cauchy--Schwarz and, on the real line,
$\|F^{-1}-F'^{-1}\|_{L^2(0,1)}=W_2(F,F')$.
Applying \eqref{eq:L-W2-basic} cell by cell gives
\begin{align*}
\sum_{k=1}^{K_{n,m}}\widehat p_k &
	\left|
	L(\widehat F_{n|k},\widehat G_{m|k})
	-
	L(\widetilde F_{n|k},\widetilde G_{m|k})
	\right|\\
	\le &
		\sum_{k=1}^{K_{n,m}}\widehat p_k
	W_2(\widehat F_{n|k},\widetilde F_{n|k})
	\left(
	\int x^2\,d\widehat G_{m|k}(x)
	\right)^{1/2} \\
	 & +
	\sum_{k=1}^{K_{n,m}}\widehat p_k
	\left(
	\int x^2\,d\widetilde F_{n|k}(x)
	\right)^{1/2}
	W_2(\widehat G_{m|k},\widetilde G_{m|k}).
\end{align*}
Cauchy--Schwarz across cells, together with
\eqref{eq:resid-W2-Y}--\eqref{eq:weighted-second-all}, therefore yields
\begin{equation}
	\sum_k\widehat p_k
	\left|
	L(\widehat F_{n|k},\widehat G_{m|k})
	-
	L(\widetilde F_{n|k},\widetilde G_{m|k})
	\right|
	=o_P(1).
	\label{eq:resid-L-diff}
\end{equation}

\medskip
\emph{(b) Truncation of the true residuals.}
Put $\rho:=2+\eps$ and, for $M>0$, define
\[
\phi_M(x):=(-M)\vee x\wedge M.
\]
For any probability law $H$, let $H^M$ denote the law of $\phi_M(X)$ when
$X\sim H$. We first derive explicitly the truncation inequality that will
be used for both the population and empirical laws.

Let $X=F^{-1}(U)$, $Y=G^{-1}(U)$, and $U\sim\mathcal U[0,1]$, so that
$L(F,G)=E[XY]$. Since $\phi_M$ is nondecreasing,
$(\phi_M(X),\phi_M(Y))$ is the comonotone coupling of $F^M$ and $G^M$.
Using
\[
XY-\phi_M(X)\phi_M(Y)
=
X\{Y-\phi_M(Y)\}
+
\phi_M(Y)\{X-\phi_M(X)\},
\]
together with $|Y-\phi_M(Y)| \le |Y|\ind{|Y|>M}$, $|\phi_M(Y)|\le M$, 
we obtain
\begin{align}
	&|L(F,G)-L(F^M,G^M)| \notag  \\
	& \le
	E\!\left[
	|X||Y|\ind{|Y|>M}
	\right]
	+
	M E\!\left[
	|X|\ind{|X|>M}
	\right].
	\label{eq:trunc-decomp}
\end{align}
Denoting by $A_F:=\int|x|^\rho\,dF(x)$ and $A_G:=\int|x|^\rho\,dG(x)$,
then Cauchy--Schwarz and Markov's inequality give
\begin{align*}
	E\!\left[
	|X||Y|\ind{|Y|>M}
	\right]
	&\le
	E[X^2]^{1/2}
	E[Y^2\ind{|Y|>M}]^{1/2}
	\nonumber\\
	&\le
	A_F^{1/\rho}
	\left(
	M^{-\eps}A_G
	\right)^{1/2}
	=
	M^{-\eps/2}
	A_F^{1/\rho}A_G^{1/2}, \\
	M E\!\left[
	|X|\ind{|X|>M}
	\right]
	&\le
	M^{-\eps}A_F.
\end{align*}
Hence
\begin{equation}
	|L(F,G)-L(F^M,G^M)|
	\le
	M^{-\eps/2}
	A_F^{1/\rho}A_G^{1/2}
	+
	M^{-\eps}A_F.
	\label{eq:trunc-error-general}
\end{equation}

For the population conditional laws, Assumption~\ref{hyp:consistency_ghat}(2b) gives a deterministic constant $D<\infty$ such that for every $W$-measurable cell $C$ with positive probability,
\begin{align*}
	E[|\nu_Y|^\rho\mid C]
	&=
	E\!\left[
	E(|\nu_Y|^\rho\mid W)
	\mid C
	\right]
	\le D,
	\nonumber\\
	E[|\nu_d|^\rho\mid C]
	&\le D.
\end{align*}
Therefore \eqref{eq:trunc-error-general} implies, uniformly in $k$,
\begin{equation*}
	|L(F_{|k},G_{|k})-L(F_{|k}^M,G_{|k}^M)|
	\le
	C M^{-\eps/2},
\end{equation*}
for a constant $C$ independent of $k,M,n,m$. Since
$\sum_{k=1}^{K_{n,m}}\widehat p_k=1$,
\begin{equation*}
\sum_{k=1}^{K_{n,m}}\widehat p_k
	|L(F_{|k},G_{|k})-L(F_{|k}^M,G_{|k}^M)|
	\le
	C M^{-\eps/2}.
\end{equation*}

For the empirical laws of the true residuals, define
\[
A_{Y,k}
:=
\int_{\R}|x|^\rho\,d\widetilde F_{n|k}(x)
=
\frac1{n_k}
\sum_{i:W_i^{(1)}\in C_k}
|\nu_{Yi}|^\rho,
\]
and
\[
A_{d,k}
:=
\int_{\R}|x|^\rho\,d\widetilde G_{m|k}(x)
=
\frac1{m_k}
\sum_{j:W_j^{(2)}\in C_k}
|\nu_{dj}|^\rho.
\]
Using \eqref{eq:weight-ratio-BN},
\begin{align}
\sum_{k=1}^{K_{n,m}}\widehat p_k A_{Y,k}
	&\le
	B_N
\sum_{k=1}^{K_{n,m}}\frac{n_k}{n}A_{Y,k}
	=
	B_N\frac1n\sum_{i=1}^n|\nu_{Yi}|^\rho
	=
	O_P(1),
	\label{eq:weighted-rho-Y}\\
\sum_{k=1}^{K_{n,m}}\widehat p_k A_{d,k}
	&=
	O_P(1).
	\label{eq:weighted-rho-d}
\end{align}
The last equalities follow from the law of large numbers and
Assumption~\ref{hyp:consistency_ghat}(2b).

Applying \eqref{eq:trunc-error-general} to each empirical cell and summing,
\begin{align}
	&\sum_{k=1}^{K_{n,m}}\widehat p_k
	\left|
	L(\widetilde F_{n|k},\widetilde G_{m|k})
	-
	L(\widetilde F_{n|k}^M,\widetilde G_{m|k}^M)
	\right| \notag \\
 &\le
	M^{-\eps/2}
\sum_{k=1}^{K_{n,m}}
	\widehat p_k
	A_{Y,k}^{1/\rho}A_{d,k}^{1/2}
	+
	M^{-\eps}
	\sum_{k=1}^{K_{n,m}}\widehat p_kA_{Y,k}.
	\label{eq:weighted-trunc-before-holder}
\end{align}
Because $1/\rho+1/2<1$,
generalized H\"older's inequality with respect to the probability weights
$\{\widehat p_k\}_k$ gives
\begin{equation}
	\sum_{k=1}^{K_{n,m}}
	\widehat p_k
	A_{Y,k}^{1/\rho}A_{d,k}^{1/2}
	\le
	\left(
\sum_{k=1}^{K_{n,m}}\widehat p_kA_{Y,k}
	\right)^{1/\rho}
	\left(
\sum_{k=1}^{K_{n,m}}\widehat p_kA_{d,k}
	\right)^{1/2}.
	\label{eq:holder-across-cells}
\end{equation}
Combining
\eqref{eq:weighted-rho-Y}--\eqref{eq:holder-across-cells},
\begin{equation*}
\sum_{k=1}^{K_{n,m}}\widehat p_k
	\left|
	L(\widetilde F_{n|k},\widetilde G_{m|k})
	-
	L(\widetilde F_{n|k}^M,\widetilde G_{m|k}^M)
	\right|
	=
	O_P(M^{-\eps/2}).
\end{equation*}

\medskip
\emph{(c) Comparing the truncated empirical and population functionals.}
 By the following identity on the real line \citep[see (2.48), for a modern treatment]{villani2021topics}: for any two cdfs $F_1, F_2$ on $\R$ with finite first moments, 
\begin{equation}
	\int_0^1
	|F_1^{-1}(u)-F_2^{-1}(u)|\,du
	=
	\int_{\R}
	|F_1(y)-F_2(y)|\,dy.
	\label{eq:KR-generic}
\end{equation}
Also, truncation cannot increase the Kolmogorov distance:
\begin{equation}
	\|F_1^M-F_2^M\|_\infty
	\le
	\|F_1-F_2\|_\infty.
	\label{eq:trunc-cdf-contraction}
\end{equation}
Since all quantiles of a truncated law lie in $[-M,M]$, we have
\begin{align}
	&\left|
	L(\widetilde F_{n|k}^M,\widetilde G_{m|k}^M)
	-
	L(F_{|k}^M,G_{|k}^M)
	\right| \notag \\
	&\quad\le
	M
	\int_0^1
	\left|
	(\widetilde F_{n|k}^M)^{-1}
	-
	(F_{|k}^M)^{-1}
	\right|du
	+
	M
	\int_0^1
	\left|
	(\widetilde G_{m|k}^M)^{-1}
	-
	(G_{|k}^M)^{-1}
	\right|du
	\nonumber\\
	&\quad=
	M\int_{\R}
	|\widetilde F_{n|k}^M(y)-F_{|k}^M(y)|\,dy
	+
	M\int_{\R}
	|\widetilde G_{m|k}^M(y)-G_{|k}^M(y)|\,dy
	\nonumber\\
	&\quad\le
	2M^2
	\left(
	\|\widetilde F_{n|k}-F_{|k}\|_\infty
	+
	\|\widetilde G_{m|k}-G_{|k}\|_\infty
	\right),
	\label{eq:L-Lipschitz}
\end{align}
where the equality uses \eqref{eq:KR-generic}, and the last line uses
\eqref{eq:trunc-cdf-contraction} and the fact that the integrands vanish
outside $[-M,M]$.

Because the weights add to one, Step~1c and \eqref{eq:L-Lipschitz} yield
\begin{align}
	&
	\sum_{k=1}^{K_{n,m}}\widehat p_k
	\left|
	L(\widetilde F_{n|k}^M,\widetilde G_{m|k}^M)
	-
	L(F_{|k}^M,G_{|k}^M)
	\right|
	\nonumber\\
	&\qquad
	\le
2M^2
\max_k\left(
\|\widetilde F_{n\mid k}-F_{\mid k}\|_\infty+
\|\widetilde G_{m\mid k}-G_{\mid k}\|_\infty
\right)
=
O_p(M^2\sqrt{a_N}). \notag 
\end{align}

Since $a_N\to0$, choose, for example, $ M_N:=a_N^{-1/8}$. Then $M_N\to\infty$, $M_N^2\sqrt{a_N}=a_N^{1/4}\to0$, $M_N^{-\eps/2}=a_N^{\eps/16}\to0$.
Using the triangle inequality,
\begin{align}
	&
\sum_{k=1}^{K_{n,m}}\widehat p_k
	\left|
	L(\widetilde F_{n|k},\widetilde G_{m|k})
	-
	L(F_{|k},G_{|k})
	\right|
	\nonumber\\
	&\quad\le
	\underbrace{
	\sum_{k=1}^{K_{n,m}} \widehat p_k
		\left|
		L(\widetilde F_{n|k},\widetilde G_{m|k})
		-
		L(\widetilde F_{n|k}^{M_N},\widetilde G_{m|k}^{M_N})
		\right|
	}_{O_P(M_N^{-\eps/2})}
	\nonumber\\
	&\qquad+
	\underbrace{
	\sum_{k=1}^{K_{n,m}}
	\widehat p_k
		\left|
		L(\widetilde F_{n|k}^{M_N},\widetilde G_{m|k}^{M_N})
		-
		L(F_{|k}^{M_N},G_{|k}^{M_N})
		\right|
	}_{O_P(M_N^2\sqrt{a_N})}
	\nonumber\\
	&\qquad+
	\underbrace{
	\sum_{k=1}^{K_{n,m}}\widehat p_k
		\left|
		L(F_{|k}^{M_N},G_{|k}^{M_N})
		-
		L(F_{|k},G_{|k})
		\right|
	}_{O(M_N^{-\eps/2})}
	=o_P(1).
	\label{eq:L-weighted-cons}
\end{align}
Finally, combining \eqref{eq:resid-L-diff} and
\eqref{eq:L-weighted-cons} once more by the triangle inequality gives
\begin{equation}
	\sum_k\widehat p_k
	\left|
	L(\widehat F_{n|k},\widehat G_{m|k})
	-
	L(F_{|k},G_{|k})
	\right|
	=o_P(1).
	\label{eq:L-weighted-cons-feasible}
\end{equation}

\medskip
\noindent\emph{Combining.}
Set $L_k:=L(F_{|k},G_{|k})$, $\widehat L_k:=L(\widehat F_{n|k},\widehat G_{m|k})$.
The term appearing in \eqref{eq:cell-plug} can be decomposed exactly as
\begin{align*}
	\sum_{k=1}^{K_{n,m}}\widehat p_k\widehat L_k-\sum_{k=1}^{K_{n,m}}p_kL_k
	&=
	\underbrace{
		\sum_{k=1}^{K_{n,m}}\widehat p_k(\widehat L_k-L_k)
	}_{(I)}
	+
	\underbrace{
	\sum_{k=1}^{K_{n,m}}(\widehat p_k-p_k)L_k
	}_{(II)}.
\end{align*}
For the first term,
\begin{equation}
	|(I)|
	\le
	\sum_{k=1}^{K_{n,m}}\widehat p_k
	|\widehat L_k-L_k|
	=o_P(1)
	\label{eq:combining-I}
\end{equation}
by \eqref{eq:L-weighted-cons-feasible}. For the second term, Assumption~\ref{hyp:consistency_ghat}(2b) implies a uniform bound on the conditional second moments. Indeed, with
\[
D_Y:=\sup_{w\in\Supp(W)}
E[|\nu_Y|^{2+\eps}\mid W=w],
\qquad
D_d:=\sup_{w\in\Supp(W)}
E[|\nu_d|^{2+\eps}\mid W=w],
\]
Lyapunov's inequality and iterated expectations imply, for every realized
$W$-measurable cell $C_k$, that  $E[\nu_Y^2\mid C_k]\le D_Y^{2/(2+\eps)}$ and $E[\nu_d^2\mid C_k]\le D_d^{2/(2+\eps)}$.
Thus, by Cauchy--Schwarz,
\begin{equation*}
	|L_k|
	\le
	\left(
	E[\nu_Y^2\mid C_k]
	E[\nu_d^2\mid C_k]
	\right)^{1/2}
	\le C_L
\end{equation*}
for a deterministic $C_L<\infty$, uniformly in $k$ and in the realized
partition. Moreover,
\[
\widehat p_k
=
\frac{n}{n+m}\frac{n_k}{n}
+
\frac{m}{n+m}\frac{m_k}{m},
\]
so the two uniform cell-probability bounds from Step~1b imply
\begin{equation*}
	\sup_k|\widehat p_k-p_k|
	=
	O_P\!\left(
	\sqrt{\frac{qK_{n,m}\log K_{n,m}}{N}}
	\right).
\end{equation*}
Consequently,
\begin{align}
	|(II)|
	&\le
	C_L
	\sum_{k=1}^{K_{n,m}}
	|\widehat p_k-p_k|
	\nonumber\\
	&\le
	C_L K_{n,m}
	\sup_k|\widehat p_k-p_k|
	\nonumber\\
	&=
	O_P\!\left(
	\sqrt{
		\frac{qK_{n,m}^{3}\log K_{n,m}}{N}
	}
	\right)
	=o_P(1),
	\label{eq:combining-II}
\end{align}
where the last equality follows from Assumption~\ref{hyp:consistency_ghat}(3).

Equations \eqref{eq:combining-I} and \eqref{eq:combining-II} prove
\eqref{eq:cell-plug}. Since the remaining terms in
$\widehat{\overline b}{}_d^{\,\widehat g_{K_{n,m}}}$ are finite-dimensional
OLS plug-in quantities and
$\widehat E[\eta_d^2]\convP E[\eta_d^2]>0$, Slutsky's theorem yields
\[
\widetilde S_{n,m}
=
\widehat{\overline b}{}_d^{\,\widehat g_{K_{n,m}}}
-
\overline b_d^{\,\widehat g_{K_{n,m}}}
\convP0.
\]

\paragraph{Step 2: $\widetilde B_{n,m} \to 0$.}
Recall that $\widetilde B_{n,m}=\overline b_d^{\,g^*_{K_{n,m}}}-\overline b_d^{g_0}$, where $g^*_{K_{n,m}}$ denotes the population K-means partition of $V=g_0(W)$, built from the population $L^2$-optimal centers $\mathbf v^*=(v_1^*,\ldots,v_{K_{n,m}}^*)$.

We first record a consequence of Assumption~\ref{hyp:consistency_ghat}(1) that will also be used in Step~3. The following argument applies to every finitely valued partition of the form
\begin{equation}
g(W)=h(V),\qquad V=g_0(W),
\label{eq:V-measurable-partition}
\end{equation}
where $h$ may itself be random; in the latter case the argument is applied conditional on the realized map $h$. Let $ C_k^g:=\{g(W)=g_k\}$, $p_k:=P(C_k^g)$,
and denote by $F_{|k,g}$ and $G_{|k,g}$ the distributions of $\nu_Y$ and $\nu_d$ conditional on $C_k^g$. Thus $\widetilde J_k^g=L(F_{|k,g},G_{|k,g})$.
Let
\[
M_Y:=\operatorname*{ess\,sup}_{v\in\Supp(V)}\E[|\nu_Y|^2\mid V=v]^{1/2},\qquad
M_d:=\operatorname*{ess\,sup}_{v\in\Supp(V)}\E[|\nu_d|^2\mid V=v]^{1/2},
\]
where $\operatorname*{ess\,sup}$ denotes the essential supremum, 
and set $A_0:=M_dL_Y+M_YL_d<\infty$. Finiteness follows from Assumption~\ref{hyp:consistency_ghat}(2b), since
\[
\E[|\nu_Y|^{2+\eps}\mid V]
=\E\!\left[\E(|\nu_Y|^{2+\eps}\mid W)\mid V\right]
\]
is essentially bounded, and Holder's inequality then gives the same for the conditional second moments. We have a similar result for $\nu_d$.

Fix a cell $C_k^g$. Because $C_k^g\in\sigma(V)$,
\begin{equation}
F_{|k,g}=\int F_{\nu_Y\mid V=v'}\,dP(v'\mid C_k^g),\qquad
G_{|k,g}=\int F_{\nu_d\mid V=v'}\,dP(v'\mid C_k^g).
\label{eq:valid-cell-mixture}
\end{equation}
Let $V'$ have distribution $\mathcal L(V\mid C_k^g)$. For any fixed $v$ in the conditional support of V given $C_k^g$, the convexity of $P \mapsto W^2_2(P,Q)$ together with \eqref{eq:valid-cell-mixture} gives
\begin{align*}
W_2^2(F_{|k,g},F_{\nu_Y\mid V=v})
&\le \E\!\left[W_2^2(F_{\nu_Y\mid V=V'},F_{\nu_Y\mid V=v})\mid C_k^g\right]
\le L_Y^2\E[\|V'-v\|^2\mid C_k^g],\\
W_2^2(G_{|k,g},F_{\nu_d\mid V=v})
&\le L_d^2\E[\|V'-v\|^2\mid C_k^g].
\end{align*}
Therefore, by Cauchy--Schwarz,
\begin{align*}
	\big|\widetilde J_k^g-J(v)\big|
	&\le W_2(F_{|k,g},F_{\nu_Y\mid V=v})\,\|G^{-1}_{|k,g}\|_{L^2(0,1)}
	+\|F^{-1}_{|k,g}\|_{L^2(0,1)}\,W_2(G_{|k,g},F_{\nu_d\mid V=v})\\
	&\le A_0\left(\E[\|V'-v\|^2\mid C_k^g]\right)^{1/2}.
\end{align*}
Integrating the preceding inequality with respect to the conditional
distribution of $V$ given $C_k^g$, and letting $V,V'$ be conditionally
i.i.d. given $C_k^g$, we obtain
\begin{align*}
	|\widetilde J_k^g-\E[J(V)\mid C_k^g]|
	&=
	\left|\E[\widetilde J_k^g-J(V)\mid C_k^g]\right|\\
	&\le
	\E[|\widetilde J_k^g-J(V)|\mid C_k^g]\\
	&\le
	A_0\E\left[
	\left(
	\E[\|V'-V\|^2\mid V,C_k^g]
	\right)^{1/2}
	\Bigm|C_k^g
	\right]\\
	&\le
	A_0\left(
	\E[\|V'-V\|^2\mid C_k^g]
	\right)^{1/2},
\end{align*}
where the last inequality follows from Jensen's inequality and the law
of iterated expectations. Finally, since $V$ and $V'$ are conditionally
i.i.d. given $C_k^g$,
\[
\E[\|V'-V\|^2\mid C_k^g]
=
2\E\!\left[
\|V-\E[V\mid C_k^g]\|^2
\mid C_k^g
\right].
\]
Therefore,
\[
|\widetilde J_k^g-\E[J(V)\mid C_k^g]|
\le
\sqrt{2}A_0
\left(
\E[
\|V-\E[V\mid C_k^g]\|^2
\mid C_k^g
]
\right)^{1/2}.
\]
Consequently, for arbitrary cell representatives $c_k\in\R^{2}$,
\begin{equation}
|\widetilde J_k^g-\E[J(V)\mid C_k^g]|
\le \sqrt2A_0\left(\E[\|V-c_k\|^2\mid C_k^g]\right)^{1/2}.
\label{eq:cell-mixture-W2}
\end{equation}
Using \eqref{eq:upper:sigma_w1} and the law of total expectation,
\begin{align*}
	\overline b_d^{\,g}-\overline b_d^{g_0}
	&=
	\frac{1}{\E[\eta_d^2]}
	\left\{
	\sum_{k=1}^{K_{n,m}}p_k\widetilde J_k^g
	-
	\E[J(V)]
	\right\}=
	\frac{1}{\E[\eta_d^2]}
	\sum_{k=1}^{K_{n,m}}p_k
	\left\{
	\widetilde J_k^g
	-
	\E[J(V)\mid C_k^g]
	\right\},
\end{align*}
since $\E[J(V)]
=
\sum_{k=1}^{K_{n,m}}
p_k\E[J(V)\mid C_k^g]$.
Therefore, by the preceding cellwise bound,
\begin{align*}
	|\overline b_d^{\,g}-\overline b_d^{g_0}|
	&\le
	\frac{\sqrt{2}A_0}{\E[\eta_d^2]}
	\sum_{k=1}^{K_{n,m}}
	p_k
	\left(
	\E[\|V-c_k\|^2\mid C_k^g]
	\right)^{1/2}.
\end{align*}
By Cauchy--Schwarz and $	\sum_{k=1}^{K_{n,m}}p_k=1$,
\begin{align*}
	\sum_{k=1}^{K_{n,m}}
	p_k
	\left(
	\E[\|V-c_k\|^2\mid C_k^g]
	\right)^{1/2}
	&\le
	\left(
	\sum_{k=1}^{K_{n,m}}
	p_k
	\E[\|V-c_k\|^2\mid C_k^g]
	\right)^{1/2}\\
	&=
	\left(
	\E[\|V-c_{k(W)}\|^2]
	\right)^{1/2}.
\end{align*}
Hence, for every partition satisfying
\eqref{eq:V-measurable-partition},
\begin{equation}
	|\overline b_d^{\,g}-\overline b_d^{g_0}|
	\le
	C_0
	\left(
	\E[\|V-c_{k(W)}\|^2]
	\right)^{1/2},
	\qquad
	C_0:=
	\frac{\sqrt{2}A_0}{\E[\eta_d^2]}.
	\label{eq:generic-W2-partition-bound}
\end{equation}

The population K-means partition $g^*_{K_{n,m}}$ is a function of $V$, so \eqref{eq:generic-W2-partition-bound} applies with $c_k=v_k^*$. By Zador's theorem \citep[Theorem~6.2]{GrafLuschgy2000},
\begin{equation}\label{eq:zador1}
\E[\|V-v^*_{k^*(W)}\|^2]=O(K_{n,m}^{-1}),
\end{equation}
and therefore  $
|\widetilde B_{n,m}|=O(K_{n,m}^{-1/2})\to0$ .Thus $\widetilde B_{n,m}\to0$ as $n,m \to \infty$.

\medskip

\paragraph{Step 3: $\widetilde F_{n,m} \xrightarrow{p} 0$.}
Recall that $\widetilde F_{n,m}=\overline b_d^{\,\widehat g_{K_{n,m}}}-\overline b_d^{\,g^*_{K_{n,m}}}$. 
As before, let $\widetilde g_{K_{n,m}}$ denote the empirical K-means partition built from the true index values $g_0(W_i)$ (the ``oracle'' empirical partition), and write
\[
\widetilde F_{n,m}
=
\underbrace{\big(\overline b_d^{\,\widehat g_{K_{n,m}}}-\overline b_d^{\,\widetilde g_{K_{n,m}}}\big)}_{=:~\widetilde F_{n,m}^{(1)}}
+
\underbrace{\big(\overline b_d^{\,\widetilde g_{K_{n,m}}}-\overline b_d^{\,g^*_{K_{n,m}}}\big)}_{=:~\widetilde F_{n,m}^{(2)}}.
\]

\medskip

\noindent\emph{The term $\widetilde F_{n,m}^{(2)}$.}
Conditional on its realized centers, $\widetilde g_{K_{n,m}}$ is a function of $V=g_0(W)$. Hence \eqref{eq:generic-W2-partition-bound} applies to $g=\widetilde g_{K_{n,m}}$ with representatives $\widetilde v_k$, and to $g=g^*_{K_{n,m}}$ with representatives $v_k^*$. By the triangle inequality,
\begin{equation}
|\widetilde F_{n,m}^{(2)}|
\le C_0(A_{n,m}+B_{n,m}),
\label{eq:Astar-bound}
\end{equation}
where $A_{n,m}:=\left(\E[\|V-\widetilde v_{\widetilde k(W)}\|^2]\right)^{1/2}$, $B_{n,m}:=\left(\E[\|V-v^*_{k^*(W)}\|^2]\right)^{1/2}$. By Step~2, $B_{n,m}=O(K_{n,m}^{-1/2})$. Let $
D_V(\mathbf v)
:=
\E\!\left[
\min_{1\le k\le K}
\|V-v_k\|^2
\right],$
and let
\[
\mathbf v^*
\in
\arg\min_{\mathbf v}D_V(\mathbf v).
\]
By definition of the assignments $\widetilde k(W)$ and $k^*(W)$, and associated cluster centers $\widetilde{\mathbf v}$ and $\mathbf v^*$, 
$A_{n,m}^2=D_V(\widetilde{\mathbf v})$, $B_{n,m}^2=D_V(\mathbf v^*)$.
Since $\mathbf v^*$ minimizes $D_V$, $D_V(\widetilde{\mathbf v})-D_V(\mathbf v^*)\ge0$.
Moreover, the finite-sample argument in \citet[Section~4]{Linder2002}
gives, under compact support of $V$,
\begin{equation}
	D_V(\widetilde{\mathbf v})-D_V(\mathbf v^*)
	=
	O_P\!\left(
	\sqrt{\frac{2K_{n,m}\log K_{n,m}}{N}}
	\right).
	\label{eq:LLZ-slow2}
\end{equation}
Indeed, Lemma~5 and the proof of Theorem~3 therein imply that the
expectation of the left-hand side is bounded, up to a constant depending
only on the support bound, by
\[
\sqrt{
	\frac{3 K_{n,m}
		\log(3K_{n,m})}{N}
},
\]
and the stated $O_P$ bound follows from Markov's inequality since the dimension of $g_0(W)$
is fixed.
Therefore,
\begin{align*}
A_{n,m}
&=\left[D_V(\mathbf v^*)+
\left\{D_V(\widetilde{\mathbf v})-D_V(\mathbf v^*)\right\}\right]^{1/2}\\
&\le D_V(\mathbf v^*)^{1/2}
+\left[D_V(\widetilde{\mathbf v})-D_V(\mathbf v^*)\right]^{1/2}\\
&=B_{n,m}
+O_P\!\left(\left(\frac{2K_{n,m}\log K_{n,m}}{N}\right)^{1/4}\right)\\
&=O(K_{n,m}^{-1/2})
+O_P\!\left(\left(\frac{2K_{n,m}\log K_{n,m}}{N}\right)^{1/4}\right)
=o_P(1),
\end{align*}
where the inequality uses $\sqrt{x+y}\le\sqrt{x}+\sqrt{y}$ for $x,y\ge0$. Hence, by \eqref{eq:Astar-bound}, $|\widetilde F_{n,m}^{(2)}|=o_P(1)$.

\medskip

\noindent\emph{The term $\widetilde F_{n,m}^{(1)}$.}
Let $\widehat h_K$ denote the Voronoi assignment map generated by the actual empirical centers $\widehat{\mathbf v}$, so that
\[
\widehat g_{K_{n,m}}(W)=\widehat h_K(\widehat g_0(W)).
\]
Introduce only for this part of the proof the same-center true-index partition
$ \check g_{K_{n,m}}(W):=\widehat h_K(V)$, where $V=g_0(W)$. Then
\begin{align}
|\widetilde F_{n,m}^{(1)}|
&\le |\overline b_d^{\,\widehat g_{K_{n,m}}}-\overline b_d^{\,\check g_{K_{n,m}}}|
+|\overline b_d^{\,\check g_{K_{n,m}}}-\overline b_d^{g_0}|
+|\overline b_d^{\,\widetilde g_{K_{n,m}}}-\overline b_d^{g_0}|.
\label{eq:F1-check-decomp}
\end{align}
The last term is at most $C_0A_{n,m}=o_P(1)$ by \eqref{eq:generic-W2-partition-bound}. The middle term can also be handled by \eqref{eq:generic-W2-partition-bound}, because $\check g_{K_{n,m}}$ is a function of $V$ conditional on the realized centers. Specifically,
\begin{equation*}
|\overline b_d^{\,\check g_{K_{n,m}}}-\overline b_d^{g_0}|
\le C_0\check A_{n,m},
\qquad
\check A_{n,m}:=\left(\E\left[\min_k\|V-\widehat v_k\|^2\right]\right)^{1/2}.
\end{equation*}

We first show that $\check A_{n,m}=o_P(1)$. Recall that
$g_0(w)=w'\varsigma$ and $\widehat g_0(w)=w'\widehat\varsigma$.
Under the standard first-stage conditions for the finite-dimensional OLS
estimators entering $\widehat\varsigma$, root-sample-size asymptotic
normality gives $\|\widehat\varsigma-\varsigma\|=O_P(N^{-1/2})$, 
see, for example, \citet[Theorem~3.1; see also p.~2146 for least squares]{newey1994large}. Since the support of $W$ is compact,
\begin{align*}
\|\widehat g_0-g_0\|_\infty
&=\sup_{w\in\Supp(W)}
\|w'(\widehat\varsigma-\varsigma)\| \notag\\
&\le
\sup_{w\in\Supp(W)}\|w\|\,
\|\widehat\varsigma-\varsigma\|
=O_P(N^{-1/2}). 
\end{align*}

For a random vector $Z$, write
\[
D_Z(\mathbf v)
:=\E\!\left[\min_{1\le k\le K}\|Z-v_k\|^2\right],
\qquad
D^*_{Z,K}:=\inf_{\mathbf v}D_Z(\mathbf v),
\]
where the infimum may be restricted to a fixed compact ball containing the support of $Z$. Thus $D^*_{Z,K}$ is simply the smallest population value of
the squared $K$-means criterion using $K$ centers. In particular, $D^*_{\widehat g_0,K}$ is the infimum of $D_{\widehat g_0}(\mathbf v)$, holding the realized
$\widehat\varsigma$ fixed.  We apply Lemma~\ref{lem:l2-shift} with
$\widetilde g_0=\widehat g_0$: compactness of $\Supp(W)$ and $\|\widehat g_0-g_0\|_\infty=o_P(1)$ imply that suprema  $\sup_{w\in\Supp(W)}\|g_0(w)\|$ and $\sup_{w\in\Supp(W)}\|\widehat g_0(w)\|$ are $O_P(1)$. Since empirical K-means centers lie in the convex hull of the transformed sample, $\max_j\|\widehat v_j\|=O_P(1)$ as well. Hence Lemma~\ref{lem:l2-shift}, $O_P(\|\widehat g_0-g_0\|_\infty)=O_P(N^{-1/2})$, and taking
$\mathbf v=\widehat{\mathbf v}$ give
\begin{equation}
D_V(\widehat{\mathbf v})
=D_{\widehat g_0}(\widehat{\mathbf v})+O_P(N^{-1/2}).
\label{eq:dist-shift}
\end{equation}

We next make explicit the empirical-process step. Let
\[
\widehat D_{\widehat g_0}(\mathbf v)
:=\widehat P_{n,m}\!\left[
\min_{1\le k\le K}
\|\widehat g_0(W)-v_k\|^2
\right]
\]
be the empirical $K$-means criterion. By Lemma~\ref{lem:kmeans-vc},
\[
r_N:=
\sup_{\widetilde\varsigma\in\mathcal N}
\sup_{\mathbf v\in B(0,R)^K}
\left|
\widehat P_{n,m}\ell_{\widetilde\varsigma,\mathbf v}
-P\ell_{\widetilde\varsigma,\mathbf v}
\right|
=O_P\!\left(
\sqrt{\frac{K_{n,m}\log K_{n,m}}{N}}
\right),
\]
where $
\ell_{\widetilde\varsigma,\mathbf v}(w)
:=
\min_{1\leq k\leq K}
\|w'\widetilde\varsigma-v_k\|^2$, $\mathbf v\in B(0,R)^K$. Note that we also have 
$$ r_N = \underset{\mathbf v}{\sup} \left| \widehat D_{\widehat g_0}(\mathbf v) -  D_{\widehat g_0}(\mathbf v)\right|.$$
Because this bound is uniform in $\widetilde\varsigma$, it applies at the
random value $\widehat\varsigma$. Let
$\mathbf v^*_{\widehat g_0}$ minimize
$D_{\widehat g_0}(\mathbf v)$. Since
$\widehat{\mathbf v}$ minimizes the empirical criterion,
\begin{align*}
D_{\widehat g_0}(\widehat{\mathbf v})
&\le
\widehat D_{\widehat g_0}(\widehat{\mathbf v})+r_N\\
&\le
\widehat D_{\widehat g_0}(\mathbf v^*_{\widehat g_0})+r_N\\
&\le
D_{\widehat g_0}(\mathbf v^*_{\widehat g_0})+2r_N
=D^*_{\widehat g_0,K}+2r_N.
\end{align*}
Since $D^*_{\widehat g_0,K}$ is the infimum of the population criterion,
the left-hand difference is nonnegative. Consequently,
\begin{equation}
0\le
D_{\widehat g_0}(\widehat{\mathbf v})-D^*_{\widehat g_0,K}
\le 2r_N
=O_P\!\left(
\sqrt{\frac{K_{n,m}\log K_{n,m}}{N}}
\right).
\label{eq:hat-kmeans-excess}
\end{equation}

Finally, Lemma~\ref{lem:l2-shift} is used twice. First,
\eqref{eq:dist-shift} and \eqref{eq:hat-kmeans-excess} imply
\begin{align*}
D_V(\widehat{\mathbf v})
&\le
D^*_{\widehat g_0,K}
+O_P\!\left(
\sqrt{\frac{K_{n,m}\log K_{n,m}}{N}}+N^{-1/2}
\right).
\end{align*}
By definition of $D^*_{\widehat g_0,K}$, $D^*_{\widehat g_0,K}
\le D_{\widehat g_0}(\mathbf v^*)$, where $\mathbf v^*$ minimizes $D_V$. Applying
Lemma~\ref{lem:l2-shift} a second time, now at $\mathbf v^*$, gives
\[
D_{\widehat g_0}(\mathbf v^*)
\le D_V(\mathbf v^*)+O_P(N^{-1/2}).
\]
Therefore,
\begin{align*}
D_V(\widehat{\mathbf v})
&\le D_V(\mathbf v^*)
+O_P\!\left(
\sqrt{\frac{K_{n,m}\log K_{n,m}}{N}}+N^{-1/2}
\right)\\
&=O(K_{n,m}^{-1})+o_P(1)
=o_P(1),
\end{align*}
where by Zador's theorem applied to $V$ (see \eqref{eq:zador1}) gives $D_V(\mathbf v^*)=O(K_{n,m}^{-1})$ and
Assumption~\ref{hyp:consistency_ghat}(3) implies
$K_{n,m}\log K_{n,m}/N\to0$. Thus
$\check A_{n,m}=D_V(\widehat{\mathbf v})^{1/2}=o_P(1)$.

It remains to compare $\widehat g_{K_{n,m}}$ and $\check g_{K_{n,m}}$. Let $ r_{n,m}:=P(\widehat g_{K_{n,m}}(W)\neq\check g_{K_{n,m}}(W))$.
By Lemma~\ref{lem:voronoi-stability}, applied conditionally on the realized empirical centers and using the density bound for $V$ in Assumption~\ref{hyp:consistency_ghat}(2a),
\begin{equation}
r_{n,m}
=O_P\!\left(K_{n,m}^2\|\widehat g_0-g_0\|_\infty\right)
=O_P(K_{n,m}^2N^{-1/2})
=o_P(1),
\label{eq:partition-disagreement}
\end{equation}
where the last equality follows from Assumption~\ref{hyp:consistency_ghat}(3). Lemma~\ref{lem:partition-bound-stability} then implies
\begin{equation}
|\overline b_d^{\,\widehat g_{K_{n,m}}}-\overline b_d^{\,\check g_{K_{n,m}}}|
=o_P(1).
\label{eq:actual-check-bound}
\end{equation}
Combining \eqref{eq:F1-check-decomp}--\eqref{eq:actual-check-bound} gives
$|\widetilde F_{n,m}^{(1)}|=o_P(1)$, and hence
$\widetilde F_{n,m}=o_P(1)$.

\medskip

\paragraph{Conclusion.} From Steps~1, 2, and 3, $\widehat{\overline b}{}_d^{\,\widehat g_{K_{n,m}}} - \overline b_d^{g_0} = \widetilde S_{n,m} + \widetilde F_{n,m} + \widetilde B_{n,m} \xrightarrow{p} 0$. Under the additional linear location-scale condition, $\overline b_d^{g_0} = \overline b_d$ by Theorem \ref{thm:caract_id_set_common}. $\hfill\square$

\subsection{Additional lemmas} 
\label{sec:additional_lemmas}

The proof of Theorem \ref{thm:asym_Xo} relies on two lemmas, which we state and prove below. Note that Lemma \ref{lem:borne_var} is similar to Corollary 2.12 in \cite{boucheron2015} but handles variables taking negative values. Also, Lemma \ref{lem:A.1.modif} is similar to Lemma A.1 in \cite{del2019central} but holds under slightly weaker conditions. 

\begin{lem}\label{lem:change_var}
	For any cdfs $F, G$, $Y=F^{-1}(U)$ and $U\sim\mathcal{U}[0,1]$, we have
$$\int_0^1 [\ind{U\le t} - t]G^{-1}(t)dF^{-1}(t) = \int_{-\infty}^{\infty} [\ind{Y \le u} -F(u)] G^{-1} \circ F(u)  du.$$
\end{lem}

\begin{lem}
	Suppose Assumptions \ref{hyp:reg} hold. Then, for all $C>0$ and  $x\in\R$ continuity point of $G$,
	\begin{align}
		\max_{j \in \{1,\dots,m\}} |\widehat{\eta}_{dj}-\eta_{dj}|\convP &\; 0, \label{eq:conv_unif_eta} \\
		\max_{j:|\eta_{dj}|\le C} |\widehat{G}(\widehat{\eta}_{dj})- G(\eta_{dj})| \convP &\; 0,  \label{eq:conv_unif_Ghat} \\ \widehat{G}(x) \convP & \; G(x).\label{eq:weak_conv_Ghat}
	\end{align}
	Moreover, if $G$ is continuous, \eqref{eq:conv_unif_Ghat} still holds if we replace $\widehat{G}(\widehat{\eta}_{dj})$ by $\widehat{G}(\widehat{\eta}_{dj}^-)$.
\label{lem:eta}
\end{lem}

\begin{lem}
	Suppose that $T$ has a cdf $F$, survival function $S$ and a positive density $f$. Then, for all $i\in\{1,...,n\}$,
	$$V(T_{(i)}) \le \frac{32}{i\wedge (n+1-i)}E\left[2\left(\frac{F(T_{(i)})S(T_{(i)})}{f(T_{(i)})}\right)^2 + T_{(i)}^2\right].$$
	\label{lem:borne_var}
\end{lem}

\begin{lem}\label{lem:A.1.modif}
For $f\in\{\overline{c}_2,\widetilde{c}_1,\widetilde{c}_2\}$, the function $(F,G)\mapsto f(\cdot,F,G)$ is continuous in $L^2[0,1]$ ($L^1[0,1]$ for $f=\widetilde{c}_2$) with respect to the metric
	$$d[(F,G),(F',G')]=\mathcal{W}_4(F',F)+\mathcal{W}_4(G',G),$$
	at any $(F_0,G_0)$ (and also at $(G_0,F_0)$ when $f=\overline{c}_2$) satisfying the same restrictions as $(F,G)$ in Assumption \ref{hyp:reg}-(ii) or (iii). The same holds if in the metric $d$, we replace $\mathcal{W}_4(F',F)$ by $\mathcal{W}_2(F',F)$, provided that $(F_0, G_0)$ satisfy the same restrictions as $(F,G)$ in Assumption \ref{hyp:reg}-(i). 
\end{lem}

\begin{lem}\label{lem:l2-shift}
    Let $\widetilde g_0:\Supp(W)\to\R^{2}$ and $\mathbf v=(v_1,\ldots,v_{K_{n,m}})$. Suppose that, for some $R<\infty$,
    \[
    \sup_{w\in\Supp(W)}\|g_0(w)\|\vee
    \sup_{w\in\Supp(W)}\|\widetilde g_0(w)\|\vee
    \max_j\|v_j\|\le R.
    \]
    Then
    \[
    \left|\E\!\left[\min_j\|g_0(W)-v_j\|^2\right]
    -\E\!\left[\min_j\|\widetilde g_0(W)-v_j\|^2\right]\right|
    \le 4R\,\|\widetilde g_0-g_0\|_\infty.
    \]
\end{lem}

\begin{lem}[Stability with respect to a change of partition]
\label{lem:partition-bound-stability}
Suppose Assumption~\ref{hyp:consistency_ghat}(2b) holds. Let $g$ and $h$ be two finitely valued $W$-measurable partitions with the same labels, and set
\[
r(g,h):=P(g(W)\neq h(W)).
\]
Then, for $\kappa:=\eps/[2(2+\eps)]$, there is a constant $C<\infty$, independent of the partitions and of $K$, such that
\begin{equation}
|\overline b_d^{\,g}-\overline b_d^{\,h}|
\le Cr(g,h)^\kappa.
\label{eq:partition-bound-stability}
\end{equation}
The result also holds pathwise for random partitions, holding the realized maps fixed.
\end{lem}

\begin{lem}[Voronoi assignment stability]
\label{lem:voronoi-stability}
Let $Z\in\R^{2}$ have a density bounded by $\overline f$ on a
compact support of diameter at most $\overline D$. Let 
$h_{\mathbf v}$ be the nearest-center map generated by $K$ centers,
with a fixed tie-breaking rule. If $Z'$ is coupled with $Z$ and
$\|Z'-Z\|\le\delta$ almost surely, then
\begin{equation}
P\bigl(h_{\mathbf v}(Z')\neq h_{\mathbf v}(Z)\bigr)
\le 2\overline f\,\overline D\,K^{2}\,\delta,
\label{eq:voronoi-stability}
\end{equation}
uniformly in the centers.
\end{lem}

\begin{lem}[Empirical balance]\label{lem:pop-balance}
Let \(N:=\min(n,m)\). Suppose Assumption \ref{hyp:consistency_ghat}. Then, there exists \(c>0\), independent of $n$, $m$, such that
\[
P\left(
\min_k\frac{n_k}{n}\geq cK_{n,m}^{-2},
\quad
\min_k\frac{m_k}{m}\geq cK_{n,m}^{-2}
\right)\longrightarrow1.
\]
\end{lem}

\begin{lem}[VC empirical-process bound for the $K$-means criterion]
\label{lem:kmeans-vc}
Let \(\widehat P_{n,m}\) denote the pooled empirical distribution of the
$W$'s from the two samples, and define
$\ell_{\widetilde\varsigma,\mathbf v}(w):=
\min_{1\leq k\leq K}\|w'\widetilde\varsigma-v_k\|^2$. Under
Assumption~\ref{hyp:consistency_ghat}, for any compact $\mathcal N$ and
any $R<\infty$,
\[
\sup_{\widetilde\varsigma\in\mathcal N}
\sup_{\mathbf v\in B(0,R)^K}
\left|
\widehat P_{n,m}\ell_{\widetilde\varsigma,\mathbf v}
-P\ell_{\widetilde\varsigma,\mathbf v}
\right|
=O_P\!\left(\sqrt{\tfrac{K\log K}{N}}\right).
\]
\end{lem}

\subsubsection{Proof of Lemma \ref{lem:change_var}} 
\label{ssub:proof_of_lemma_ref_lem_change_var}

Note that $F$ is a generalized inverse of $F^{-1}$ \citep[see, e.g.,][p.7]{SW86}. Then, by, e.g., Eq. (1) in \cite{falkner2012substitution},
$$\int_0^1 [\ind{U\le t} - t]G^{-1}(t)dF^{-1}(t) = \int_{-\infty}^\infty [\ind{U\le F(u)} - F(u)]G^{-1}\circ F(u)du.$$
The result follows by noting that $U\le F(u)$ if and only if $Y\le u$ \citep[see, e.g., Lemma 21.1 in][]{van2000asymptotic}.


\subsubsection{Proof of Lemma \ref{lem:eta}} 
\label{ssub:proof_of_lemma_eta}

First, 
\begin{align*}
\max_{i=1,...,m}|\widehat{\eta}_{di}-\eta_{di}| & = \max_{i=1,...,m}\left|T_{-1i}'(\widehat{\gamma}-\gamma_0)\right|\\
& \le \left[\max_{i=1,...,m}\left\|T_{-1i}\right\|\right] \left\|\widehat{\gamma}-\gamma_0\right\|\\
& = o_P(m^{1/2}) \times O_P(m^{-1/2}) \\
& = o_P(1). 
\end{align*}
The second equality follows since $E[\|T_{-1i}\|^2]<\infty$, see e.g. Exercise 4 in Section 2.3 of \cite{VdV_Wellner}.

\medskip
Let us turn to \eqref{eq:conv_unif_Ghat}. First, assume that (i) holds in Assumption \ref{hyp:reg}. Given that $|\Supp(\eta_d)|<\infty$, it suffices to prove that for all $u\in \Supp(\eta_d)$,
\begin{equation}
\max_{j:\eta_{dj}=u} \left|\frac{1}{m}\sum_{k=1}^m \ind{\widehat{\eta}_{dk}=\widehat{\eta}_{dj}}- P(\eta_d=u)\right|\convP 0.	
	\label{eq:conv_G_discret}
\end{equation}
Fix such a $u\in \Supp(\eta_d)$. Since $|\Supp(\eta_d)|=|\Supp(X)|$, there exists $x\in\Supp(X)$ such that for all $j$, $\eta_{dj}=u \Leftrightarrow X_j=x$. Also, because of \eqref{eq:conv_unif_eta}, there exists $u_m\convP u$ such that $X_j=x\Leftrightarrow \widehat{\eta}_{dj}=u_m$. Hence, for sufficiently large $m$,
$$\frac{1}{m}\sum_{k=1}^m \ind{\widehat{\eta}_{dk}=\widehat{\eta}_{dj}} = \frac{1}{m}\sum_{k=1}^m \ind{\eta_{dk}=\eta_{dj}}.$$
The result follows from the law of large numbers. 

\medskip
Now, assume that (ii) or (iii) holds in Assumption \ref{hyp:reg}. Fix $\delta>0$. Since $G$ is continuous, there exists $\eps>0$ such that for all $(x,y)\in[-C-2\eps,C+2\eps]^2$, $|x-y|\le 2\eps$ implies $|G(y)-G(x)|<\delta$. By \eqref{eq:conv_unif_eta}, with probability approaching one, $|\widehat{\eta}_{dj}-\eta_{dj}|\le \eps$ for all $j$. Under this event, $\widehat{G}(x)\in[G_m(x-\eps),G_m(x+\eps)]$ for all $x\in\R$. Thus,
\begin{align*}
	& \max_{j:|\eta_{dj}|\le C} |\widehat{G}(\widehat{\eta}_{dj}) - G(\eta_{dj})| \\
	\le & \max_{j:|\eta_{dj}|\le C} \max\left\{|G_m(\eta_{dj}+2\eps) - G(\eta_{dj})|,|G_m(\eta_{dj}-2\eps) - G(\eta_{dj})|\right\} \\
	\le & \sup_{x\in[-C-2\eps,C+2\eps]} |G_m(x)-G(x)| + \sup_{(x,y)\in[-C-2\eps,C+2\eps]^2:|x-y|\le 2\eps}|G(x)-G(y)| \\
	\le & \sup_{x\in\R} |G_m(x)-G(x)| + \delta.
\end{align*}
By Glivenko-Cantelli theorem, $\sup_{x\in\R} |G_m(x)-G(x)|<\delta$ with probability approaching one. Eq. \eqref{eq:conv_unif_Ghat} follows since $\delta>0$ was arbitrary. To see that Eq. \eqref{eq:conv_unif_Ghat} still holds with $\widehat{G}(\widehat{\eta}_{dj})$ replaced by $\widehat{G}(\widehat{\eta}_{dj}^-)$, remark that there exists $(\widetilde{\eta}_{dj})_{j=1,...,m}$ such that $\widehat{G}(\widetilde{\eta}_{dj})=\widehat{G}(\widehat{\eta}_{dj}^-)$ and $|\widetilde{\eta}_{dj}-\eta_{dj}|\le \eps/2$. Then, the same proof as above applies, once we remark that with probability approaching one, $|\widehat{\eta}_{dj}-\eta_{dj}|\le \eps/2$ for all $j$. 

\medskip
Finally, we prove \eqref{eq:weak_conv_Ghat}. If (i) holds in Assumption \ref{hyp:reg}, a continuity point $x$ of $G$ is  such that either $x<\min(\Supp(\eta_d))$, $x>\max(\Supp(\eta_d))$ or there exists $(u,v)\in\Supp(\eta_d)^2$ such that $u<x<v$ (and $G(x)=G(u)$). In the first case, wpao, $x<\min_j \widehat{\eta}_j$ and thus $\widehat{G}(x)=0$. The reasoning with $x>\max(\Supp(\eta_d))$ is the same. In the third case, wpao, 
$$\underline{x}:=\max\{\widehat{\eta}_{dj}:\, \eta_{dj}=u\}<x<\overline{x}:=\min\{\widehat{\eta}_{dj}:\, \eta_{dj}=v\}$$
and there is no $j$ such that $\widehat{\eta}_{dj}\in (\underline{x},\overline{x})$. Hence, under this event, $$\widehat{G}\left(\max\{\widehat{\eta}_{dj}:\, \eta_{dj}=u\}\right)= \widehat{G}(x).$$ 
By \eqref{eq:conv_unif_Ghat}, the left-hand side converges in probability to $G(u)$. The result follows.

\medskip
Now, assume that (ii) or (iii) holds in Assumption \ref{hyp:reg}. Fix $\delta>0$ and let $\eps>0$ be such that $|y-x|<\eps$ implies $|G(y)-G(x)|<\delta$. We proved above that $\widehat{G}(x)\in[G_m(x-\eps),G_m(x+\eps)]$ wpao. By the law of large numbers, $G_m(x-\eps)\convP G(x-\eps)$ and $G_m(x+\eps)\convP G(x+\eps)$. Hence, wpao, $|\widehat{G}(x)-G(x)|<2\delta$. The result follows since $\delta>0$ was arbitrary.


\subsubsection{Proof of Lemma \ref{lem:borne_var}} 
\label{ssub:proof_of_lemma_borne_var}

First, note that
\begin{equation}
V(T_{(i)}) \le 2\left[V(T_{(i)} F(T_{(i)}))+V(T_{(i)} S(T_{(i)}))\right].	
	\label{eq:ineg_var}
\end{equation}
Remark that $T_{(i)}=F^{-1}(1-\exp(-E_{(i)}))$, where $(E_1,...,E_n)$ are iid, Exponential variables of parameter 1.
Then, by R\'enyi's  representation of order statistics for such variables,
$$	V(T_{(i)} F(T_{(i)}))  = V\left[F^{-1}\left(1-e^{-\sum_{k=n+1-i}^n E_k/k}\right)\left(1-e^{-\sum_{k=n+1-i}^n  E_k/k}\right)\right].$$
Let us define
$$g(x_{n+1-i},...,x_n)=F^{-1}\left(1-e^{-\sum_{k=n+1-i}^n x_k/k}\right)\left(1-e^{-\sum_{k=n+1-i}^n x_k/k}\right).$$
Then, by Poincare's inequality for exponential variables \citep[see, e.g., Proposition 2.10 in][]{boucheron2015}, we have
$$V(T_{(i)} F(T_{(i)}))\le 4 E\left[\sum_{k=n+1-i}^n \Deriv{g}{x_k}(E_{n+1-i},...,E_n)^2\right].$$
Remark that for all $j\in\{n+1-i,...,n\}$,
\begin{align*}
	\Deriv{g}{x_j}(x_{n+1-i},...,x_n) = \frac{1}{j}& \left[\frac{1-e^{-\sum_{k=n+1-i}^n x_k/k}}{h\circ F^{-1}\left(1-e^{-\sum_{k=n+1-i}^n x_k/k}\right)}\right. \\
	& \; \left. + e^{-\sum_{k=n+1-i}^n x_k/k}F^{-1}\left(1-e^{-\sum_{k=n+1-i}^n x_k/k}\right)\right].	
\end{align*}
Thus,
\begin{align}
	V(T_{(i)} F(T_{(i)}))  \le & 4 E\left[\sum_{k=n+1-i}^n \Deriv{g}{x_k}(E_{n+1-i},...,E_n)^2\right] \notag \\
	= & 4 \left[\sum_{j=n+1-i}^n \frac{1}{j^2}\right]E\left[\left(\frac{F(T_{(i)})}{h(T_{(i)})} + S(T_{(i)})T_{(i)}  \right)^2\right] \notag  \\
	\le & \frac{16}{n+1-i}E\left[\left(\frac{F(T_{(i)})S(T_{(i)})}{f(T_{(i)})}\right)^2 + S(T_{(i)})^2 T_{(i)}^2\right]. \label{eq:ineg_F}
\end{align}
To deal with $V(T_{(i)} S(T_{(i)}))$, we use $T_{(i)} = F^{-1}(\exp(-E_{(n+1-i)}))$ and reason exactly as above. This yields:
\begin{equation}
V(T_{(i)} S(T_{(i)}))  \le \frac{16}{i}E\left[\left(\frac{F(T_{(i)})S(T_{(i)})}{f(T_{(i)})}\right)^2 + F(T_{(i)})^2 T_{(i)}^2\right].	
	\label{eq:ineg_S}	
\end{equation}
By combining \eqref{eq:ineg_var}, \eqref{eq:ineg_F}, \eqref{eq:ineg_S} and $x^2+(1-x)^2\le 1$ for $0\le x\le 1$, we finally obtain
$$V(T_{(i)}) \le \frac{32}{i\wedge (n+1-i)}E\left[2\left(\frac{F(T_{(i)})S(T_{(i)})}{f(T_{(i)})}\right)^2 + T_{(i)}^2\right].$$


\subsubsection{Proof of Lemma \ref{lem:A.1.modif}} 
\label{proof:lem:A.1.modif}

We mostly focus on $f=\overline{c}^2_2$, as the reasoning is the same when $f\in\{\widetilde{c}_1,\widetilde{c}_2\}$. As in \eqref{eq:psi_3}, we assume without loss of generality that $a$ is a continuity point of $G_0^{-1}$ and $F_0^{-1}$. Consider a sequence $(F_n,G_n)_{n\ge 1}$ converging to $(F_0,G_0)$ satisfying Assumption \ref{hyp:reg}-(ii) or (iii). We first show that $c_2(t,F_n,G_n)\to c_2(t,F_0,G_0)$ for almost every $t\in (0,1)$. Then, we prove that $c_2(\cdot,F_n,G_n)\to c_2(\cdot,F,G)$ in $L_2[0,1]$. The continuity result at $(F_0,G_0)$ follows. Then, we show how to adapt the reasoning to prove continuity at $(G_0,F_0)$ instead of $(F_0,G_0)$. Next, we prove the result with the alternative metric when $(F_0,G_0)$ satisfy Assumption \ref{hyp:reg}-(i). Finally, we show how to adapt the argument when $f\in\{\widetilde{c}_1,\widetilde{c}_2\}$.

\paragraph{1. $c_2(t,F_n,G_n)\to c_2(t,F_0,G_0)$ for almost every $t\in (0,1)$.} 
\label{par:step1_conv_ae_c2}
Assume without loss of generality that $t\in (a,1)$, and that it is a continuity point of $G_0^{-1}$. Suppose also that $s$ (a) is a continuity point of $G_0$; (b) is such that $G_0(s)$ is a continuity point of $F_0^{-1}$; (c) satisfies $s\not\in\{G_0^{-1}(a),G_0^{-1}(t)\}$. Note that by, e.g., Theorem 6.9 of \cite{villani2009optimal} and Lemma 21.2 in \cite{van2000asymptotic}, $G_n(s')\to G_0(s')$ for all continuity points $s'$ of $G_0$, and $F_n^{-1}(u)\to F_0^{-1}(u)$ for all continuity points of $F_0$. Fix $\delta>0$ and let $\eps>0$ be such that $|u- G_0(s)|\le \eps$ implies  $|F_0^{-1}(u)-F_0^{-1}(G_0(s))|\le \delta$ and $G_0(s)-\eps$ and $G_0(s)+\eps$ are continuity points of $F_0^{-1}$. Such an $\eps$ exists since the set of discontinuity points of $F_0^{-1}$ is at most countable. Now, for all $n$ large enough, $|G_n(s)-G_0(s)|\le \eps$, which implies by monotonicity that $F_n^{-1}(G_n(s))\in [F_n^{-1}(G_0(s)-\eps), F_n^{-1}(G_0(s)+\eps)]$. Then, by construction, 
$$F_n^{-1}(G_0(s)-\eps) \to F_0^{-1}(G_0(s)-\eps)\ge F_0^{-1}(G_0(s)) -\delta,$$
    and similarly for $F_n^{-1}(G_0(s)+\eps)$. Since $\delta$ was arbitrary, we obtain $F_n^{-1}(G_n(s))\to F_0^{-1}(G_0(s))$. In turn, because $s\not\in\{G_0^{-1}(a),G_0^{-1}(t)\}$,
{\small \begin{equation}
F_n^{-1}(G_n(s))\ind{G_n^{-1}(a) \le s\le G_n^{-1}(t)} \to F_0^{-1}(G_0(s)) \ind{G_0^{-1}(a) \le s\le G_0^{-1}(t)}.	
	\label{eq:conv_Fn_Gn}
\end{equation}}
Now remark that under Assumption \ref{hyp:reg}-(ii) or (iii), Conditions (a), (b) and (c) above hold for almost every $s$. Hence, \eqref{eq:conv_Fn_Gn} holds for almost all $s$. Moreover, for all $n$ large enough, $s\mapsto |F_n^{-1}(G_n(s))|\ind{G_n^{-1}(a) \le s\le G_n^{-1}(t)}$ is bounded by some $K>0$. Then, by the dominated convergence theorem, $c_2(t,F_n,G_n)\to c_2(t,F_0,G_0)$. Since almost all $t$ are continuity point of $G_0^{-1}$, Point 1 follows.

\paragraph{2. $c_2(\cdot,F_n,G_n)\to c_2(\cdot,F,G)$ in $L_2[0,1]$.} Given Step 1, it suffices to prove, by Lebesgue–Vitali theorem \citep[see e.g., Theorem 4.5.4 in][]{bogachev2007measure}, that
\begin{equation}\label{eq:uniform_c}
	\lim_{M\to\infty}\sup_{n\ge 1}\int_0^1 c_2(t,F_n,G_n)^2\ind{c_2(t,F_n,G_n)^2>M} dt=0.
\end{equation}
First, remark that for all $s<G_n^{-1}(t)$, $G_n(s)\le t$. Hence, for such $s$, $F_n^{-1}(G_n(s))\le F_n^{-1}(t)$. Similarly, for $s\ge G_n^{-1}(a)$, $F_n^{-1}(G_n(s))\ge F_n^{-1}(a)$. As a result, 
\begin{align}
	|c_2(t,F_n,G_n)| \le & \left|G_n^{-1}(t) -G_n^{-1}\left(a\right)  \right| \left(\left| F_n^{-1}(t)\right| +  \left| F_n^{-1}\left(a\right)  \right|\right) \notag \\
	\le & q(G_n,t)\times q(F_n,t), \label{eq:ineg_c2}
\end{align} 
where $q(F,t):=|F^{-1}(t)|+|F^{-1}(a)|$ for any cdf $F$. Then,
$$\ind{c_2(t,F_n,G_n)^2>M} \le \ind{q(G_n,t)>\sqrt{M}} +  \ind{q(F_n,t)>\sqrt{M}}.$$
As a result, by Cauchy-Schwarz inequality,
\begin{align*}
& \int_0^1 c_2(t,F_n,G_n)^2\ind{c_2(t,F_n,G_n)^2>M} dt \\
\le & \left[\int_0^1 q(F_n,t)^4dt \int_0^1 q(G_n,t)^4\ind{q(G_n,t)>\sqrt{M}} dt\right]^{1/2} \\
+ &  \left[\int_0^1 q(F_n,t)^4 \ind{q(F_n,t)>\sqrt{M}}dt \int_0^1 q(G_n,t)^4dt \right]^{1/2}.
\end{align*}
We have $G_n^{-1}(a)\to G_0^{-1}(a)$. Also, because $\mathcal{W}_4(G_n,G_0)\to 0$, $G_n^{-1}\to G^{-1}$ in $L_4[0,1]$. Then, we also have $q(G_n,\cdot)\to q(G_0,\cdot)$ in $L_4[0,1]$. By Lebesgue–Vitali theorem again, $\lim_{M\to\infty}\sup_{n\ge 1}\int_0^1 q(G_n,t)^4\mathds{1}\{q(G_n,t)  $ $>\sqrt{M}\} dt=0$. The same holds with $F_n$ instead of $G_n$. Equation \eqref{eq:uniform_c} follows.

\paragraph{3. Continuity at $(G_0,F_0)$.} The reasoning is the same as above. Step 2 holds as is. In Step 1, we just need to check that Condition (b) on $s$ still holds for almost every $s$ when exchanging the roles of $F_0$ and $G_0$. This is true under Assumption \ref{hyp:reg}-(ii), since $G_0^{-1}$ is continuous on $(0,1)$. This also holds under Assumption \ref{hyp:reg}-(iii): $F_0$ is strictly increasing on its support (since $F_0^{-1}$ is continuous), and the set of discontinuity points of $G_0^{-1}$ is at most countable. 

\paragraph{4. Continuity with respect to the alternative metric.} The proof is very similar as above. Step 1 is as above, once we note that Condition (b) still holds for almost every $s$ under Assumption \ref{hyp:reg}-(i). Regarding Step 2, start from \eqref{eq:ineg_c2} and use instead that $q(G_n,t)$ is bounded for $t\in (0,1)$. Since $q(F_n,\cdot)^2$ is uniformly integrable, $c_2(\cdot,F_n,G_n)^2$ is uniformly integrable as well. The result follows.

\paragraph{5. Adaptation to $f\in\{\widetilde{c}_1,\widetilde{c}_2\}$.} The reasoning in Part 1 above is the same. For Part 2, let us just consider $f=\widetilde{c}_2$; with $f=\widetilde{c}_1$, we simply have to adjust some exponents and the use of H\"older's inequality below. The reasoning is the same as above but we use instead the inequality $|\widetilde{c}_2(t,F_n,G_n)| \le q(G_n,t)^3\times q(F_n,t)$. Then,
$$\ind{\widetilde{c}_2(t,F_n,G_n)>M} \le \ind{q(G_n,t)>M^{1/3}} +  \ind{q(F_n,t)>M}.$$
As a result, by Hölder's inequality with exponents $4/3$ and $4$,
\begin{align*}
& \int_0^1 \widetilde{c}_2(t,F_n,G_n)\ind{\widetilde{c}_2(t,F_n,G_n)>M} dt \\
\le & \left[\int_0^1  q(F_n,t)^4dt \int_0^1 q(G_n,t)^4\ind{q(G_n,t)>M^{1/3}} dt\right]^{3/4} \\
+ &  \left[\int_0^1 q(F_n,t)^4 \ind{q(F_n,t)>M}dt \int_0^1 q(G_n,t)^4dt \right]^{1/4}.
\end{align*}
We conclude as above.



\subsubsection{Proof of Lemma \ref{lem:l2-shift}}
For any $a,b\in\R^{2}$ and $\mathbf v$, let $j_b^*\in\arg\min_j\|b-v_j\|^2$. Then
\begin{align*}
\min_j\|a-v_j\|^2-\min_j\|b-v_j\|^2
&\le \|a-v_{j_b^*}\|^2-\|b-v_{j_b^*}\|^2\\
&=(a-b)'(a+b-2v_{j_b^*}).
\end{align*}
Swapping $a$ and $b$ and applying Cauchy--Schwarz therefore gives
\[
\left|\min_j\|a-v_j\|^2-\min_j\|b-v_j\|^2\right|
\le \|a-b\|\left(\|a\|+\|b\|+2\max_j\|v_j\|\right).
\]
Applying this pointwise with $a=g_0(W)$ and $b=\widetilde g_0(W)$ yields
\[
\left|\min_j\|g_0(W)-v_j\|^2-\min_j\|\widetilde g_0(W)-v_j\|^2\right|
\le 4R\,\|g_0(W)-\widetilde g_0(W)\|.
\]
Taking expectations proves the result. $\hfill\square$

\medskip

\subsubsection{Proof of Lemma \ref{lem:partition-bound-stability}.}
Let $C_k\triangle D_k$ denote the symmetric difference of the two sets \(C_k\) and \(D_k\): $$ C_k\triangle D_k := (C_k\setminus D_k)\cup(D_k\setminus C_k). $$
We also denote the total variation distance between two measures $P$ and $Q$ by 
$ \|P-Q\|_{\rm TV} = \sup_A |P(A) - Q(A)|$. 
It is the set of observations that belong to exactly one of the two cells.  Write $C_k=\{g(W)=k\}$, $D_k=\{h(W)=k\}$,
$p_k=P(C_k)$, $q_k=P(D_k)$, $r_k=P(C_k\triangle D_k)$ and $s_k=p_k\wedge q_k$. Assumption~\ref{hyp:consistency_ghat}(2b) implies that the conditional $(2+\eps)$ moments of $\nu_Y$ and $\nu_d$ are bounded uniformly over every $W$-measurable event of positive probability.\\

The conditional second-moment bounds implied by Assumption~\ref{hyp:consistency_ghat}(2b) 
give, uniformly over all cells of positive probability,
\begin{equation}
\|F_{|C_k}^{-1}\|_{L^2(0,1)}
\vee
\|G_{|C_k}^{-1}\|_{L^2(0,1)}
\vee
\|F_{|D_k}^{-1}\|_{L^2(0,1)}
\vee
\|G_{|D_k}^{-1}\|_{L^2(0,1)}
\le C.
\label{eq:cellwise-L2-bound}
\end{equation}
In particular, both $L(F_{|C_k},G_{|C_k})$ and $L(F_{|D_k},G_{|D_k})$ are 
uniformly bounded by Cauchy--Schwarz.

If two probability measures $P,Q$ on $\R$ have uniformly bounded
$(2+\eps)$ moments, the weighted-total-variation bound for Wasserstein
distance in \citet[Theorem~6.15]{villani2009optimal}, followed by H\"older's
inequality, gives
\begin{equation}
W_2(P,Q)\le C\|P-Q\|_{\rm TV}^{\kappa},
\qquad
\kappa=\frac{\eps}{2(2+\eps)}.
\label{eq:TV-W2-interpolation}
\end{equation}
Indeed, writing $|P-Q|$ for the variation measure and absorbing into $C$
the harmless factor depending on the normalization of total variation,
Theorem~6.15 of \citet{villani2009optimal} gives
\[
W_2^2(P,Q)
\le C\int_{\R}|x|^2\,d|P-Q|(x).
\]
With $\rho:=2+\eps$, H\"older's inequality yields
\begin{align*}
\int_{\R}|x|^2\,d|P-Q|(x)
&\le
\left(\int_{\R}|x|^{\rho}\,d|P-Q|(x)\right)^{2/\rho}
|P-Q|(\R)^{1-2/\rho}\\
&\le
C\|P-Q\|_{\rm TV}^{\eps/(2+\eps)},
\end{align*}
because $|P-Q|\le P+Q$ and the $\rho$-moments are uniformly bounded.
Taking square roots proves \eqref{eq:TV-W2-interpolation}. 

Moreover, direct conditioning yields, for $s_k>0$,
\[
\|P_{\nu_Y|C_k}-P_{\nu_Y|D_k}\|_{\rm TV}
\vee
\|P_{\nu_d|C_k}-P_{\nu_d|D_k}\|_{\rm TV}
\le 1\wedge\frac{2r_k}{s_k}.
\]
Indeed, for any measurable $A$, writing
$P_{X|C_k}(A) - P_{X|D_k}(A)$ with common denominator $P(C_k)\,P(D_k)$
gives
\begin{align*}
&P_{X|C_k}(A) - P_{X|D_k}(A) \\
& = 
\frac{[P(D_k)-P(C_k)]\,P(X\in A,C_k)+P(C_k)\,[P(X\in A,C_k)-P(X\in A,D_k)]}{P(C_k)\,P(D_k)}.    
\end{align*}
Since
$|P(C_k)-P(D_k)|
=|P(C_k\setminus D_k)-P(D_k\setminus C_k)|
\le r_k$
and, by the same decomposition,
$|P(X\in A,C_k)-P(X\in A,D_k)|
=|P(X\in A,C_k\setminus D_k)-P(X\in A,D_k\setminus C_k)|
\le r_k$,
one obtains, using $P(X\in A,C_k)\le P(C_k)$,
\[
\bigl|P_{X|C_k}(A) - P_{X|D_k}(A)\bigr|
\le \frac{2r_k}{P(D_k)}.
\]
By symmetry the same bound holds with $P(D_k)$ replaced by $P(C_k)$,
and hence by $s_k=P(C_k)\wedge P(D_k)$. Taking the supremum over $A$
gives $\|P_{X|C_k}-P_{X|D_k}\|_{\rm TV}\le 2r_k/s_k$, applied with
$X\in\{\nu_Y,\nu_d\}$.

Using \eqref{eq:L-W2-basic}, \eqref{eq:TV-W2-interpolation} and 
\eqref{eq:cellwise-L2-bound}, we therefore obtain
\begin{align*}
&|L(F_{|C_k},G_{|C_k})-L(F_{|D_k},G_{|D_k})| \\
&\le
W_2(F_{|C_k},F_{|D_k})\,
\|G_{|C_k}^{-1}\|_{L^2(0,1)}
+
\|F_{|D_k}^{-1}\|_{L^2(0,1)}\,
W_2(G_{|C_k},G_{|D_k}) \\
&\le C\left(
\|P_{\nu_Y|C_k}-P_{\nu_Y|D_k}\|_{\rm TV}^{\kappa}
\|G_{|C_k}^{-1}\|_{L^2(0,1)}
+
\|F_{|D_k}^{-1}\|_{L^2(0,1)}\,
\|P_{\nu_d|C_k}-P_{\nu_d|D_k}\|_{\rm TV}^{\kappa}
\right) \\
&\le C\left(1\wedge\frac{r_k}{s_k}\right)^\kappa,
\end{align*}
where the factor $2^\kappa$ has been absorbed into $C$.\\

We next compare the cell contributions including their probabilities. If, for instance, $p_k\ge q_k$, then $s_k=q_k$ and
\begin{align*}
&|p_kL(F_{|C_k},G_{|C_k})-q_kL(F_{|D_k},G_{|D_k})|\\
&\le
q_k|L(F_{|C_k},G_{|C_k})-L(F_{|D_k},G_{|D_k})|
+(p_k-q_k)|L(F_{|C_k},G_{|C_k})|\\
&\le
Cs_k\left(1\wedge\frac{r_k}{s_k}\right)^\kappa
+C|p_k-q_k|.
\end{align*}
The same bound holds when $q_k\ge p_k$. Since $|p_k-q_k|\le P(C_k\triangle D_k)=r_k$ and
\[
s_k\left(1\wedge\frac{r_k}{s_k}\right)^\kappa
\le s_k\left(\frac{r_k}{s_k}\right)^\kappa
=s_k^{1-\kappa}r_k^\kappa,
\]
we obtain, for $s_k>0$,
\begin{equation}
|p_kL(F_{|C_k},G_{|C_k})-q_kL(F_{|D_k},G_{|D_k})|
\le C(s_k^{1-\kappa}r_k^\kappa+r_k).
\label{eq:weighted-cell-stability}
\end{equation}
If $s_k=0$, the same inequality holds after interpreting the first term as zero: indeed, one of $p_k,q_k$ is zero, the other is at most $r_k$, and the corresponding $L$-functional is uniformly bounded.

Finally, since $s_k=p_k\wedge q_k\le p_k$ for every $k$, $\sum_{k=1}^K s_k\le \sum_{k=1}^K p_k=1$. Moreover,
\[
\sum_{k=1}^K r_k
=
\sum_{k=1}^K P(C_k\triangle D_k)
=
2P(g(W)\neq h(W))
=2r(g,h),
\]
because whenever $g(W)\neq h(W)$ the observation belongs to exactly two of the symmetric differences: the cell it leaves and the cell it enters. H\"older's inequality with conjugate exponents $1/(1-\kappa)$ and $1/\kappa$ then gives
\begin{align*}
\sum_{k=1}^K s_k^{1-\kappa}r_k^\kappa
&\le
\left(\sum_{k=1}^K s_k\right)^{1-\kappa}
\left(\sum_{k=1}^K r_k\right)^\kappa\\
&\le
(2r(g,h))^\kappa.
\end{align*}
Summing \eqref{eq:weighted-cell-stability} over $k$ therefore gives
\[
\sum_{k=1}^K
|p_kL(F_{|C_k},G_{|C_k})-q_kL(F_{|D_k},G_{|D_k})|
\le C\left(r(g,h)^\kappa+r(g,h)\right).
\]
Since $0\le r(g,h)\le1$ and $0<\kappa<1$, $r(g,h)\le r(g,h)^\kappa$; thus the right-hand side is bounded by $Cr(g,h)^\kappa$. Noting that the remaining term in \eqref{eq:upper:sigma_w1} does not depend on the partition proves \eqref{eq:partition-bound-stability}. $\hfill\square$

\subsubsection{Proof of Lemma \ref{lem:voronoi-stability}.}

Let $\mathcal X:=\Supp(Z)$, so that $\diam(\mathcal X)\le\overline D$
by hypothesis. Write $\operatorname{Leb}_{2}$ (resp.\
$\operatorname{Leb}_{1}$) for the Lebesgue measure on $\R^{2}$ (resp.\
on $\R$).

\medskip
\noindent\emph{Reduction to nearness to a bisecting line.} Let 
$j:=h_{\mathbf v}(Z)$, $k:=h_{\mathbf v}(Z')$. If $j=k$, there is 
nothing to prove; suppose therefore $j\neq k$. Since $j$ is a nearest 
center to $Z$, $\|Z-v_j\|^{2}\le \|Z-v_k\|^{2}$; since $k$ is a 
nearest center to $Z'$, $\|Z'-v_k\|^{2}\le \|Z'-v_j\|^{2}$.

For $t\in[0,1]$, define $Z_t:=(1-t)Z+tZ'$ and the continuous function
$\phi_{jk}(t):=\|Z_t-v_j\|^{2}-\|Z_t-v_k\|^{2}$. The two preceding 
inequalities imply $\phi_{jk}(0)\le0$ and $\phi_{jk}(1)\ge0$, so by
continuity there exists $t^{*}\in[0,1]$ with $\phi_{jk}(t^{*})=0$. 
Thus $Z_{t^{*}}\in H_{jk}$, where 
$H_{jk}:=\{z\in\R^{2}:\|z-v_j\|=\|z-v_k\|\}$. Moreover,
\[
\|Z-Z_{t^{*}}\|=t^{*}\|Z'-Z\|\le\delta,
\]
so $\operatorname{dist}(Z,H_{jk})\le\delta$. It follows that
\begin{equation}
	\{h_{\mathbf v}(Z')\neq h_{\mathbf v}(Z)\}
	\subseteq
	\bigcup_{1\le j<k\le K}
	\{\operatorname{dist}(Z,H_{jk})\le\delta\}.
	\label{eq:voronoi-disagreement-inclusion}
\end{equation}

\medskip
\noindent\emph{Strip-area bound.} For $v_j\neq v_k$, the set $H_{jk}$ 
is an affine line, since $\|z-v_j\|^{2}=\|z-v_k\|^{2}$ is equivalent 
to $2z'(v_k-v_j)=\|v_k\|^{2}-\|v_j\|^{2}$. If $v_j=v_k$, the pair 
can be ignored, as the two centers coincide and the fixed tie-breaking 
rule does not generate a genuine separating boundary. For an affine 
line $H\subset\R^{2}$, denote by
\[
H^\delta
:=
\{z\in\R^{2}:\operatorname{dist}(z,H)\le\delta\}
\]
the closed strip of half-width $\delta$ around $H$.

Since $\diam(\mathcal X)\le\overline D$, for any fixed 
$z_0\in\mathcal X$, $\mathcal X\subseteq B(z_0,\overline D)$. After a 
translation and an orthogonal change of coordinates, we may write 
$H=\{z_1=0\}$ and $H^\delta=\{|z_1|\le\delta\}$. For each fixed 
$t\in[-\delta,\delta]$, the section 
$B(z_0,\overline D)\cap\{z_1=t\}$ is either empty or a closed line 
segment of length at most $2\overline D$. Hence, by Fubini's theorem,
\begin{align}
\operatorname{Leb}_{2}\bigl(\mathcal X\cap H^\delta\bigr)
& \le
\operatorname{Leb}_{2}\bigl(B(z_0,\overline D)\cap H^\delta\bigr) \notag \\
& = 
\int_{-\delta}^{\delta}
	\operatorname{Leb}_{1}\bigl(B(z_0,\overline D)\cap\{z_1=t\}\bigr)\,dt  \notag  \\
& \le
4\overline D\,\delta.
\label{eq:hyperplane-slab-volume}
\end{align}
Since the density of $Z$ is bounded by $\overline f$, 
\eqref{eq:hyperplane-slab-volume} implies, for every pair $j<k$ with 
$v_j\neq v_k$,
\[
P\!\left(\operatorname{dist}(Z,H_{jk})\le\delta\right)
=
\int_{\mathcal X\cap H_{jk}^\delta} f_Z(z)\,dz
\le
\overline f\,\operatorname{Leb}_{2}\bigl(\mathcal X\cap H_{jk}^\delta\bigr)
\le
4\overline f\,\overline D\,\delta.
\]

\medskip
\noindent\emph{Union bound.} Using 
\eqref{eq:voronoi-disagreement-inclusion},
\[
P(h_{\mathbf v}(Z')\neq h_{\mathbf v}(Z))
\le
\sum_{1\le j<k\le K}
	P\!\left(\operatorname{dist}(Z,H_{jk})\le\delta\right)
\le
\binom{K}{2} 4\overline f\,\overline D\,\delta
\le
2\overline f\,\overline D\,K^{2}\,\delta,
\]
which proves \eqref{eq:voronoi-stability}.
\hfill$\square$

\subsubsection{Proof of Lemma \ref{lem:pop-balance}.}
Write \(K:=K_{n,m}\), and, denote by $\conv(\mathcal X)$ the convex hull of $\mathcal X$, whose diameter equals $\diam(\mathcal X)$.  Under Assumption \ref{hyp:consistency_ghat}-(1), $W$ has compact support and we denote by $D_W := \operatorname{diam}(\operatorname{supp}(W)) < \infty$. Combined with Assumption \ref{hyp:consistency_ghat}-(2a), there exist common constants $0<\overline f,\overline D<\infty$ such that, uniformly over $\widetilde\varsigma\in\mathcal N$, the density of $Z_{\widetilde\varsigma}=W'\widetilde\varsigma$ is bounded by $\overline f$ and the diameter of its compact support is bounded by $\overline D$. The proof has three parts.

\medskip
\noindent
\emph{Part 1: a generic balance bound for near-optimal \(K\)-means configurations.}
Let \(Z\in\R^{2}\) admit a density \(f_Z\) satisfying $\|f_Z\|_\infty\leq\overline f$
and let its compact support \(\mathcal X\) satisfy $\diam(\mathcal X)\leq\overline D$.
For
\(\mathbf v=(v_1,\ldots,v_K)\in\conv(\mathcal X)^K\), define
\[
D_Z(\mathbf v)
:=
\E\!\left[\min_{1\leq j\leq K}\|Z-v_j\|^2\right],
\qquad
D^*_{Z,K}
:=
\inf_{\mathbf u\in\conv(\mathcal X)^K}D_Z(\mathbf u),
\]
so that $D^*_{Z,K}$ is the smallest population mean squared distance to the
nearest of $K$ centers. Let \(V_k(\mathbf v)\) denote the \(k\)-th Voronoi cell, with an arbitrary
fixed rule for ties.

We claim that there exist constants \(c_0,c_1>0\), depending only on
\(\overline f,\overline D\), such that
$D_Z(\mathbf v)-D^*_{Z,K} \leq c_1K^{-2}$ implies  
\begin{equation}\min_{1\leq k\leq K}P\!\left(Z\in V_k(\mathbf v)\right)
\geq
c_0K^{-2}
\label{eq:generic-near-opt-balance}
\end{equation}
for all sufficiently large \(K\). To prove the claim, suppose that for some \(k_0\),
\[
p_{k_0}
:=
P\!\left(Z\in V_{k_0}(\mathbf v)\right)
<
\delta K^{-2},
\]
where \(\delta>0\) will be fixed below. Let $\mathbf v_{-k_0}$ denote the collection of centers obtained after
removing $v_{k_0}$, and define the resulting increase in population
distortion by
\begin{align*}
	\Delta_{\rm cost}
	&:=
	D_Z(\mathbf v_{-k_0})-D_Z(\mathbf v)\\
	&=
	\E\!\left[
	\left(
	\min_{\ell\neq k_0}\|Z-v_\ell\|^2
	-
	\|Z-v_{k_0}\|^2
	\right)
	\ind{Z\in V_{k_0}(\mathbf v)}
	\right].
\end{align*}
Since all points and centers lie in a set $\mathcal X$ of diameter at most
$\overline D$, and all remaining centers belong to
\(\conv(\mathcal X)\), whose diameter equals \(\diam(\mathcal X)\), the integrand is bounded above by $\overline D^2$.
Therefore,
\begin{equation}
	\Delta_{\rm cost}
	\le
	\overline D^2 p_{k_0}
	<
	\overline D^2\delta K^{-2}.
	\label{eq:remove-cost-prob}
\end{equation}

Let \(\widetilde k\) be a cell of largest probability and set $p:=P\!\left(Z\in V_{\widetilde k}(\mathbf v)\right)$, so that \(p\geq1/K\) since the cells partition $\R^2$. For all sufficiently large \(K\),
\(\widetilde k\neq k_0\), since
\(\delta K^{-2}<1/K\). Let $r
:=
\left( p/(2\overline f\,\omega_{2})
\right)^{1/2}$,
where $\omega_{2}$ denotes the volume of $B(0,1)$. Since $P\!\left(Z\in B(v_{\widetilde k},r)\right)
\leq
\overline f\,\omega_{2}r^{2}
=p/2$,
we have
\[
P\!\left(
Z\in V_{\widetilde k}(\mathbf v)
\setminus B(v_{\widetilde k},r)
\right)
\geq\frac p2.
\]
For every
\(x\in V_{\widetilde k}(\mathbf v)\setminus B(v_{\widetilde k},r)\),
there exist \(j\in\{1,2\}\) and \(s\in\{-1,+1\}\) such that
\[
s(x_j-v_{\widetilde k,j})\geq\frac r{\sqrt{2}}.
\]
Hence, among the \(4\) sets $A_{j,s}
:=
\left\{
x\in V_{\widetilde k}(\mathbf v):
s(x_j-v_{\widetilde k,j})\geq r/\sqrt{2}
\right\}$,
which cover  $V_{\widetilde k}(\mathbf v)
\setminus B(v_{\widetilde k},r)$, there is one, denoted \(A_{s_0,j_0}\), satisfying $P(Z\in A_{s_0,j_0})\geq p/8$. 	Let $\overline v
:=
v_{\widetilde k}+ s_0re_{j_0}/\sqrt{2} $,
and let \(\overline v^\pi\) denote its Euclidean projection onto
\(\conv(\mathcal X)\). 
Then, for every $x\in A$,
\begin{align*}
	\|x-v_{\widetilde k}\|^2-\|x-\overline v\|^2
	&=
	\|x-v_{\widetilde k}\|^2
	-
	\left\|
	x-v_{\widetilde k}
	-s_0\frac{r}{\sqrt{2}}e_{j_0}
	\right\|^2\\
	&=
	2s_0\frac{r}{\sqrt{2}}
	(x_{j_0}-v_{\widetilde k,j_0})
	-\frac{r^2}{2}\\
	&\ge	\frac{r^2}{2}.
\end{align*}
Since \(x\in\mathcal X\subset\conv(\mathcal X)\), the metric projection
property gives $\|x-\overline v^\pi\|
\leq
\|x-\overline v\|$.
Let $\mathbf v_{-k_0}$ denote the collection of the $K-1$ centers
remaining after $v_{k_0}$ has been removed, and let $\mathbf v^+
:=
\left(\mathbf v_{-k_0},\overline v^\pi\right)$ denote the collection of $K$ centers obtained by adding
$\overline v^\pi$ to these remaining centers. Define the reduction in
population distortion generated by this addition as
\begin{equation}
	\begin{aligned}
		\Delta_{\rm gain}
		&:=
		D_Z(\mathbf v_{-k_0})-D_Z(\mathbf v^+)\\
		&=
		\E\!\left[
		\min_{\ell\neq k_0}\|Z-v_\ell\|^2
		-
		\min\left\{
		\min_{\ell\neq k_0}\|Z-v_\ell\|^2,\,
		\|Z-\overline v^\pi\|^2
		\right\}
		\right]\\
		&=
		\E\!\left[
		\left(
		\min_{\ell\neq k_0}\|Z-v_\ell\|^2
		-
		\|Z-\overline v^\pi\|^2
		\right)_+
		\right],
	\end{aligned}
	\label{eq:def-split-gain}
\end{equation}
where $x_+:=\max\{x,0\}$. Recall that $A_{s_0,j_0}$ is contained in the Voronoi cell
$V_{\widetilde k}(\mathbf v)$ and that $\widetilde k\neq k_0$.
Hence $v_{\widetilde k}$ is still present in $\mathbf v_{-k_0}$.
Moreover, for every $x\in A_{s_0,j_0}$,
\[
\min_{\ell\neq k_0}\|x-v_\ell\|^2
=
\|x-v_{\widetilde k}\|^2.
\]
The preceding argument also showed that, for every $x\in A_{s_0,j_0}$,
\[
\|x-v_{\widetilde k}\|^2
-
\|x-\overline v^\pi\|^2
\ge
\frac{r^2}{2}.
\]
Therefore, restricting the expectation in
\eqref{eq:def-split-gain} to the event $\{Z\in A_{s_0,j_0}\}$ gives
\begin{align*}
	\Delta_{\rm gain}
	&\ge
	\E\!\left[
	\left\{
	\min_{\ell\neq k_0}\|Z-v_\ell\|^2
	-
	\|Z-\overline v^\pi\|^2
	\right\}
	\ind{Z\in A_{s_0,j_0}}
	\right]\\
	&=
	\E\!\left[
	\left\{
	\|Z-v_{\widetilde k}\|^2
	-
	\|Z-\overline v^\pi\|^2
	\right\}
	\ind{Z\in A}
	\right]\\
	&\ge
	\frac{r^2}{2}P(Z\in A_{s_0,j_0}).
\end{align*}
Since, by construction of $A_{s_0,j_0}$, $P(Z\in A_{s_0,j_0})\ge p/8$,
we obtain
\begin{align}
	\Delta_{\rm gain}
	&\ge
	\frac{p}{8}\frac{r^2}{2}
	\nonumber\\
	&=
	\frac{1}{16}
	\left(
	\frac{1}{2\overline f\,\omega_{2}}
	\right)
	p^{2}
	\nonumber\\
	&\ge
	C_bK^{-2},
	\label{eq:split-gain-prob}
\end{align}
where $C_b
:=
(16)^{-1}
\left(2\overline f\,\omega_{2}\right)^{-1}
>0$. The last inequality follows from $p\ge K^{-1}$.

Use the center freed by removing $v_{k_0}$ to add
$\overline v^\pi$, and denote the resulting $K$-center configuration by $\mathbf v^+
:= (\mathbf v_{-k_0},\overline v^\pi)$. By definition, $\Delta_{\rm cost}
=
D_Z(\mathbf v_{-k_0})-D_Z(\mathbf v)$,
whereas $\Delta_{\rm gain}
=
D_Z(\mathbf v_{-k_0})-D_Z(\mathbf v^+)$.
Therefore, 	$D_Z(\mathbf v^+)-D_Z(\mathbf v)
	=
	\Delta_{\rm cost}-\Delta_{\rm gain}$. 
Using \eqref{eq:remove-cost-prob} and
\eqref{eq:split-gain-prob},
\begin{align*}
	D_Z(\mathbf v^+)-D_Z(\mathbf v)
	&\le
	\overline D^2\delta K^{-2}
	-
	C_bK^{-2}\\
	&=
	-
	\left(
	C_b-\overline D^2\delta
	\right)
	K^{-2}.
\end{align*}
Hence
\begin{equation}
	D_Z(\mathbf v^+)
	\le
	D_Z(\mathbf v)
	-
	\left(
	C_b-\overline D^2\delta
	\right)
	K^{-2}.
	\label{eq:remove-split-net-gain}
\end{equation}

Take $\delta:=C_b/(2\overline D^2)$, $c_1:=C_b/4$. Then $C_b-\overline D^2\delta
= C_b/2$, so \eqref{eq:remove-split-net-gain} becomes
\[
D_Z(\mathbf v^+)
\le
D_Z(\mathbf v)
-
\frac{C_b}{2}K^{-2}.
\]

If $D_Z(\mathbf v)-D^*_{Z,K}
\le
c_1K^{-2}
= C_bK^{-2}/4$,
then $D_Z(\mathbf v)
\le
D^*_{Z,K}
+ C_bK^{-2}/4$.
Substituting this into the preceding inequality yields
\begin{align*}
	D_Z(\mathbf v^+)
	&\le
	D^*_{Z,K}
	+
	\frac{C_b}{4}K^{-2}
	-
	\frac{C_b}{2}K^{-2}\\
	&\le
	D^*_{Z,K}
	-
	\frac{C_b}{4}K^{-2},
\end{align*}
which contradicts the definition $D^*_{Z,K}
:=
\inf_{\mathbf u}
D_Z(\mathbf u)$, since $\mathbf v^+$ is itself a $K$-center configuration. Thus \eqref{eq:generic-near-opt-balance} holds with $c_0:=C_b/(2\overline D^2)$.

\medskip
\noindent
\emph{Part 2: population balance for the actual empirical centers.}
Since \(\widehat\varsigma\to_P\varsigma\), $P(\widehat\varsigma\in\mathcal N)\to1$.
On this event, the law of $Z_{\widehat\varsigma}=\widehat g_0(W)$
satisfies the same constants \(\overline f,\overline D\) as in Part~1.

We next show that the actual empirical \(K\)-means centers are
sufficiently close to population optimality. Because \(W\) has compact
support and \(\mathcal N\) can be taken bounded, there exists
\(R<\infty\) such that, for all
$w\in\Supp(W)$, $\widetilde\varsigma\in\mathcal N$, 
\[
\|w'\widetilde\varsigma\|\leq R.
\]
Both empirical and population \(K\)-means centers may therefore be
restricted without loss to the compact ball \(B(0,R)\). Moreover, the
empirical minimization may be restricted without loss of generality to
centers in the convex hull of the transformed sample. Let
$$ \mathcal C_{n,m}
:=
\conv\left(
\{\widehat g_0(W_i^{(1)}):1\leq i\leq n\}
\cup
\{\widehat g_0(W_j^{(2)}):1\leq j\leq m\}
\right) $$
denote the convex hull of the pooled transformed sample.
If a center $v_k$ lies outside $\mathcal C_{n,m}$, let
$v_k^\pi:=\Pi_{\mathcal C_{n,m}}(v_k)$ denote its Euclidean projection
onto $\mathcal C_{n,m}$. Since $\mathcal C_{n,m}$ is closed and convex,
$\|z-v_k^\pi\|\le \|z-v_k\|$ for every $z\in\mathcal C_{n,m}$, so
replacing $v_k$ by $v_k^\pi$ cannot increase the empirical $K$-means
criterion. Hence an empirical minimizer $\widehat{\mathbf v}$ may be
chosen so that $\widehat v_k\in\mathcal C_{n,m}$, $k=1,\ldots,K$.

Since $\widehat g_0(W_i)\in\mathcal X_{\widehat\varsigma}:=\Supp(W'\widehat\varsigma)$
almost surely, we have $\mathcal C_{n,m}\subseteq\conv(\mathcal X_{\widehat\varsigma})$,
and therefore $\widehat v_k\in\conv(\mathcal X_{\widehat\varsigma})$ for
$k=1,\ldots,K$, as required to apply Part~1.

By Lemma~\ref{lem:kmeans-vc} applied to the class
\(\{\ell_{\widetilde\varsigma,\mathbf v}:\widetilde\varsigma\in\mathcal N,\ \mathbf v\in B(0,R)^K\}\)
with $\ell_{\widetilde\varsigma,\mathbf v}(w):=\min_{1\leq k\leq K}\|w'\widetilde\varsigma-v_k\|^2$,
\begin{equation}
r_N
:=
\sup_{\widetilde\varsigma\in\mathcal N}
\sup_{\mathbf v\in B(0,R)^K}
\bigl|\widehat P_{n,m}\ell_{\widetilde\varsigma,\mathbf v}
-P\ell_{\widetilde\varsigma,\mathbf v}\bigr|
=
O_P\!\left(\sqrt{\frac{K\log K}{N}}\right).
\label{eq:kmeans-uniform-risk}
\end{equation}

Let \(\widehat{\mathbf v}\) minimize the empirical \(K\)-means criterion
based on \(\widehat g_0\), and let $\mathbf v^*_{\widehat g_0}$ minimize
the corresponding population criterion. We choose both configurations
in \(B(0,R)^K\). On \(\{\widehat\varsigma\in\mathcal N\}\), empirical optimality of $\widehat{\mathbf v}$ (over the compact admissible set) and the uniform empirical-process bound \eqref{eq:kmeans-uniform-risk} applied first to $\widehat{\mathbf v}$ and then to $\mathbf v^*_{\widehat g_0}$ give
\begin{align*}
D_{\widehat g_0}(\widehat{\mathbf v})
&\leq
\widehat D_{\widehat g_0}(\widehat{\mathbf v})+r_N\\
&\leq
\widehat D_{\widehat g_0}(\mathbf v^*_{\widehat g_0})+r_N\\
&\leq
D^*_{\widehat g_0,K}+2r_N.
\end{align*}
Therefore
\begin{equation}
D_{\widehat g_0}(\widehat{\mathbf v})
-
D^*_{\widehat g_0,K}
=
O_P\!\left(\sqrt{\frac{K\log K}{N}}\right).
\label{eq:kmeans-pop-near-opt}
\end{equation}
By the tuning condition,
$\sqrt{(K\log K)/N}=o(K^{-2})$ because $K^{5}\log K/N\to0$. Hence
\[
P\!\left(
D_{\widehat g_0}(\widehat{\mathbf v})
-
D^*_{\widehat g_0,K}
\leq
\frac{C_b}{4}K^{-2},\ 
\widehat\varsigma\in\mathcal N
\right)
\longrightarrow1.
\]
Applying Part~1 on this event yields
\begin{equation}
\min_{1\leq k\leq K}
P\!\left(\widehat g_0(W)\in\widehat V_k\right)
\geq
c_0K^{-2}
\label{eq:actual-cell-pop-balance}
\end{equation}
with probability tending to one.

\medskip
\noindent
\emph{Part 3: transfer to empirical counts.}
For each actual empirical Voronoi cell define $C_k
:=
\left\{
w:
\widehat g_0(w)\in\widehat V_k
\right\}$. To make the uniformity explicit, for
$\varsigma\in\mathcal N$,
$\mathbf v=(v_1,\ldots,v_K)$, and $1\le k\le K$, define
\[
C_k(\varsigma,\mathbf v)
:=
\left\{
w:
\|w'\varsigma-v_k\|
\le
\|w'\varsigma-v_\ell\|,
\quad\forall \ell\neq k
\right\},
\]
with the fixed tie-breaking rule used throughout. For each
$\ell\neq k$, the pairwise comparison can be written as
\[
\|w'\varsigma-v_k\|^2
\le
\|w'\varsigma-v_\ell\|^2
\iff
2w'\varsigma(v_\ell-v_k)
\le
\|v_\ell\|^2-\|v_k\|^2.
\]
Thus, for every value of $(\varsigma,\mathbf v)$, each pairwise
comparison is an affine halfspace in $w\in\R^q$, and
\[
C_k(\varsigma,\mathbf v)
=
\bigcap_{\ell\neq k}
\left\{
w:
2w'\varsigma(v_\ell-v_k)
\le
\|v_\ell\|^2-\|v_k\|^2
\right\}.
\]
Accordingly, the enlarged class $\mathcal C_K
:=
\left\{
C_k(\varsigma,\mathbf v):
\varsigma\in\mathcal N,\,
\mathbf v\in\mathcal V_K,\,
1\le k\le K
\right\}$
is contained in the class of intersections of at most $K-1$ affine
halfspaces in $\R^q$. Since affine halfspaces in $\R^q$ have VC
dimension $q+1$, and intersections of at most $K$ members of a
VC class of dimension $d$ have VC dimension $O(dK\log K)$, the VC dimension of $C_K$ is $O(qK\log K)$ (see Step~1a of the proof of Theorem \ref{thm:cons} for both
bounds and the corresponding references). In particular, this bound is uniform over both the projection
$\varsigma$ and the locations of the $K$ centers. Therefore
\[
\sup_{C\in\mathcal C_K}
\left|
\widehat P_{n,1}(C)-P(C)
\right|
=
O_P\!\left(
\sqrt{\frac{qK\log K}{n}}
\right)
=
o_P(K^{-2}),
\]
and similarly $\sup_{C\in\mathcal C_K}
\left|
\widehat P_{m,2}(C)-P(C)
\right|
=
o_P(K^{-2})$.
Since $\widehat P_{n,1}(C_k)=n_k/n$, $\widehat P_{m,2}(C_k)=m_k/m$,  $P(C_k)=P(\widehat g_0(W)\in\widehat V_k)$, combining these bounds with \eqref{eq:actual-cell-pop-balance} gives
\[
\min_k\frac{n_k}{n}
\geq
\frac{c_0}{2}K^{-2},
\qquad
\min_k\frac{m_k}{m}
\geq
\frac{c_0}{2}K^{-2}
\]
with probability tending to one. Taking \(c:=c_0/2\) proves the claimed empirical balance bound. $\hfill\square$

\subsubsection{Proof of Lemma~\ref{lem:kmeans-vc}.}

Write $K:=K_{n,m}$. Consider the class
\[
\mathcal L_K
:=
\bigl\{
\ell_{\widetilde\varsigma,\mathbf v}\,:\,
\widetilde\varsigma\in\mathcal N,\ 
\mathbf v\in B(0,R)^K
\bigr\},
\qquad
\ell_{\widetilde\varsigma,\mathbf v}(w)
:=
\min_{1\leq k\leq K}
\|w'\widetilde\varsigma-v_k\|^2.
\]
Since $W$ has compact support and $\mathcal N$ is compact, there exists
a finite constant $M$ (depending only on $\Supp(W)$, $\mathcal N$ and $R$)
such that $0\leq \ell_{\widetilde\varsigma,\mathbf v}\leq M$ on
$\Supp(W)$ for every 
$(\widetilde\varsigma,\mathbf v)\in\mathcal N\times B(0,R)^K$. So $\mathcal L_K$ is
uniformly bounded.

\medskip
\noindent
\emph{VC dimension.}
For every $t\in\R$,
\[
\bigl\{w:\ell_{\widetilde\varsigma,\mathbf v}(w)\leq t\bigr\}
=
\bigcup_{k=1}^{K}
\bigl\{w:\|w'\widetilde\varsigma-v_k\|^2\leq t\bigr\}.
\]
For a single $k$, $\|w'\widetilde\varsigma-v_k\|^2$ is a polynomial of
degree at most two in $w$. The vector space of such polynomials is
finite-dimensional (dimension bounded by a constant depending only on
$q=\dim(W)$), and its sublevel sets form a VC class with VC dimension
bounded by a constant $d_q$ depending only on $q$ 
\citep[Lemma~2.6.15]{VdV_Wellner}. By \citet[Lemma~3.2.3]{BlumerEtAl1989},
the class of unions of at most $K$ sets drawn from a VC class of
dimension $d_q$ has VC dimension at most $2d_qK\log(3K)$. Hence
$V(\mathcal L_K)\leq C_q\,K\log K$ for a finite constant $C_q$ depending
only on $q$.

\medskip
\noindent
\emph{Per-sample maximal inequality.} Let 
$\widehat P_n^{(1)}$ and $\widehat P_m^{(2)}$ denote the empirical
distributions of $W$ in $\S_1$ and $\S_2$ respectively. By the 
sharpened Vapnik-Chervonenkis inequality for uniformly bounded 
VC-subgraph classes (see, e.g., \citealp[Theorem~3.4]{BoucheronBousquetLugosi2005};  see also 
the proof of Theorem~3 in \citet{Linder2002} for the analogous 
squared $K$-means class),
\begin{equation}
\E\!\left[
\sup_{f\in\mathcal L_K}
\bigl|\widehat P_n^{(1)}f-Pf\bigr|
\right]
\leq
C\,M\,\sqrt{\frac{V(\mathcal L_K)}{n}}
=
O\!\left(\sqrt{\frac{K\log K}{n}}\right),
\label{eq:vc-max-S1}
\end{equation}
with $C$ a universal constant, and analogously for $\widehat P_m^{(2)}$.

\medskip
\noindent
\emph{Pooling.} The pooled empirical distribution satisfies
\[
\widehat P_{n,m}
=
\frac{n}{n+m}\,\widehat P_n^{(1)}
+
\frac{m}{n+m}\,\widehat P_m^{(2)},
\]
so for every $f\in\mathcal L_K$,
\[
\widehat P_{n,m}f-Pf
=
\frac{n}{n+m}\bigl(\widehat P_n^{(1)}f-Pf\bigr)
+
\frac{m}{n+m}\bigl(\widehat P_m^{(2)}f-Pf\bigr).
\]
Taking suprema over $f\in\mathcal L_K$, the triangle inequality gives
\[
\sup_{f\in\mathcal L_K}\bigl|\widehat P_{n,m}f-Pf\bigr|
\leq
\sup_{f\in\mathcal L_K}\bigl|\widehat P_n^{(1)}f-Pf\bigr|
+
\sup_{f\in\mathcal L_K}\bigl|\widehat P_m^{(2)}f-Pf\bigr|,
\]
because $n/(n+m)+m/(n+m)=1$. Combining with \eqref{eq:vc-max-S1} and
its $\S_2$ analogue, and using Markov's inequality to convert the
expectation bound into a rate in probability, we obtain
\begin{align*}
    \sup_{\widetilde\varsigma\in\mathcal N}
\sup_{\mathbf v\in B(0,R)^K}
\bigl|\widehat P_{n,m}\ell_{\widetilde\varsigma,\mathbf v}
-P\ell_{\widetilde\varsigma,\mathbf v}\bigr|
& =
O_P\!\left(\sqrt{\frac{K\log K}{n}}\right)
+
O_P\!\left(\sqrt{\frac{K\log K}{m}}\right) \\
& =
O_P\!\left(\sqrt{\frac{K\log K}{N}}\right).
\end{align*} The claim under deterministic weights bounded
away from zero and one follows by the same convex-combination argument
with the two weights in place of $n/(n+m)$ and $m/(n+m)$.  $\hfill\square$

\end{document}